\documentclass[11pt]{article}
\usepackage{mydef2col}
\usepackage{markArticle}

\title{SciPhy Reinforcement Learning for Portfolio Optimization}
\shorttitle{SciPhyRL for Portfolio Optimization}
\author{
\authorstyle{Igor Halperin\textsuperscript{1}\thanks{e-mail: \url{ighalp@gmail.com}} \,
and Andrey Itkin\textsuperscript{2}\thanks{e-mail: \url{aitkin@nyu.edu}}
}
\newline\newline
\textsuperscript{1}
\institution{Fidelity Investments, USA.}\\
\textsuperscript{2}
\institution{FRE Department, Tandon School of Engineering, New York University, USA.}
}

\date{\today}

\begin{document}

\maketitle

\lettrineabstract{This paper introduces a dynamic portfolio optimization framework for large institutional investors using Scientific Physics-Informed Reinforcement Learning (SciPhyRL). Formulated in continuous time over an extended state space that includes explicit cumulative costs, the approach leverages offline historical data to learn optimal, distribution-aware strategies. A core innovation reduces the optimization challenge to solving an HJB equation by projecting it onto observed trajectories as a pathwise Hamilton-Jacobi equation. This is solved directly from data using PINN in a single offline sweep, eliminating the need for traditional value or policy iteration. To make the method effective at practical short horizons, the control variable is recast from a continuous trading rate to a discrete target holding. This ensures signal-implied positions are reached immediately, while execution costs are evaluated against a microstructure-grounded quadratic price impact model. Evaluated on a $14$-asset ETF universe using an engineered oracle signal, the learned Gibbs policy yields substantial out-of-sample Sharpe ratio improvements over static and myopic baselines. The results demonstrate that the proposed framework successfully translates known signal quality into a robust, multi-period, and cost-aware allocation mechanism with strictly controlled volatility and turnover.
}

\section{Introduction}

Reinforcement learning (RL) offers a data-driven, learning-based framework for optimal control problems, \cite{Sutton_Barto_2018}. In contrast to classical control theory (e.g., \cite{Bertsekas_2019}), RL is characterized by its heavy reliance on data and a less stringent requirement for an explicit dynamics model. Today, the most prominent RL methods in fields like robotics and gaming are based on deep reinforcement learning, which combines RL with deep neural networks (NN) for flexible function approximators. While deep RL has demonstrated remarkable successes, such as achieving super-human performance in Go \cite{Silver_2017}, a significant drawback is its typically high demand for training data. This requirement often renders it impractical for applications with only small to moderate datasets, a common scenario in financial contexts.

While deep reinforcement learning normally works in a discrete-time formulation, for many applications of practical interest it proves very useful to consider a corresponding continuous-time version of dynamics. It is well known that a continuous time analysis brings valuable insights in classical control problems, where reducing a control problem to solving a partial differential equation (PDE) such as the Hamilton-Jacobi-Bellman (HJB) equation opens doors to a plethora of available numerical methods developed for PDEs. In particular, a burgeoning field of Scientific Machine Learning (SciML) is a new sub-field of machine learning research that applies deep neural networks (DNNs) to solving partial differential equations (PDEs) and other classical problems of applied mathematics and physics, \cite{SciML}.

A recent paper by one of the authors, \cite{Halperin2023}, introduces a new approach for offline reinforcement learning called SciPhyRL, which leverages methods from Scientific Machine Learning (SciML), particularly Physics-Informed Neural Networks (PINNs). This method addresses the problem of Distributional Offline Continuous-Time Reinforcement Learning (DOCTR-L), where the goal is to learn an optimal risk-sensitive policy $\pi$ from a fixed dataset generated by a behavioral policy
$\pi_0$. In this setting, the optimal value function is governed by a semilinear partial differential equation derived from the corresponding HJB equation.

In SciPhyRL, the equation is solved using data generated by the underlying dynamics rather than an explicit dynamics model. Both the optimal policy and the value function are then recovered from the equation and the data alone, solving the HJB equation directly from the behavioral data.

The Deep DOCTR-L algorithm, introduced in \cite{Halperin2023}, provides a general method for solving optimal control problems for a wide class of high-dimensional diffusion processes using offline data. A key feature of this approach is its foundation in distributional RL, which facilitates a computable assessment of learned policies, evaluating both their expected returns and the associated uncertainties. Furthermore, by directly encoding the HJB equation into the loss function, the method achieves a powerful PDE regularization. This regularization significantly reduces the amount of data required for training compared to conventional deep RL methods applied to similar high-dimensional control problems.

This paper applies the SciPhyRL methodology to the dynamic portfolio optimization problem, a classical challenge in quantitative finance originally pioneered by Merton \cite{Merton_alloc}. While Merton's foundational work permitted an analytical solution by restricting the portfolio to two assets, the methodology presented here excels in high-dimensional settings. Our approach can be qualified as a a goal-based portfolio optimization method \cite{Dasetal2020,Halperin2024}, and it shares some elements such as G-learning/MaxEnt with previous RL approaches
\cite{Dixon_Halperin_2021,Halperin_Liu_2022} for the discrete-time goal-based portfolio management.

We build upon the general framework introduced in \cite{Halperin2023}, but the present work is designed to be entirely self-contained. Throughout this paper, we use \emph{SciPhyRL} exclusively to denote the overarching methodology of solving an HJB-derived PDE from data. We reserve the term \emph{Deep DOCTR-L} for the specific neural-network solver used in \cite{Halperin2023}, which we do not employ here. Instead, this work is distinguished by its rich financial modeling and the introduction of model-specific solvers.

Specifically, we retain the approach of \cite{Halperin2023} that is based on projecting the full stochastic PDE onto the observed trajectory and relying on a path-wise Hamilton-Jacobi equation (HJ) solved with the PINN using offline data, while diverging from \cite{Halperin2023} in the following concrete respects:
\begin{enumerate}

\item \emph{A semi-analytical solver with convergence analysis.} Alongside the PINN approach, \cref{pideSolver} constructs a novel operator-splitting / generalized Duhamel / fixed-point scheme for the resulting PDE, featuring closed-form sub-steps and a Fast Gauss Transform. \cref{convergPicard} establishes sufficient conditions for the convergence of these iterations. We leave a low-dimensional numerical comparison between the PINN and semi-analytical solvers (intended to substantiate the control-variate idea in a combined method) for future work.

\item \emph{Quadratic-in-action dynamics.} Through the price-impact channel, both the running cost and the asset-price drift become quadratic in the action $\ba_t$ (\cref{cost_quadratic,mu_sigma_SDE}). This moves the problem outside the standard linear-quadratic regulator (LQR) class, which requires dynamics to be linear in the control.

\item \emph{A three-variable state with stochastic prices.} We work with the extended state $\by_t = (\bx_t, \bS_t, C_t)$, where $\bx_t$ denotes the number of shares held, $\bS_t$ is the vector of asset prices, and $C_t$ is the realized cost. By treating prices as endogenous stochastic drivers and explicitly tracking cumulative cost, rather than treating returns as exogenous inputs, this richer representation allows us to derive the corresponding likelihood ratio over the entire price path (\cref{sect_path_probs}).

\item \emph{A microstructure-grounded price-impact model.} \cref{appImpact} develops a multi-component impact model (incorporating temporary, permanent-convex, cross-impact, and memory terms). Its parameters are tied to observable market quantities, supported by an explicit program that reduces the parameter count from thirteen to five.
\end{enumerate}

The remainder of this paper is organized as follows. \Cref{sect_portfolio_optimization} presents our model of portfolio dynamics. In \cref{Soft_HJB_control}, we adopt a probabilistic distributional reinforcement learning framework to derive an HJB equation for the cost function, which we subsequently reduce to a nonlinear PDE that we call (the \emph{Hamilton-Induced} (HI) PDE). This PDE is then simplified by approximating the partition function using a Gaussian mixture model, resulting in another semilinear PDE. In
\cref{sec:deep_solver}, we develop two efficient approaches for solving this PDE.
The first is a semi-analytical solver, accompanied by a comprehensive convergence
analysis presented in \cref{pideSolver,convergPicard}. The second, our primary
approach, solves the PDE directly from offline data by projecting it onto
observed trajectories and reducing it to a first-order pathwise HJ equation.
Numerical experiments, based on a discrete approximation of the model described
in \cref{app:jump_control}, are reported in \cref{sect_Experiments}. Finally,
\cref{sect_Discussion,sect_Summary} provide discussion and concluding remarks.

\section{Background and related work} \label{sect_Related_work}

The problem of offline RL has been approached from multiple methodological directions. This section surveys key developments in the literature, focusing on methods that are most relevant to our proposed framework. We first review classical and contemporary offline RL algorithms, then examine the emerging use of Scientific Machine Learning for solving high-dimensional PDEs, which provides the foundational machinery for our approach.

\myparagraph{MaxEnt RL and G-learning.}
While classical approaches to problems of optimal control based on dynamic programming always deal with deterministic decision policies, reinforcement learning methods extend the class of possible decision policies by allowing for {\it stochastic} policies, where the decision variable $\ba_t $ at time $t$ when the system is in state $\bx_t$ is determined by some {\it probability distribution} (or probability density, for continuous actions) $ \pi_{\theta} ({\bf a}_t|\bx_t,t) $. Since any deterministic policy $\ba_{\theta} (\bx_t,t) $ can be represented as a Dirac-function probability density $ \pi_{\theta}({\bf a}_t |\bx_t, t) = \delta \left( {\bf a}_t - \ba_{\theta} (\bx_t,t) \right) $, stochastic policies naturally embed deterministic policies as a special case.

Maximum Entropy (MaxEnt) RL has emerged as a popular paradigm for stochastic optimal control in discrete time \cite{Levine_2018}. It formulates an entropy-regularized, sample-based approach to solving Bellman optimality equations, replacing the deterministic policies of classical dynamic programming with stochastic policies $\pi_{\theta}(\mathbf{a}_t | \bx_t, t)$. While many MaxEnt approaches use Shannon entropy for regularization, G-learning, \cite{G_learning}, regularizes by the KL-divergence $KL[\pi || \pi^{(0)}]$ from a prior policy $\pi^{(0)}$. This generalizes standard entropy regularization (recovered with a uniform prior) and allows incorporation of constraints or domain knowledge. In the following, we refer to both MaxEnt RL and G-learning collectively as \emph{MaxEnt RL}.

MaxEnt RL offers several advantages. It naturally quantifies uncertainty in optimal actions, provides a principled mechanism for exploration, enables offline learning from suboptimal demonstrations, and aligns closely with methods in inverse reinforcement learning. Computationally, it often simplifies optimization by replacing a max over actions with an integral.

\myparagraph{Reinforcement learning in continuous time.}
While reinforcement learning is most often formulated in discrete time and, for value-based RL methods, amounts to solving discrete-time Bellman optimality equations, continuous time RL operates with data-driven methods of solving continuous-time limits of Bellman optimality equations, which are known as the Hamilton-Jacobi-Bellman (HJB) equations \cite{Doya_2000}. HJB equations are nonlinear PDEs that should be solved numerically in a vast majority of use cases of practical interest.

While the classical HJB equation deals with deterministic policies, its probabilistic extension can be obtained by considering a continuous-time limit of MaxEnt RL discussed above. In particular, recent work has developed soft relaxations of the classical HJB equation that accommodate stochastic policies within the risk-neutral MaxEnt framework \cite{Wang_2020}. Extensions to practical continuous-time MaxEnt RL algorithms, specifically tailored for deterministic nonlinear systems via soft HJB relaxations, have also been explored \cite{Kim_2020}.

\myparagraph{Offline reinforcement learning.}
For many potential applications of RL, accessing a real or simulated environment to try different policies might be too expensive or unfeasible. Offline RL (also known as batch-mode RL) assumes that only a fixed dataset collected under some unknown behavioral policy is available. Developing reliable methods for such an offline RL setting has generated considerable interest in the recent literature   \cite{Levine_Offline_RL_2020}. The main challenge with offline RL is that while a fixed dataset may not cover some combinations of states and actions that produce high rewards, a model should rely on some sort of extrapolation in the state-action space. With Time-Difference (TD) methods commonly used in RL, rewards are evaluated at actions where there is no data, and propagated through the Bellman equation, potentially bootstrapping errors arising due to such extrapolation. Without proper safeguards, this can produce suggested state-actions combinations producing high rewards in a model, that would be very different from actual data used for training. This is known as the \emph{extrapolation problem} of offline RL \cite{Fujimoto_2019}. A number of approaches focused on constraining policies to not deviate too much from the actual behavioral data were recently proposed in the literature \cite{Fujimoto_2019, Siegel_2020,Levine_Offline_RL_2020}.

\myparagraph{Distributional RL and risk-sensitive RL.} Traditional RL focuses on policies that minimize the expected total return $Z_T$, as estimated at the current time $t$, without controlling for its higher moments (e.g., variance, skewness). Because these moments govern the risk (uncertainty) of future returns, such methods are often termed risk-neutral RL.

Distributional RL, \cite{Bellemare_2017} and risk-sensitive RL, \cite{Shen_2014} extend this paradigm by explicitly modeling higher-order statistics, or even the full conditional distribution, of $Z_T$. Risk-sensitive RL typically employs a specific risk measure (e.g., Conditional Value-at-Risk, CVaR, \cite{Shen_2014}), tailoring the algorithm accordingly. In contrast, distributional RL models the return distribution without pre-specifying a utility function; however, most distributional methods still derive a policy by maximizing the expected value of that distribution, thereby partially forfeiting the benefits of a full distributional perspective. A distributional offline actor-critic method for discrete-time problems was introduced in \cite{Urpi_2021}.

\myparagraph{Scientific Machine Learning and Physics-Informed Neural Networks.}
Scientific Machine Learning (SciML) is an emerging sub-field that applies deep neural networks (DNNs) to solving partial differential equations (PDEs) and other problems in applied mathematics and physics \cite{SciML}. By incorporating the structure of PDEs directly into their loss functions, these methods use neural networks to approximate solutions. This approach leverages automatic differentiation to provide mesh-free algorithms that can, in principle, mitigate the curse of dimensionality \cite{Poggio_2017, Grohs_2018}.

Encoding a PDE into a DNN offers several key advantages. First, it imposes the PDE as a structural constraint, effectively regularizing the solution. This enforces fundamental conservation laws (e.g., of energy, momentum, or probability) and ensures solution smoothness - a powerful form of “regularization by theory". Second, it reduces the original infinite-dimensional problem to a finite-dimensional optimization over network parameters, which can often be estimated with a moderate number of samples, leading to sample-efficient schemes. Third, and most critically for our context, a PDE-encoded model can predict system behavior for arbitrary inputs, unlike conventional DNNs limited to interpolating available data. This extrapolative capacity is especially valuable for offline RL, which must contend with the challenge of evaluating state-action pairs not present in the fixed dataset.

Beyond the general paradigm of using neural networks to solve PDEs, SciML methodologies can be categorized into three major classes (see \cite{Blechschmidt_2021, Beck_2020} for reviews). The first, known as Physics-Informed Neural Networks (PINNs), parameterizes the solution with a single network and encodes the PDE (including its derivatives via automatic differentiation) as a soft constraint imposed on a set of collocation points \cite{PINNs} (see also \cite{DeepXDE} for a review, applications, and a dedicated software implementation). The term \emph{physics-informed} here refers not solely to physical applications, but to the network’s direct use of derivative information, allowing it to capture notions of smoothness and local behavior through the structure of the Taylor expansion. PINNs perform well for complex, nonlinear PDEs in low dimensions \cite{PINNs,Blechschmidt_2021,cuomo2023scientific} but often become inefficient in higher-dimensional settings.

The remaining two classes incorporate additional mathematical structure to address high-dimensional problems more effectively. The second class exploits the Feynman–Kac formula, which expresses the solution of a linear backward Kolmogorov PDE as an expectation over trajectories of a forward stochastic differential equation (SDE). A neural network parameterizes the solution at a fixed time, and the method relies on Monte Carlo simulation of the SDE for training \cite{Beck_2018, Beck_2020}.

The third class targets semilinear and quasi-linear PDEs via forward-backward stochastic differential equations (FBSDEs). These methods provide pathwise approximations to the stochastic dynamics associated with the PDE. The Deep BSDE solver \cite{Han_2018} uses a network to approximate certain gradient terms, training on simulated paths to produce the solution at a fixed point in time and state, similar in spirit to the Feynman–Kac approach but distinct from PINNs, which approximate the full space-time solution. Extensions that yield solutions for arbitrary inputs have also been developed \cite{Raissi_2018,Zhang_2020,e2021algorithms}.

\section{A Framework for Portfolio Modeling} \label{sect_portfolio_optimization}

Consider a portfolio that was established at time $t_0 = 0$  and has been traded up to the current time $t > 0$. We aim to examine its evolution over the future interval $s \in (t, T]$, where $T$  is the portfolio's maturity while the period $s \in [0,t]$ covers the portfolio history. The portfolio consists of  $N+1$  assets with prices $S_{i,t}$ indexed by $i = 0,\ldots,N$. The first asset, $S_{0,t} = B_t = F e^{rt}$, is a risk-free money market account paying a constant instantaneous risk-free rate $r$ with a face value $F$. We assume that the evolution of the remaining $N$ risky assets with price vector $\bm{S}_t = [S_{1,t},\ldots,S_{N,t}]$ follows the stochastic differential equations (SDEs)
\begin{equation} \label{SDE_stocks}
dS_{i,t} = S_{i,t} \left[ r + \hat{r}(t,x_{i,t}) \right] dt + \sigma_i S_{i,t} \, dW_{i,t},
\end{equation}
where  $\bm{W}_t$ is a vector of correlated Brownian motions with $\langle dW_{i,t} \, dW_{j,t} \rangle = \rho_{i,j} dt$, $\sigma_i$ are the log-normal volatilities, $\hat{r}_i(t,x_{i,t})$  is the excess return which depends on the holdings $x_{i,t}$ (in shares) of a large portfolio manager (e.g., a pension fund, a mutual fund, or the agent whose behavior is modeling)\footnote{ For tractability, this framework assumes constant log-normal volatilities $\sigma_i$. To capture empirical smiles, a natural extension is to use a local volatility function $\sigma(t, S_t)$, see \cite{ItkinLocalVol} and references therein for detailed fitting approaches.}.

\myparagraph{Temporal evolution of portfolio value.}

Below we consider a continuous-time evolution of the portfolio and its assets. At the beginning of the interval $[t, t+\Delta t]$ (denoted $t_{+}$), the portfolio is rebalanced. Asset holdings are adjusted by $\mathbf{a}_t \Delta t$, giving the new instantaneous holdings:
\begin{equation} \label{eq:post_rebalance_values}
\bx_t^{+} = \bx_t + \ba_t \Delta t.
\end{equation}
Here, $\ba_t \in \mathbb{R}^N$ represents the rate of change of the portfolio holdings per unit time. In the context of RL, $\ba_t$ is known as the \emph{action} (or control) vector. Taking the limit $\Delta t \to 0$, we obtain the continuous-time relation $\dot{\bx}_t = \ba_t$. The total value of the portfolio at time $t$ is
\begin{equation} \label{portfolio_value}
\Pi_t = \sum_{i=0}^{N} x_{i,t} S_{i,t} = \sum_{i=1}^{N} x_{i,t} S_{i,t} + x_{0,t} B_t.
\end{equation}

The change in portfolio value $d\Pi_t$ can be expressed as
\begin{align} \label{dPi}
d\Pi_t &= \sum_{i=0}^{N} x_{i,t}\,dS_{i,t} + \sum_{i=0}^{N} S_{i,t}\,dx_{i,t}  - \mathrm{TC}(\bm{S}_t, \bm{a}_t) dt \\
&= \left[ \spr{\bm{S}_t, \ba_t} + \spr{\bm{S}_t \circ \bx_t, r \bm{1} + \hat{\bm{r}}_t } - \mathrm{TC}(\bm{S}_t, \bm{a}_t) \right] dt  + \spr{ {\bf 1}, (\bm{S}_t \circ \bx_t) \circ (\boldsigma d{\bm W}_t)}, \nonumber
\end{align}
where $\circ$ and $\spr{\cdot,\cdot}$ denote the element-wise (Hadamard) and dot products. The first term in the first line represents passive growth due to changes in asset prices. The second and third terms arise from active portfolio management: the second captures changes in asset holdings, while the third reflects transaction costs incurred over $dt$.

This equation motivates the definition of the continuous-time instantaneous cost function:
\begin{equation} \label{cost_fun}
c_t(\bm{S}_t, \bx_t, \ba_t) dt = - \CEt{d\Pi_t} + \Lambda \CEt{d\Pi_t^2}, \quad (\bm{S}_t, \bx_t, \ba_t) \in {\bm \Omega}_c: \Big\{ \mathbb{R^+}^N \times \mathbb{R}^N \times \calA\, |\, \ba_t \in \calA(\bx_t) \Big\},
\end{equation}
where $\calA(\bx_t)$ is the set of admissible actions given state $\bx_t$,
$\Lambda$ is a risk-aversion parameter, and $\mathcal{F}_t$ is the natural filtration of the Brownian motion at time $t$.

The cost function in \eqref{cost_fun} is a continuous-time analogue of the negative Markowitz reward. Minimizing this cost is therefore equivalent to maximizing the portfolio's risk-adjusted return over an infinitesimal time step $dt$. Furthermore, by rearranging \eqref{cost_fun} and integrating, the expected terminal portfolio wealth can be expressed as
\begin{equation} \label{W_T_exp}
\CEt{\Pi_T} =  \Pi_0 + \int_{0}^{T} \left( -  c_t(\bm{S}_t,\bx_t, \ba_t) dt + \Lambda \CEt{d\Pi_t^2} \right).
\end{equation}
Therefore, minimizing the expected total cost over $[0, T]$ is equivalent to maximizing the expected terminal wealth penalized by its accumulated variance, i.e., a \emph{risk-adjusted} terminal wealth rather than the raw terminal wealth.

Let us also specify the variance of the portfolio increments which follows from \Ito lemma
\begin{equation} \label{var_dPi}
(d\Pi_t)^2 = \sum_{i,j=1}^{N} \rho_{i,j} \sigma_i \sigma_j x_{i,t} x_{j,t} S_{i,t} S_{j,t} dt = (\bm{S}_t \circ \bx_t)^T \Sigma (\bm{S}_t \circ \bx_t) dt,
\end{equation}
where $\Sigma$ is the covariance matrix of asset returns.

The trading term $\spr{\bm{S}_t, \ba_t}$ in \eqref{dPi} is the net cash paid into
the risky assets at rebalancing. It is a transfer between the money market account
and the risky book, not a source of profit. Under a strictly self-financing
strategy the money market leg absorbs that transfer exactly, the two flows cancel,
and the term disappears from $d\Pi_t$. We do not impose the cancellation as an identity on the control. A hard self-financing rule confines the action to the hyperplane $\spr{\bm{S}_t, \ba_t} = 0$, on which a Gaussian mixture approach (see \cref{Soft_HJB_control} in more detail) places no mass, and it is precisely the
Gaussian mixture structure of the policy that makes the Gibbs step of
\cref{Soft_HJB_control} analytic. Small departures from the budget are also
economically meaningful in their own right, since they represent cash entering or
leaving the fund rather than an infeasible trade.

We therefore retain the term in the running cost and enforce the budget softly.
Soft here means that the constraint is priced rather than imposed. The admissible
action set stays full dimensional and every trade remains feasible. What changes is
the cost. The running cost carries an additional term, quadratic in the deviation
of the post-trade notional from its target, so a deviation is charged in proportion
to its square and the optimal control is pulled back toward the budget rather than
confined to it. Because the penalty is quadratic in the action, it enters the Gibbs
exponent as a rank one update of the quadratic coupling and leaves the Gaussian
mixture form of the optimal policy intact. Any mismatch that survives the penalty is
booked as an explicit external cash flow in the cumulative cost $C_t$, so that the
wealth identity holds exactly at every step. The penalty, its effect on the
couplings, its calibration, and the cash flow accounting are given in
\cref{sect_self_financing}.

Substituting \eqref{dPi} into \eqref{cost_fun}, the running cost is obtained in the
form
\begin{equation} \label{running_cost}
c(\bm{S}_t, \bm{x}_t, \bm{a}_t) =
- \left[  \spr{\bm{S}_t, \ba_t} + \spr{\bm{S}_t \circ \bx_t, r \bm{1} + \hat{\bm{r}}_t } - \mathrm{TC}(\bm{S}_t, \bm{a}_t) \right]  + \Lambda (\bm{S}_t \circ \bx_t)^T \Sigma (\bm{S}_t \circ \bx_t).
\end{equation}
The product $\bm{S}_t \circ \bx_t$ is the vector of dollar positions of the portfolio
assets. The realized cost $C_t$ of \eqref{running_cost1} is the discounted accumulation of \eqref{running_cost} along a trajectory, and it is carried as a component of the extended state in \cref{Soft_HJB_control}.

\myparagraph{Asset returns with price impact.}
Here, we consider the following model of expected asset returns $\hat{\bf r}_t$
\begin{equation} \label{r_t_one_more}
\hat{\bf r}_t  = {\bf w} {\bm \zeta}_t  + {\bf f} (\ba_t),
\end{equation}
where $r$ is the instantaneous risk-free rate, ${\bm \zeta}_t$ is a column vector of predictive signals, and ${\bf w}$ is a matrix of coefficients. The final term, ${\bf f}(\ba_t)$, represents the price impact of the trades $\ba_t$. This is a crucial consideration because our setting involves a large institutional portfolio whose trades $\ba_t \Delta t$ may be substantial enough to move the market. Thus, the vector of equity returns is defined as a combination of the risk-free rate, the weighted predictors, and this market impact factor.

Although this paper models the portfolio of a single large institutional investor, we can reasonably assume that its investment philosophy is similar to that of its peers, leading to positive correlations in their trading activity. This collective behavior can create price impacts that are initially positive but may later reverse. For instance, stocks with higher momentum tend to attract more investors. In the short run, this concentrated buying can further inflate returns, reinforcing the momentum. However, this effect is temporary. Once a stock becomes "saturated" or "crowded", meaning too many participants hold large simultaneous positions, further allocations are likely to yield diminishing returns.

We therefore adopt a price impact model that captures the "dumb money" effect described by \cite{Frazzini_2008}. This theory posits that while initial inflows into a stock temporarily boost its expected returns, a sustained buildup of inflows (or "crowding") eventually leads to diminishing long-term returns. To capture such saturation effects, the impact function should depend on the history of past inflows. A model of this form is described in \cref{appImpact}.

This model proposes a comprehensive, multi-component framework for estimating the price impact of trading activity across multiple assets. The core equation decomposes the total return impact on an asset into four key parts: (1) a \emph{temporary linear self-impact} term, capturing immediate costs like the bid-ask spread and short-term liquidity consumption; (2) a \emph{permanent convex (quadratic) self-impact} term, designed to capture the non-monotonic, long-term price effects from sustained order flow; (3) a \emph{linear cross-impact} term, where trading in one asset affects the price of another, weighted by relative liquidity; and (4) a \emph{memory impact} term, which incorporates the decaying influence of past trades, capable of modeling both momentum (reinforcement) and increased market sensitivity (magnitude) effects. The model's parameters, such as the linear impact coefficient, cross-impact matrix, and memory persistence, are not set as simple constants but are instead structurally derived from observable market microstructure variables like bid-ask spread, volatility, average daily volume (ADV), market capitalization, and turnover. This design allows the model to dynamically reflect differences in liquidity and market structure between assets.

Within the framework of this model we obtain the following representation
\begin{equation} \label{quadratic_impact_vec}
f({\bm a}_t) = {\bm f}_t^{(0)} + {\bm f}_t^{(1)} \circ \ba_{t} + {\bm f}_t^{(2)} \circ \ba_t^{\circ 2},
\end{equation}
where $\ba_t^{\circ 2} = \ba_t \circ \ba_t$ and ${\bm f}_t^{(0)}, {\bm f}_t^{(1)}, {\bm f}_t^{(2)}$ are defined in \eqref{fim_terms}.

In the data-generating dynamics the convex term retains the sign factor of \eqref{impactModel}. The plain square in \eqref{quadratic_impact_vec} is used only inside the analytic Gibbs step, where a quadratic-in-a exponent is required.

\myparagraph{Cost model.}

To complete the cost model specification, the transaction cost term $\mathrm{TC}(\bm{S}_t, \ba_t)$ appearing in \eqref{dPi} is defined as
\begin{equation} \label{TC_fun}
\mathrm{TC}(\bm{S}_t,\ba_t) = \eta \, \bm{S}^T_t \ba_t^{\circ 2},
\end{equation}
with $\eta$ a constant parameter. The cost is therefore homogeneous in $\bm{S}_t$. This linear scaling property ensures that costs move proportionally with prices - a widely adopted assumption in the literature.

With this definition, \eqref{running_cost} after applying some matrix algebra, can be re-written in the form
\begin{equation} \label{running_cost1}
c(\bm{S}_t, \bm{x}_t, \bm{a}_t) =
\spr{\bm{S}_t, -\left(\ba_t - \eta \, \ba_t^{\circ 2} \right)}
- \spr{\bm{S}_t \circ \bx_t,\; r{\bm{1}} + \hat{\bm{r}}_t}
+ \Lambda \bm{S}_t^T (\bx \bx^T \circ \Sigma) \bm{S}_t.
\end{equation}
Here the first inner product collects the terms driven by the trading rate (the cash exchanged at rebalancing and the transaction cost), while the second is the expected return earned on the held position, weighted by the holdings $\bx_t$. Combining \cref{running_cost,r_t_one_more,quadratic_impact_vec} with this specification, the cost $c(\bm{S}_t,\bx_t, \ba_t)$ can be written as a quadratic form of $\ba_t$
\begin{gather} \label{cost_quadratic}
 c(\bm{S}_t,\bx_t, \ba_t) = {\bf C}_0(\bm{S}_t,\bx_t) + \bm{C}_1(\bm{S}_t,\bx_t) \ba_t +  {\bf C}_2 (\bm{S}_t,\bx_t) \ba_t^{\circ 2}, \\
 \begin{align}
 {\bf C}_0(\bm{S}_t,\bx_t) &=  -\spr{\bm{S}_t \circ \bx_t,\; r{\bm 1} + {\bf w} {\bm \zeta}_t + {\bm f}_t^{(0)}}
 + \Lambda \bm{S}_t^T (\bx \bx^T \circ \Sigma) \bm{S}_t, \nonumber \\
  {\bf C}_1(\bm{S}_t,\bx_t) &= - \left[ \mathbf{S}_t \circ \left(1 + {\bm f}_t^{(1)} \circ \bx_t \right) \right]^T, \qquad
 {\bf C}_2(\bm{S}_t,\bx_t) =  \text{diag}\!\left(\eta \bS_t - \bm{S}_t \circ \bx_t \circ {\bm f}_t^{(2)}\right). \nonumber
\end{align}
\end{gather}

Also, to ease notation, further we denote the cost function as $c(\bm{S}_t, \bx_t, \ba_t)$ rather than of $c_t(\bm{S}_t, \bx_t, \ba_t)$, even though it depends explicitly on time $t$ through its dependence on signals ${\bm \zeta}_t$ and past flows in \eqref{cost_quadratic}.

With the chosen impact function in \eqref{quadratic_impact_vec}, the SDE in \eqref{SDE_stocks} takes the form
\begin{equation} \label{SDE_2}
d\bm{S}_t = \bm{S}_t \circ \left[ \boldmu(\bx_t, \ba_t) dt + \boldsigma \circ d \bm{W}_t \right],
\end{equation}
where the drift is determined as
\begin{gather} \label{mu_sigma_SDE}
\boldmu(\bx_t, \ba_t) = \boldmu_0  + \boldmu_1  \left( \bx_t \right) \circ \ba_t + \boldmu_2  \circ \ba_t^{\circ 2}, \\
\begin{align*}
\boldmu_0 \left(\bx_t \right) &= r {\bf 1} + {\bf w} {\bm \zeta}_t + {\bf f}_t^{(0)}, \qquad \boldmu_1 \left(\bx_t \right) =  {\bf 1} +  {\bf f}_t^{(1)} \circ \bx_t, \qquad \boldmu_2 \left(   \bx_t \right) = {\bm f}_t^{(2)}. \nonumber
\end{align*}
\end{gather}

Note that while our cost function \eqref{cost_quadratic}) is quadratic in the actions $\ba_t$, the drift $\boldmu ( \bx_t, \ba_t)$ in \eqref{SDE_2} is also quadratic in $\ba_t$ due to the impact function ${\bf f} (\ba_t)$. Consequently, our problem setting differs from the classical Linear Quadratic Regulator (LQR), \cite{anderson_moore_1990}, which requires the dynamics to be linear in the control variable. A further distinction in our formulation is that we aim to control the entire distribution of the total cost, rather than merely its expectation, as in standard LQR.

\section{Portfolio optimization via HJB equation}  \label{Soft_HJB_control}

We consider a system with a running cost $c(\bm{S}_t,\bx_t, \mathbf{a}_t)$ as in \eqref{cost_quadratic}. At any time $0 < t \leq T$, the total cumulative cost of the portfolio from time $0$ to $T$ is a sum of the future cost from time $t$ onward, discounted back to time $t$, and the realized cost $C_t$ with
\begin{align} \label{Z_T}
\calZ_t &= \int_{t}^{T} e^{-r (k-t)}  c(\bm{S}_k,\bx_k, \mathbf{a}_k) dk, \qquad
C_t =  \int_{0}^{t} e^{-r (k-t)} c(\bm{S}_k,\bx_k, \ba_k)dk.
\end{align}
In the absence of constraints, such as budget, leverage, trading, and position limits, and for a non-self-financing strategy, we have $C_t \in \mathbb{R}$. The discount factor $e^{-r(k-t)}$ converts the cost incurred at a future time $k$ to its equivalent present value at time $t$. Consequently, $\calZ_t$ is a random variable, as it depends on future price-state-action triples $(\bm{S}_k,\bx_k, \ba_k), \, t < k \leq T$. At the terminal time $t = T$, the random variable $C_T$ becomes fully observed and $\calZ_T = C_T$.

This formulation of the total cost motivates the definition of an extended state vector $\mathbf{y}_t = (\bm{S}_t,\mathbf{x}_t, C_t)$. The inclusion of the running cost $C_t$ renders the dynamics of the joint process $(\mathbf{y}_t, \mathbf{a}_t)$ Markovian, a technique commonly employed for path-dependent derivatives such as Asian options.

\subsection{Probabilistic distributional RL formulation} \label{sect_cont_time_RL}

Traditional RL optimizes the expected total cost
\begin{equation} \label{totalCost}
J(\mathbf{y}_t,t) = \mathbb{E} \left[ \calZ_t \mid \mathbf{y}_t \right].
\end{equation}
Our approach, however, is inspired by distributional and risk-averse RL, \cite{dabney2018distributional}, which seek to manage the entire distribution of returns, not just its mean. Consequently, we focus on the dynamics of the conditional distribution of the total return $\calZ(t)$ as seen at time $t$ given a state $ {\bf y}_t = (\bx_t, \bS_t, C_t)$
\begin{equation} \label{P_pi}
P^{\pi} (z| {\bf y}_t, t) := P^{\pi} \left( \left. \calZ(t) = z \right|  {\bf y}_t, t \right),
\end{equation}
where the superscript $\pi$ emphasizes the dependence of this conditional probability on the policy $\pi$ (see below). Given that $\calZ(T) = C_T$, we obtain the terminal condition for $P^{\pi}(z | \mathbf{y}_t, t)$
\begin{equation} \label{P_pi_T}
P^{\pi}(z | \mathbf{y}_T, T) = \delta(z - C_T).
\end{equation}

Note that, unlike the expected value $\mathbb{E} \left[ \calZ(t) \mid \mathbf{y}_t \right]$, the dependence of the conditional probability $P^{\pi}(z | \mathbf{y}_t, t)$ on $C_t$ is non-trivial (in fact, it obeys a backward Kolmogorov equation, see below).

With this approach, instead of \eqref{totalCost}, the cost of policy $ \pi $ is determined by the following cost functional
\begin{equation} \label{J_t}
J_0^{\pi} ({\bf y}_t,t) = \int U(z) P^{\pi} (z | {\bf y}_t, t) dz = \int dz\, U(z) \int d\ba_t\, \pi(\ba_t| {\bf y}_t) P^{\pi} (z | {\bf y}_t, \ba_t, t).
\end{equation}
The function $U(z)$ is a convex 'negative utility' that quantifies the undesirability of the cumulative cost $z$. This cost is accrued along a realized trajectory of state-action pairs over the time horizon $[t,T]$.

The choice of a negative utility function dictates the risk sensitivity of the policy. The linear case $U(z) = z$ recovers the standard risk-neutral RL formulation, which minimizes the expected total cost. To model risk aversion, one employs a convex nonlinear function; a canonical example is the quadratic utility \begin{equation} \label{quadratic_U}
U(z) = (z - z_{\rm tg})^2,
\end{equation}
where the target cost $z_{\rm tg}$ represents a benchmark terminal cost. In our framework the terminal cost $z = C_T$ is the accumulated running cost of \eqref{Z_T}, so that profit corresponds to \emph{negative} cost. A portfolio with a target risk profile on the Capital Market Line (CML) \cite{elton2020modern} reaches the wealth level $\Pi_{\rm tg} = \Pi_0 \exp(r_{\rm tg} T)$; since $z$ is a cost, the matching target is the negative of that gain:
\begin{equation} \label{z_tg_CML}
z_{\rm tg} = \Pi_0 \left(1 - \exp(r_{\rm tg} T)\right), \qquad r_{\rm tg} = r_f + \left(\mathbb{E}[r_m] - r_f\right) \frac{\sigma_m}{\sigma_{\rm tg}}.
\end{equation}
This is negative for $r_{\rm tg} > 0$ (a profit target). In \eqref{z_tg_CML} $r_f$ is the risk-free rate, while $\mathbb{E}[r_m]$ and $\sigma_m$ denote the expected return and volatility of the market portfolio, respectively. The target growth rate $r_{\rm tg}$ is determined by the target volatility $\sigma_{\rm tg}$, which is a free parameter expressing the modeler's risk appetite: it selects the point on the CML that the agent targets, with $\sigma_{\rm tg} = \sigma_m$ recovering the market portfolio. We emphasize that $\sigma_{\rm tg}$ here plays a different role from, and should not be conflated with, the variance-penalty (risk-aversion) parameter $\Lambda$ in the single-step cost function \eqref{cost_fun}; the two enter the problem through distinct channels (the terminal target versus the running cost). This anchors the agent's objective to a theoretically sound financial benchmark.

The minimum of this utility function, attained at $z = z_{\rm tg}$, represents the successful achievement of the risk-adjusted target wealth. The quadratic form is symmetric, so it penalizes overshooting the target as much as falling short of it, and therefore caps the upside by construction. This is a deliberate property of a tracking mandate rather than an oversight, its quantitative consequences for the learned policy are examined in \cref{results}, and an asymmetric alternative that removes the cap is proposed in \cref{sect_Summary}. It is important to note that our methodological approach is general. The specific choice of the negative utility function $U(z)$ is flexible; any convex function can be used and will only alter the terminal condition of the problem, as discussed later.

As in MaxEnt RL, \cite{dabney2018distributional,haarnoja2018softicml} we prefer to work with a regularized version of the cost functional
\begin{equation} \label{J_t_G}
J^{\pi}({\bf y}_t,t) =  J_0^{\pi} ({\bf y}_t,t) +  \frac{1}{\beta} \mathcal{R}^{\pi}(\bx_t,t),
\end{equation}
where $ \mathcal{R}^{\pi}(\bx_t,t)  $ is a time-integrated expected Kullback-Leibler (KL) divergence of policy $\pi$
\begin{equation} \label{regularizer}
\mathcal{R}^{\pi}(\bx_t,t) = \int_{t}^{T} e^{-r (s-t)} \mathbb{E}_{x,t} \left[ \mathcal{D}_{KL}[\pi || \pi_0] (\bx_s,s) \right] ds, \enspace
\mathcal{D}_{KL}[\pi || \pi_0] (\bx_s,s) = \int \pi(\ba_s | \bx_s) \log \frac{ \pi(\ba_s| \bx_s)}{ \pi_0 (\ba_s | \bx_s) } d \ba_s.
\end{equation}
The second term in \eqref{J_t_G} acts as a regularizer, penalizing significant deviations of the policy $\pi$ from the behavioral prior $\pi_0$. The strength of this penalty is controlled by the 'inverse temperature' parameter $\beta > 0$. Although a specific method for constructing $\pi_0$ will be introduced later, the relationships derived in this section hold for an arbitrary choice of $\pi_0$.

It is worth mentioning that parameter $\beta$ is a crucial hyperparameter that controls the fundamental trade-off between two competing objectives: a) maximizing the expected return (the standard RL goal), and b) maximizing the policy's entropy (the "randomness" or exploration bonus). The {\it temperature} analogy comes from statistical mechanics and thermodynamics, where it describes the behavior of particles in a system. When the temperature is high, particles have high kinetic energy, the system is disordered and random, and the particles explore all states freely. When the temperature is low, particles have low kinetic energy, the system is ordered and deterministic, and the particles settle into the lowest energy states.

This parameter has the same meaning as the one that comes into play when the hard minimum in an HJB equation is replaced with a soft minimum better known as the LogSumExp or softmax operator. When $\beta$ is high, soft HJB recovers the original hard HJB/Bellman equation, and this is the case in our context as well.

The optimal cost function $ J ({\bf y}_t,t) $ satisfies the following relation
\begin{equation} \label{J_t_opt}
J({\bf y}_t,t) = \min_{\pi} \left\{ J_0^{\pi} ({\bf y}_t,t) +  \frac{1}{\beta} \mathcal{R}^{\pi}(\bx_t,t) \right\}
\end{equation}
subject to the terminal condition
\begin{equation} \label{J_T}
J({\bf y}_T,T) = J^{\pi}({\bf y}_T,T) =  U(C_T).
\end{equation}

Note that while the terminal condition for both the optimal cost functional $J(\mathbf{y}_T, T)$ and the fixed-policy functional $J^{\pi}(\mathbf{y}_T, T)$  is given by $U(C_T)$, the {\it realized} value of its argument $C_t$  is policy-dependent. Consequently, in simulations with identical random disturbances but different action sampling $\mathbf{a}_t \sim \pi(\cdot| \bx_t)$ the realized cost $C_t$  will differ between a fixed policy $\pi$  and the optimal policy $\pi_{\star}$.

The agent's objective is to find the optimal policy $\pi_{\star}$  by computing the value of the optimal cost functional $J(y_0, 0)$ at time $t = 0$. Our method differs from existing approaches in two key ways. First, unlike risk-averse RL methods that apply a nonlinear utility function to each intermediate cost, e.g., \cite{Shen_2014}, our model applies the convex {\it cost price} $U(z)$ in \eqref{J_t} solely to a single terminal value $C_T$. Second, unlike the distributional RL approach, which operates on a distributional version of the Bellman equation, \cite{Bellemare_2017}, we adopt a probabilistic framework that directly models the {\it probabilities} of random total returns over time. Consequently, while traditional RL focuses on control {\it under} uncertainty, our approach enables control {\it of} uncertainty itself.

\subsection{Backward Kolmogorov PDE} \label{sect_BKE}

Further on, for notational simplicity, we will drop the subscript $_t$ when conditioning
on the known values at time $t$, namely, $\mathbf{y}_t$ , $\mathbf{x}_t$ and $C_t$.
As the dynamics are Markov in the pair $ ({\bf y}, \ba)$ (or, $(\bx, \ba, C)$), the conditional probability in \eqref{P_pi} can be expressed in terms of its values at a future time $s$ such that $t <  s <  T$ by inserting an integral over intermediate states ${\bf y}_s, \ba_s$
\begin{align} \label{second_eq}
 P^{\pi} \left( \left. \calZ(s) = z\right| {\bf y}, \ba, t \right) &=
 \int d{\bf y}_s d \ba_s P^{\pi} ( \calZ(s) = z, s, {\bf y}_s, \ba_s | {\bf y}, \ba, t)  \\
 &= \int d{\bf y}_s d \ba_s P^{\pi} ( \calZ(s) = z | {\bf y}_s, \ba_s)  P^{\pi} ( {\bf y}_s, \ba_s, s | {\bf y}, \ba, t). \nonumber
\end{align}
Plugging this relation into \eqref{J_t}, we obtain
\begin{align} \label{J_t_1}
J_0^{\pi} ({\bf y},t) &= \int dz\, U(z) \int d\ba \int d{\bf y}_s\, d\ba_s\, \pi(\ba| {\bf y}) P^{\pi} (z| {\bf y}_s, \ba_s)  P^{\pi} ( {\bf y}_s, \ba_s, s | {\bf y}, \ba, t) \\
&=  \int d{\bf y}_s\, J^{\pi} ({\bf y}_s,s) \int d\ba\,  \pi(\ba| {\bf y}) P^{\pi} ( {\bf y}_s | {\bf y}, \ba) = \int d{\bf y}_s\,  J_0^{\pi} ({\bf y}_s,s) P^{\pi} ( {\bf y}_s | {\bf y}). \nonumber
\end{align}

The \eqref{J_t_1} relates the cost functional $J_0^{\pi} ({\bf y},t)$ to its future values at time $s$. Using \eqref{J_t_G}, we can express this relationship in terms of the regularized  cost functional $J^{\pi} ({\bf y},t)$
\begin{equation} \label{J_t_pi}
J^{\pi} ({\bf y},t) - \frac{1}{\beta} \mathcal{R}^{\pi}(\bx,t)
 = \int d{\bf y}_s  P^{\pi} ( {\bf y}_s | {\bf y}) \left( J^{\pi} ({\bf y}_s,s) -   \frac{1}{\beta} \mathcal{R}^{\pi}(\bx_s) \right).
\end{equation}

To derive the continuous-time limit, we set $s = t + \Delta t$ for a small time step $\Delta t$ and expand the integrand in a Taylor series
\begin{align} \label{J_t_2}
J^{\pi} ({\bf y},t) &- \frac{1}{\beta} \mathcal{R}^{\pi}(\bx,t)
 = \int d{\bf y}_s  P^{\pi} ( {\bf y}_s | {\bf y})  \calA \left[ J^{\pi} ({\bf y}_s,s) - \frac{1}{\beta} \mathcal{R}^{\pi}(\bx_s,s) \right]_{s \uparrow t} + O\left(\Delta t^2\right), \\
 \calA &= \bm{1} + \fp{}{s} \Delta t + \fp{}{C_s} \Delta C + \fp{}{\mathbf{x}_s} \Delta \mathbf{x} + \fp{}{\mathbf{S}_s} \Delta \mathbf{S} + \frac{1}{2} \sop{}{\mathbf{S}_s}\circ \left(\Delta \mathbf{S}\right)^2, \nonumber
\end{align}
where the state vector is defined as ${\bf y} = (\mathbf{x}, \mathbf{S}, C)$. The operational terms for the holdings and price increments are given by
\begin{align}
\fp{J}{\mathbf{x}} \circ \Delta \mathbf{x} &= \sum{i} \fp{J}{x_i} \Delta x_i, \qquad \Delta \mathbf{x} = \mathbf{a} \Delta t + O(\Delta t^2), \nonumber \\
\fp{J}{\mathbf{S}} \circ \Delta \mathbf{S} &+ \frac{1}{2} \sop{J}{\mathbf{S}} \circ (\Delta \mathbf{S})^2 = \sum_i S_i \mu_i \fp{J}{S_i} \Delta t + \frac{1}{2} \sum_{i,j} S_i \Sigma_{ij} S_j \mop{J}{S_i}{S_j} \Delta t,
\end{align}
with $\Delta \mathbf{x}$ representing the controlled change in holdings and the price increments derived from the SDE \eqref{SDE_2}. The increment $\Delta C$ is obtained from \eqref{Z_T} as
\begin{equation} \label{delta_C_t}
\Delta C =  \left[c(\mathbf{x}, \mathbf{S}, \mathbf{a}) + r C \right] \Delta t + O \left(\Delta t^2 \right),
\end{equation}
where $c(\mathbf{x}, \mathbf{S}, \mathbf{a})$ now accounts for the cost of changing holdings at current market prices $\mathbf{S}$.

We define policy-dependent {\it effective} drift, volatility, and cost functions as follows:
\begin{alignat}{2} \label{mu_sigma_pi_formal}
\bar{\ba}(\bx, \mathbf{S}, \pi_t) &= \int \ba \, \pi(\ba | \bx, \mathbf{S}) \, d\ba,
&\quad
\bar{\boldmu}_S(\bx, \mathbf{S}, \pi_t) &= \int \boldmu_S(\mathbf{S}, \ba) \, \pi(\ba | \bx, \mathbf{S}) \, d\ba, \\
\bar{\boldsigma}_S^2(\bx, \mathbf{S}, \pi_t) &= \int \boldsigma_S^2(\mathbf{S}, \ba) \, \pi(\ba | \bx, \mathbf{S}) \, d\ba,
&\quad
\bar{c}(\bx, \mathbf{S}, \pi_t) &= \int c(\bx, \mathbf{S}, \ba) \, \pi(\ba | \bx, \mathbf{S}) \, d\ba. \nonumber
\end{alignat}

Note that \eqref{J_t_2} requires an expectation over the action $\ba$ under the policy $\pi$. Consequently, in the infinitesimal limit $\Delta t \rightarrow 0$, the effective dynamics are governed by a coupled system where the dependence on $\ba$ has been integrated out:
\begin{align} \label{CT_Langevin_controlled_pi}
d\bx_t &= \bar{a}(\bx_t, \mathbf{S}_t, \pi_t) dt, \qquad
d\mathbf{S}_t = \text{diag}(\mathbf{S}_t) \left[ \bar{\boldmu}_S(\bx_t, \mathbf{S}_t, \pi_t) dt + \bar{\boldsigma}_S(\bx_t, \mathbf{S}_t, \pi_t) d {\bf W}_t \right].
\end{align}

Taking the continuous-time limit $\Delta t \rightarrow 0$ in \eqref{J_t_2} via \crefrange{delta_C_t}{CT_Langevin_controlled_pi}, we derive the backward PDE for the cost functional $J^{\pi}$
\begin{align} \label{Kolmogorov_J}
0 &= \fp{J^\pi}{t} + \left(\bar{c}(\bx, \mathbf{S}, \pi_t) + r C \right) \fp{J^\pi}{C} + \sprL{\bar{\ba}(\bx, \mathbf{S}, \pi_t); \fp{J^\pi}{\bx}}
+ \sprL{D_S \bar{\bm{\mu}}_S(\bx, \mathbf{S}, \pi_t); \fp{J^\pi}{\mathbf{S}}} \\
&+ \frac{1}{2} \tr \left(\Xi_S \nabla\mathbf{S}^2 J^\pi\right) + \frac{1}{\beta} \mathcal{D}_{KL}[ \pi_t || \pi_0 ](\bx, \mathbf{S}, t). \nonumber
\end{align}
Here, the matrix $\Xi_S = D_S \bar{\boldsigma}_S^2 D_S$ is symmetric, $D_S = \text{diag}(\mathbf{S})$ and $\nabla\mathbf{S}^2$ is the Hessian matrix with respect to prices. The diffusion term can be represented in explicit form as
\begin{align}
\tr \left(\Xi_S \nabla_\mathbf{S}^2 J^\pi\right) &= \sum_{i=1}^n \sum_{j=1}^n \Xi_{S,ij} \frac{\partial^2 J^\pi}{\partial S_i \partial S_j} = \sum_{i=1}^n \sum_{j=1}^n S_i [\bar{\boldsigma}_S^2]_{ij} S_j \frac{\partial^2 J^\pi}{\partial S_i \partial S_j},
\end{align}
where $[\bar{\boldsigma}_S^2]_{ij}$ represents the $ij$-th entry of the policy-averaged covariance matrix of relative returns.

The conditional probability $P^{\pi}(z|{\bf y}, t)$ satisfies a similar backward Kolmogorov PDE
\begin{equation} \label{Kolmogorov_P}
0 = \fp{P^\pi}{t} + \left(\bar{c} + r C \right) \fp{P^\pi}{C} +
\sprL{\bar{\ba}; \fp{P^\pi}{\bx}} + \sprL{D_S\bar{\bm{\mu}}_S; \fp{P^\pi}{\mathbf{S}}} + \frac{1}{2} \tr \left(\Xi_S \nabla\mathbf{S}^2 P^\pi\right).
\end{equation}

Given a policy $\pi$, \eqref{Kolmogorov_P} can be solved with the terminal
condition \eqref{P_pi_T} to obtain the conditional distribution $P^{\pi}(z | {\bf
y}, t)$ of the cumulative return $\calZ_T$. This distribution quantifies the
initial uncertainty in the value of $\calZ_T$. The quality of control obtained
using an optimal policy $\pi_\star$ instead of a given policy $\pi$ can be judged
by comparing how both policies impact the solution of the backward PDE
\eqref{Kolmogorov_P} at $t = 0$.

\subsection{The HJB and HI equations} \label{sect_Soft_HJB_equation}

The optimal policy $\pi_\star$ is determined by the optimal value function $J^\star(\mathbf{y}, t)$. We define $J^\star$ by maximizing the expression for the value function in \eqref{Kolmogorov_J} over all admissible policies $\pi$. A key consequence of this definition is the HJB equation, which states that at the optimum, the time derivative of $J^\star$ is precisely the best rate achievable by any control. For ease of notation in the subsequent analysis, we will refer to the optimal value function simply as $J(\mathbf{y}, t)$ (so, by dropping the asterisk)
\begin{align} \label{HJBh}
- \fp{J}{t} -  r C  \fp{J}{C} &=  \min_{\pi} \int d\ba\, \pi(\ba| \bx, \bS) \Bigg[ c(\bx, \bS, \ba ) \fp{J}{C} + D_S \boldmu^T_S(\bx,\bS, \ba) \fp{J}{\bS}
+ \ba \fp{J}{\bx} \\
 &+  \frac{1}{2} \tr \left(\Xi_S \nabla^2_\bS J^\pi\right)
 +  \frac{1}{\beta} \log \frac{ \pi(\ba| \bx, \bS)}{ \pi_0(\ba|\bx, \bS)} \Bigg]. \nonumber
 \end{align}

It is important to underline, that in this paper we consider stochastic policies within continuous action spaces. Consequently, a policy $\pi({\bm a}_s|{\bm x}_s, \bS_s)$ is a probability density function defining the agent's behavior and representing the likelihood of taking action ${\bm a}_s$ in state $s$. A fundamental requirement is that the policy must be normalized, meaning the integral over the action space must satisfy $\int \pi({\bm a}_s|{\bm x}_s, \bS_s) d{\bm a}_s = 1$.

Therefore, the minimization problem in \eqref{HJBh} must be solved subject to this
normalization constraint. Applying the method of Lagrange multipliers (see, e.g.,
\cite{boyd2004convex} among others) and omitting the lengthy algebraic details, we
obtain the optimal policy $\pi^*$ that minimizes \eqref{HJBh} analytically in terms of
$J$ and its derivatives
\begin{equation} \label{pi_opt}
\pi^*(\ba| \bx, \bS) = \frac{\pi_0(\ba| \bx, \bS)}{Z(J,\bx, \bS)} \, e^{-\beta \calH},
\qquad
Z(J,\bx, \bS) = \int d \ba \, \pi_0(\ba| \bx, \bS) \, e^{-\beta \calH},
\end{equation}
where the Hamiltonian $\calH$ is the scalar-valued function of the state, the action, and the derivatives of $J$
\begin{equation} \label{operB}
\calH \left(\by, \ba, \nabla J, \nabla^2 J \right) = c(\bx, \ba, \bS) \fp{J}{C}
+ D_S \boldmu^T(\bx, \ba, \bS) \fp{J}{\bS}
+ \ba \cdot \fp{J}{\bx}
+ \frac{1}{2} \tr \left(\Xi_S \nabla_\bS^2 J \right).
\end{equation}
We stress that $\calH$ is a number and not a differential operator. It is
obtained by applying the derivatives to $J$ at the point $\by$, so the
exponential in \eqref{pi_opt} is the exponential of a scalar. The integral in $Z$
runs over the action, which is an auxiliary variable, and not over the state.
Consequently, no nonlocal operator acts on $J$ anywhere below.

Thus, the optimal policy \eqref{pi_opt} is the Boltzmann policy, \cite{szepesvari2010algorithms}. Plugging it back into \eqref{HJBh}, we obtain a semilinear PDE
\begin{align} \label{HJBh_nonlin}
& \frac{\partial J}{\partial t} + r C \frac{ \partial J}{\partial C}
- \frac{1}{\beta} \log Z(J,\bx, \bS) = 0.
\end{align}
The HJB equation in \eqref{HJBh} extends the classical HJB equation in two key
ways. First, by including the cumulative cost $C$ in an extended state variable
${\bf y} = (\bx, \bS, C)$, it eliminates the dependence on the utility function
$U(z)$ from the equation itself, confining it to a terminal condition. Second, it
employs a stochastic policy, which replaces the pointwise minimum over actions by
the log-integral term in \eqref{HJBh_nonlin}.

Although the forms of the drift and diffusion functions in \eqref{mu_sigma_SDE} and the cost function in \eqref{cost_quadratic} differ from those in \cite{Halperin2023}, they share a crucial feature: a quadratic dependence of both reward and dynamics on the actions $\ba$. Consequently, while the formulas presented below are tailored to our specifications, their derivation and final form bear a strong resemblance to the analysis in \cite{Halperin2023}.

Using the drift, diffusion and cost function introduced in
\cref{mu_sigma_SDE,cost_quadratic}, the Hamiltonian \eqref{operB} splits into an
action-independent part and the rest,
\begin{align} \label{pi_opt_2}
e^{-\beta \calH} &= \calC \, e^{-\beta \hat{\calH}}, \qquad
\calC = \exp \left\{ - \beta \left[ C_0 \fp{J}{C} + D_S \boldmu^T_0 \fp{J}{\bS}
+ \frac{1}{2} \tr \left(\Xi_S \nabla_\bS^2 J \right) \right] \right\}, \\
\hat{\calH} &= \ba \left( {\bf C}_1 \fp{J}{C} + \fp{J}{\bx}
+ D_S \left(\boldmu_1 \circ \fp{J}{\bS} \right) \right)
+ \ba^T \left( {\bf C}_2 \fp{J}{C}
+ D_S \mathrm{diag} \left( \boldmu_2 \circ \fp{J}{\bS} \right) \right) \ba. \nonumber
\end{align}
The factorization in \eqref{pi_opt_2} is exact, with no Baker-Campbell-Hausdorff
remainder, precisely because $\calH$ is a scalar. For brevity, we have omitted the
$(\bx, \bS)$-dependence in the functions ${\bf C}_1, \boldmu_1, {\bf C}_2, \boldmu_2$.

Since $\calC$ does not depend on $\ba$, it cancels between the numerator and the
denominator of \eqref{pi_opt} and factors out of the integral in $Z$, which
leaves the modified normalization factor $\tZ(J,\bx,\bS)$
\begin{equation} \label{tZ}
\tZ(J,\bx, \bS) = \int d\ba\, \pi_0(\ba| \bx, \bS) \, e^{- \beta \hat{\calH}},
\qquad Z = \calC \, \tZ.
\end{equation}
Note that $\tZ(J,\bx,\bS)$ is a well-known quantity in statistical mechanics, where it is referred to as a partition function \cite{huang2009statistical}. The relation
$Z = \calC \tZ$ turns the logarithm in \eqref{HJBh_nonlin} into a sum, $-\beta^{-1} \log Z = -\beta^{-1}\log \calC - \beta^{-1} \log \tZ$, and the first term is linear in the derivatives of $J$ by construction. Hence, \eqref{HJBh_nonlin} becomes
\begin{equation} \label{HJBh_nonlin_2}
\fp{J}{t} + \left( C_0(\bx,\bS) + r C \right) \fp{J}{C}
+ D_S\boldmu^T_0(\bx,\bS) \fp{J}{\bS}
+ \frac{1}{2} \tr \left(\Xi_S \nabla_\bS^2 J\right)
- \frac{1}{\beta} \log \tZ(J,\bx,\bS) = 0.
\end{equation}
This new equation can be interpreted as a probabilistic relaxation for the classical HJB approach. It extends the framework to {\it distributional learning}, where the goal is to control the entire probability distribution of returns, not just its expected value. For a deeper discussion of distributional learning, see \cite{Halperin2023}.

Equation $\eqref{HJBh_nonlin_2}$ admits several treatments (see below). To distinguish this PDE and its soft policy from the standard HJB equation, we refer to it as the \emph{Hamilton Incepted} (HI) equation - a formulation that stands as one of the primary contributions of this paper.

It is worth mentioning that \eqref{HJBh_nonlin_2} is semilinear, unlike the fully
nonlinear \eqref{HJBh_nonlin}, because the factor $\calC$, which carries the second
derivatives, has been extracted from the logarithm. The highest derivatives now appear linearly and with coefficients that do not depend on $J$, while the nonlinearity is confined to a zeroth-order term that involves $J$ and $\nabla J$ alone.

Accordingly, the representation of the optimal policy $\pi^*(\ba \mid \bx,\bS)$ in \eqref{pi_opt} becomes
\begin{equation} \label{pi_opt2}
\pi^*(\ba| \bx,\bS) = \frac{\pi_0(\ba| \bx,\bS)}{\hat{Z}(J,\bx,\bS) } e^{-\beta \hat{\calH}}.
\end{equation}
Below we provide three interpretations of the HI equation. The first identifies its nonlinear term as a Helmholtz free energy and, through the Gibbs variational principle, as an entropy-smoothed Legendre-Fenchel transform. The second reads the Hamiltonian financially, as the minimized instantaneous effective cost rate of the agent. The third places the equation in the large-deviation family and identifies the classical HJB limit as a Maslov dequantization.

\myparagraph{The Hamiltonian is a free energy.}
The nonlinear term in \eqref{HJBh_nonlin_2} is not merely reminiscent of statistical mechanics. It represents the Helmholtz free energy of a specific system. This system uses the Hamiltonian $\hat{\calH}$ as its energy function, the behavioral policy $\pi_0$ as the a priori measure over microstates $\ba$, and $\beta$ as the inverse temperature \cite{huang2009statistical}. This identification is exact. It explains why no minimization appears in \eqref{HJBh_nonlin_2}.

By the Gibbs variational principle known in probability as the Donsker-Varadhan formula \cite{dupuis1997weak}, we have
\begin{equation} \label{eq:gibbs_variational}
- \frac{1}{\beta} \log \int d \ba \, \pi_0(\ba| \bx, \bS) \, e^{- \beta \hat{\calH}}
= \min_{\pi} \left\{ \mathbb{E}_{\pi} \left[ \hat{\calH} \right]
+ \frac{1}{\beta} D_{\rm KL} \left[ \pi \| \pi_0 \right] \right\}.
\end{equation}
Thus, the free energy \emph{is} the entropy-regularized minimum over policies. The minimization in \eqref{HJBh} has not been discarded, rather, it has been performed in closed form and absorbed into the Hamiltonian.

This bookkeeping mirrors classical mechanics. In that context, the Hamiltonian is itself a supremum: $H(x,p) = \sup_v \left\{ p \cdot v - L(x,v) \right\}$. Once that transform has been formed, the Hamilton-Jacobi equation $\partial_t S + H(x, \nabla S) = 0$ displays no supremum. The free energy is a Legendre-Fenchel transform, smoothed by an entropy. Accordingly, \eqref{HJBh_nonlin_2} is a semilinear PDE where the Hamiltonian is that smoothed transform. The qualifier used in RL here has a formal definition, rather than just an intuitive one.

\myparagraph{The Hamiltonian as an effective cost rate.}
From a financial point of view, the Hamiltonian \(\calH(t, \mathbf{x}_t, \bS, \mathbf{p}_t = \nabla_{\mathbf{x}} J)\) is defined as the Legendre–Fenchel transform of the instantaneous cost function, see \cite{Itkin2018,lipton2025analytical} among others. It depends on the state $\bx, \bS$, the costate variable \(\mathbf{p}\) (the gradient of the value function \(J\)), and time \(t\). The Hamiltonian encodes the fundamental trade-off between immediate cost and future dynamics in the optimization. Economically, it represents the \emph{minimized instantaneous effective cost rate} for the agent - the sum of the immediate transaction cost and the future impact of an action, valued in present marginal utility terms.

\myparagraph{The zero-temperature limit and the large-deviation family.}
Laplace asymptotics applied to \eqref{eq:gibbs_variational} imply that $- \beta^{-1} \log \tZ \to \min_{\ba} \hat{\calH}$ as $\beta \to \infty$. This result restores the hard minimum of the classical HJB equation and represents the zero-temperature limit (see \eqref{HJBh_nonlin_GM_zeroemp} below). That limit is known as Maslov dequantization. It represents the collapse of the algebra $(+, \times)$ onto the tropical algebra $(\min, +)$ as the parameter $\hbar = 1/\beta$ tends to zero \cite{kolokoltsov1997idempotent}.

In this language, the classical HJB equation is the dequantized object. The \eqref{HJBh_nonlin_2} is its deformation at finite $\hbar$. Here, the Boltzmann policy \eqref{pi_opt} plays the role of the smeared classical trajectory. This identification places \eqref{HJBh_nonlin_2} in a wider family. A quantity of the form $- \beta^{-1} \log \mathbb{E} \left[ e^{- \beta X} \right]$ is a scaled cumulant generating function. It is Legendre conjugate to a rate function \cite{touchette2009large}. Hamilton-Jacobi equations with such a Hamiltonian
govern the quasipotential in two areas: the Freidlin-Wentzell theory of small
random perturbations and macroscopic fluctuation theory. The \eqref{HJBh_nonlin_2} belongs to this family. Consequently, the inverse temperature $\beta$ is an inverse temperature in the literal sense.

\vspace{1em}
Taken together, the three interpretations justify our proposed terminology. While the classical HJB equation explicitly displays the minimization over actions, this operation is absent from \eqref{HJBh_nonlin_2}. Nevertheless, no information is lost; via the Gibbs variational principle \eqref{eq:gibbs_variational}, the entropy-regularized minimum over policies is carried out in closed form, effectively embedding the optimization into the Hamiltonian as its free energy. Because the optimization has been incepted into the Hamiltonian, the formulation becomes a Hamilton-Jacobi rather than a Bellman-type equation. It yields a semilinear PDE lacking an explicit action operator on the right-hand side, making it directly amenable to the numerical methods presented in \cref{sec:deep_solver} and \cref{pideSolver}. The term \emph{Hamiltonian Incepted} reflects this core mechanism and distinguishes the HI equation from the original HJB framework.

\subsection{Gaussian mixture.} 

The HI PDE in \eqref{HJBh_nonlin_2} remains difficult to solve. To improve its tractability, we employ a representation of the behavioral policy $\pi_0(\ba | \bx, \bS)$ as a Gaussian Mixture (GM)
\begin{gather} \label{pi_0_GM}
\pi_0(\ba | \bx, \bS) = \sum_{k=1}^{K} \omega_k \mathcal{N} \left( \ba | {\bf u}_ k(\bx, \bS), \Omega_k^{(0)} \right), \\
0 \leq \omega_k \leq 1, \qquad \sum_{k=1}^{K} \omega_k = 1, \qquad \Omega_k^{(0)} = \sigma_k^2 {\bm I}. \nonumber
\end{gather}
Here, the mean functions ${\bf u}_k(\bx, \bS)$ are parameterized in linear form as
\begin{equation} \label{uParam}
{\bf u}_k(\bx, \bS) = {\bm A}^u_k \bx + {\bm M}^u_k \bS + {\bm B}^u_k,
\end{equation}
where ${\bm A}^u_k \in \mathbb{R}^{N \times N}$ and ${\bm M}^u_k \in \mathbb{R}^{N \times N}$ are matrices of linear coefficients, and ${\bm B}^u_k \in \mathbb{R}^N$ is a constant vector.

The quadratic dependence on $\ba$ in \eqref{pi_opt_2} implies that the optimal policy $\pi$ is also a Gaussian mixture. Using \eqref{pi_0_GM}, the integral in \eqref{tZ} can be computed analytically by applying the Gaussian identity for each component $k$. We define the following derivative-dependent terms:
\begin{align}  \label{pi_GM_params}
{\bm C} &= 2 \beta \left[ {\bf C}_2 \fp{J}{C} - D_S \mathrm{diag} \left( \boldmu_2 \circ \fp{J}{\bS} \right) \right], \qquad
{\bm D} = \beta \left[ {\bf C}_1 \fp{J}{C} + D_S \left( \boldmu_1 \circ \fp{J}{\bS} \right) \right].
\end{align}
This yields the optimized policy parameters:
\begin{align}
\pi(\ba | \bx, \bS) &= \sum_{k=1}^{K} \omega_k(J) \mathcal{N} \left(\ba | {\bf u}_k(J), {\bf \Omega}_k(J) \right), \qquad \omega_k(J) = \frac{ \omega_k e^{ - \beta \mathcal{H}_k (J)}}{ \sum_{j=1}^K  \omega_j e^{ - \beta \mathcal{H}_j (J)}}, \\
{\bf \Omega}_k(J) &= \left[ \frac{1}{\sigma_k^2} \mathbf{I} + {\bf C}\right]^{-1},  \qquad {\bf u}_k(J) = {\bf \Omega}_k(J) \left[ \frac{1}{\sigma_k^2} {\bf u}_k(\bx, \bS) - {\bm D} \right]. \nonumber
\end{align}

With our price impact model in \cref{appImpact}, this simplifies to
\begin{equation}
\bm{D} = \beta\, \bm{B}_\mu^\top[\bS\circ(\nabla_S J - \partial_C J\,\bm{1})],
\qquad
\bm{B}_\mu = \mathbf I + \mathbf f_t^{(1)}\text{diag}(\bx),
\end{equation}
and reduces to elementwise form when $\mathbf f_t^{(1)}$ is diagonal.

The Hamiltonians $\mathcal{H}_k (J) $ are now functionals of the derivatives $\partial J/ \partial \bS$ and $\partial J/ \partial C$:
\begin{align} \label{H_J}
\mathcal{H}_k (J) &= \frac{1}{2} \left[ \frac{1}{\sigma_k^2} {\bf u}_k^T(\bx, \bS) {\bf u}_k(\bx, \bS) - {\bf u}_k^T(J) {\bm \Omega}_k^{-1}(J) {\bf u}_k(J) \right] + \frac{1}{2 \beta} \log\left( \frac{|\Omega_k^{(0)}|}{|{\bm \Omega}_k(J)|} \right).
\end{align}

In the low-temperature limit ($\beta \rightarrow \infty$), the variances vanish and the policy becomes deterministic. The action is given by
\begin{align} \label{a_deterministic}
\ba_{k} &= \lim_{\beta \rightarrow \infty} {\bf u}_k (J) = - \frac{1}{2} {\bf \bar{\Omega}}_{\infty} (J) \left[{\bf C}_1(\bx, \bS) \frac{ \partial J}{ \partial C} + D_S \boldmu_1(\bx, \bS) \circ \frac{ \partial J }{ \partial \bS} \right], \\
{\bf \bar{\Omega}}_{\infty}(J) &= - \left[- {\bf C}_2(\bx, \bS) \frac{ \partial J}{ \partial C} + D_S \text{diag} \left( \boldmu_2(\bx, \bS) \circ \frac{ \partial J}{ \partial \bS} \right) \right]^{-1}. \nonumber
\end{align}
Finally, the resulting HI PDE in \eqref{HJBh_nonlin_2} takes the form:
\begin{equation} \label{HJBh_nonlin_GM}
\fp{J}{t} + \left( C_0 + r C \right) \fp{J}{C} + D_S \boldmu^T_0 \fp{J}{\bS} + \frac{1}{2} \tr \left(\Xi_S \nabla{\bS}^2 J\right) - \frac{1}{\beta} \log \sum_{k=1}^K \omega_k e^{ - \beta \mathcal{H}_k}(J) = 0.
\end{equation}

It is important to note that the linear-in-$\mathbf{a}$ part of $\hat{\mathcal{H}}$ couples $\mathbf{a}$ to $\partial J/\partial\mathbf{x}$ via the term $\mathbf{a}\,\partial/\partial\mathbf{x}$, so that $\partial J/\partial\mathbf{x}$ does enter the partition function $\tilde{Z}$ through $e^{-\beta\hat{\mathcal{H}}}(J)$. However, inspecting \eqref{pi_GM_params}, one sees that the vector $\mathbf{D}$ involves only $\partial J/\partial C$ and $\partial J/\partial \mathbf{S}$, with no explicit $\partial J/\partial\mathbf{x}$ term. This is a direct consequence of our model's parameterization: the coefficients $\mathbf{C}_1$ and $\boldsymbol{\mu}_1$ arise from the specific quadratic-in-$\mathbf{a}$ structure of the cost and drift in \eqref{mu_sigma_SDE} and \eqref{cost_quadratic}, in which the action $\mathbf{a}$ couples to $\mathbf{x}$ purely kinematically (i.e., $\dot{\mathbf{x}} \sim \mathbf{a}$), and not through cost or drift coefficients. Consequently, the coefficient multiplying $\partial J/\partial\mathbf{x}$ in the linear-in-$\mathbf{a}$ term is simply $\mathbf{a}$ itself, with no additional $(\mathbf{x}, \mathbf{S})$-dependent prefactor from $\mathbf{C}_1$ or $\boldsymbol{\mu}_1$.

More precisely, upon completing the Gaussian integral in \eqref{tZ}, the $\partial J/\partial\mathbf{x}$ contribution is absorbed into the optimal mean $\mathbf{u}_k(J)$ and hence into $\mathcal{H}_k(J)$. As can be verified by direct inspection of \eqref{H_J}, the derivative $\partial J/\partial\mathbf{x}$ then cancels identically in the explicit expression for $\mathcal{H}_k(J)$, and does not appear in the final PDE. The $\mathbf{x}$-dynamics are therefore fully encoded in the optimal policy $\pi^*$ through the optimal mean, but leave no residual derivative term in the value function equation \eqref{HJBh_nonlin_GM}.

\myparagraph{High and low temperature limits.}
Both limits can be written out for the Gaussian-mixture equation \eqref{HJBh_nonlin_GM}. As $\beta \rightarrow \infty$ the Hamiltonian \eqref{H_J}
becomes independent of $k$. With a slight abuse of notation, we write its
limiting value as $\calH_{\infty}(J)$
\begin{equation} \label{H_J_2}
\mathcal{H}_{\infty}(J) = - \frac{1}{4} \left({\bf C}_1(\bx, \bS) \frac{ \partial J}{ \partial C} + D_S \boldmu_1(\bx, \bS) \circ \frac{ \partial J }{ \partial \bS} \right)^T {\bf \bar{\Omega}}_{\infty}(J) \left({\bf C}_1 (\bx, \bS) \frac{ \partial J}{ \partial C} + D_S \boldmu_1(\bx, \bS) \circ \frac{ \partial J }{ \partial \bS} \right).
\end{equation}

Using this expression, the zero-temperature limit of \eqref{HJBh_nonlin_GM} reads:
\begin{equation} \label{HJBh_nonlin_GM_zeroemp}
\frac{\partial J}{\partial t} + \left(C_0(\bx, \bS) + r C \right) \frac{ \partial J}{\partial C} + D_S \boldmu_0^T (\bx, \bS) \frac{ \partial J}{\partial \bS} + \frac{1}{2} \tr \left(\Xi_S \nabla_{\bS}^2 J\right) + \mathcal{H}_{\infty}(J) = 0.
\end{equation}

This is the classical HJB equation, with the deterministic policy \eqref{a_deterministic} in place of the Boltzmann one. Laplace asymptotics applied to \eqref{eq:gibbs_variational} already guarantee convergence. The only distinction from the standard case is the fractional nonlinearity carried by $\mathbf{\bar{\Omega}}_{\infty}(J)$, which is the inverse of a matrix that depends on $\nabla J$. Two features of the model produce that dependence. First, the extended state $C$ multiplies the action-quadratic execution cost by $\partial J / \partial C$. Second, the quadratic price impact contributes $D_S \text{diag}(\boldsymbol{\mu}_2 \circ \partial J / \partial \mathbf{S})$. In the linear-quadratic case both features are absent. There, $\mathbf{\bar{\Omega}}_{\infty}$ is constant, and $\mathcal{H}_{\infty}$ reduces to an ordinary quadratic form in $\nabla J$.

The opposite limit is read directly from \eqref{eq:gibbs_variational}. As $\beta \rightarrow 0$, the entropic term $\beta^{-1} D_{\mathrm{KL}}[\pi \| \pi_0]$ dominates. Hence the minimizer becomes $\pi = \pi_0$, and no policy optimization takes place. The information cost of departing from the prior is prohibitive. Accordingly, \eqref{HJBh_nonlin} reduces to \eqref{Kolmogorov_J} with $\pi = \pi_0$.

\section{Deep solver for the optimal value function \texorpdfstring{$J(t,\bx,\bS,C)$}{J(t,x,S,C)}} \label{sec:deep_solver}

The existing literature offers diverse methodological approaches to this problem, ranging from those rooted in the general RL framework to entirely different paradigms. Building upon the broader overview in \cref{sect_Related_work}, below we review some technical and methodological aspects most pertinent to our approach.

\myparagraph{Classical PDE solver.} Since we derived an HJB equation \eqref{HJBh} for the value function $J$ and reduced it to the semilinear PDE in \eqref{HJBh_nonlin_GM}, this equation can be solved numerically. Moreover, in our particular case it is possible to develop a semi-analytical method of solution that is significantly less computationally expensive and accelerates calculations considerably, see \cref{pideSolver}. However, since the solution is obtained on a grid and the dimensionality of this PDE is high (we anticipate a typical portfolio size of 15 to 30 different assets), the solution may still suffer from the curse of dimensionality.

\myparagraph{A PINN approach.}

The PINN is a deep learning framework designed to solve forward and inverse
problems involving PDEs, see \cite{raissi2019physics,karniadakis2021physics,wang2021understanding} among
others.  Instead of relying on traditional discretization methods like finite
elements, a PINN approximates the unknown solution directly with a NN, whose
parameters are trained by minimizing a composite loss function. This loss
function encodes the underlying physics by penalizing the residual of the PDE at
a set of collocation points within the domain, while also satisfying initial and
boundary conditions on sampled data points. By leveraging the automatic
differentiation capabilities of modern machine learning libraries, the network
can compute exact derivatives of its output with respect to its inputs, ensuring
the PDE constraints are enforced in a mesh-free manner.

PINNs are flexible and mesh-free. They handle complex geometries, high-dimensional problems, and inverse problems within a single framework, and once trained they deliver a continuously differentiable approximation of the solution that can be evaluated at any point of the domain. The drawbacks are equally well documented. Training is expensive and can converge poorly when the PDE residual and the boundary terms are badly balanced, sharp gradients and multi-scale behavior are hard to resolve, no convergence guarantee is available, and the accuracy is sensitive to the architecture and to the sampling of collocation points. In addition to the PDE itself, the loss function must also contain terms corresponding to the boundary conditions.

\myparagraph{A combined method.} As PINNs can encounter convergence and approximation issues in regions where the solution of the PDE is non-smooth, the objective function (the PDE) can be augmented with an additional \emph{control variate} (CV) term, as in \cite{antonov2023optimal}. The CV would be constructed as the difference between two solutions: one obtained via the PINN and the other via our semi-analytical solver described in \cref{pideSolver}. Since the semi-analytical solver may suffer from the curse of dimensionality for portfolios with a large number of constituents, it would be applied only along a subset of trajectories $k < n$, where $n$ is the total number of available trajectories. The experiments reported below use the PINN alone, so this combination is described here but left for future work.

Also, the model parameters can be learned by the same PINN in addition to NN parameters simultaneously. To do that, we would need some market data to calibrate the model to, e.g., the historical data. This, however, could be again computationally challenging.

\myparagraph{A FBSDE approach.} A possible route to high-dimensional HJB problems is to solve the corresponding FBSDE with deep learning, that is, the Deep FBSDE method reviewed in \cref{sect_Related_work}. The solution is learned along simulated trajectories, so no spatial grid is required, the Brownian motion of the forward SDE supplies a natural exploration of the state space, and the value function and the optimal policy are produced simultaneously. The price is that the method requires simulation of the forward dynamics, which is time-consuming, and that it presupposes an explicit dynamics model, which is precisely what the offline setting is meant to avoid.

\myparagraph{A HJ approach.} Another approach is to derive a pathwise HJ equation along observed trajectories (the market data) that can be learned from historical data. Since the observed paths $\{x_t(\omega), S_t(\omega), dx_t(\omega), dS_t(\omega)\}$ are given, each realization can be treated as a \emph{known sample path}. The HJ equation then emerges pathwise from the stochastic dynamics by applying the chain rule for the value function along the observed trajectory, effectively \emph{cancelling} the second-order terms. The resulting first-order equation remains valid despite the noise, because it is the projection of the full stochastic PDE onto the direction of the observed trajectory. Along a single path the evolution of $J$ is driven by the realized drift rather than by the average drift, and the second-order term $\tr \left(\Xi_S \nabla_\bS^2 J\right)$ is absorbed into the martingale part of $dJ$, which cancels when $dJ$ is related to $d\bx_t$ through the observed path increments. The construction is the stochastic counterpart of the method of characteristics for PDEs, \cite{courant1989methods}, with the observed sample paths in the role of the characteristic curves.

In more detail, this algorithm consists of a few steps: a) we collect trajectories $\{(\bx_t, d\bx_t, \bS_t, d\bS_t, C_t)\}$; b) estimate the realized drift from $(\bx_t, d\bx_t, \bS_t, d\bS_t, C_t)$; c) assume a parameterized value function $J_\theta(t, \bx, \bS, C)$; d) enforce the pathwise HJ equation as a consistency loss. This learns $J$ directly from data without solving a second-order PDE, using only first-order constraints along trajectories.

Because this algorithm learns exclusively from historical data, it recovers the past \emph{behavioral} policy, not the optimal policy. Furthermore, since the observed data represents only a limited subspace of possible states, any predictions about future or unobserved states are prone to significant inaccuracy.

\myparagraph{An offline RL approach.} In \cite{Halperin2023} the RL framework of the previous section was adapted for offline learning. Their approach reformulates the HJ equation's loss function by adding a term that penalizes deviations from optimality, effectively recasting the historical data as demonstrations of the \emph{optimal} policy. Consequently, this method directly synthesizes the optimal policy and value function from the given (static) dataset, a data-driven objective central to Q-learning \cite{MLF}.

The method enforces the pathwise HJ equation on a discrete grid defined by the observed trajectories. Since RL requires optimal-policy trajectories, but only behavioral data is available, \cite{Halperin2023} introduces a pivotal technique: the same trajectories are treated as draws from the optimal policy. The transformation is governed by a likelihood ratio of the path probabilities, which is a function of the very value function being estimated. This allows for the simultaneous, data-driven learning of both the optimal policy and the optimal value function.

This approach has the advantage of learning all its information directly from data, though it also inherits the common challenges of offline RL. Most importantly, due to infeasibility of exploration of the whole space-action space beyond a fixed historical dataset, offline RL is subject to extrapolation risk, \cite{Fujimoto_2019, Levine_Offline_RL_2020}, as discussed above in Sect.~\ref{sect_Related_work}.

%



Our work partly addresses this traditional problem of offline RL by modeling the behavioral policy as a stochastic process in continuous action spaces, providing a more expressive representation. Moreover, by framing the search for the optimal value function as the solution to an HJ equation, our formulation benefits from the mathematical regularization inherent to PDE/ODEs. Specifically, the HJ equation enforces that the value function $J(t, \bx_t, C_t)$ must be a continuous and differentiable solution that satisfies the equation at every point in its domain. This smoothness constraint means that when $J$ is learned from trajectory data $(x_t, \dot{x}_t, C_t)$, its estimates cannot vary arbitrarily between nearby states, even those not directly observed in the dataset.

Consequently, the value function generalizes more predictably within the vicinity of the behavioral data's support. While this does not eliminate the fundamental problem of distributional shift for states far outside the dataset, it mitigates extrapolation error by ensuring that small deviations from observed states lead to small, structured changes in value estimates. This prevents the unstable, sharp overestimations that often cause offline RL algorithms to fail. We thus argue that the continuous-time PDE/ODE formulation provides a \emph{built-in smoothing mechanism} that regularizes the value function, making it more robust to the queries of a learned policy near, but not exactly on, the behavioral trajectories.

We proceed by deriving the corresponding HJ equation in \cref{sect_HJ} and explaining how to compute likelihood ratios for paths with different drifts in \cref{sect_path_probs}. We recall that despite in this paper we don't use our semi-analytical solution of the PDE in \eqref{HJBh_nonlin_GM}, this approach for the reference is detailed in \cref{pideSolver}.

\subsection{The Hamilton-Jacobi equation} \label{sect_HJ}

In this section we derive a first-order HJ equation along known stochastic
trajectories. The state variables are $\bx_t$ (controlled deterministically)
and $\bS_t$ (stochastic), with dynamics
\begin{align} \label{SDEs}
d\bx_t &= \ba_t\, dt, \qquad d\bS_t = \bS_t \circ \left[ \boldsymbol{\mu}(\bx_t, \ba_t, J)\, dt + \boldsymbol{\sigma}(\bx_t, \ba_t, J) \circ d\bm{W}_t \right].
\end{align}
where the optimal drift $\bm{\mu}$ and diffusion $\bm{\sigma}$ are implicitly defined by the structure of the HJB, $\bm{\sigma} \hm{\sigma}^T = \Xi_S$, and $\bm{\mu}$ is related to $\boldsymbol{\mu}_0$ and $C$.

The observed paths $\{\bx_t(\omega), \ba_t(\omega), \bS_t(\omega), d\bS_t(\omega)\}$ provide empirical data for the state variables and their increments along each
trajectory. Since $\bx_t$ is driven purely by the control $\ba_t$ with no
diffusion, all stochastic variation in the system enters through $\bS_t$.

Because, by the definition in \eqref{Z_T}, the process $C_t$ is deterministic,
from the HJB structure we infer the optimal controlled dynamics \eqref{SDEs}. Applying \Ito's lemma to $J(t, \bx_t, \bS_t, C_t)$ along the optimal path gives
\begin{align} \label{dJ}
dJ &= \Big[ \fp{J}{t} + \ba_t^T \nabla_\bx J + \boldsymbol{\mu}(\bx_t,\ba_t)^T (\bS_t \circ \nabla_\bS J) + \left( C_0(\bx) + r C \right) \fp{J}{C}
+ \frac{1}{2} \tr \left( \Xi_S \nabla_\bS^2 J \right) \Big] dt \\
&+ (\bS_t \circ \boldsymbol{\sigma} \circ \nabla_\bS J)^T d\bm{W}_t. \nonumber
\end{align}

By definition of the optimal policy, the PDE in \eqref{HJBh_nonlin_GM} implies
that the instantaneous cost exactly cancels the deterministic part of the \Ito expansion:
\begin{align}
\fp{J}{t} &+ \underbrace{ \ba_t^T \nabla_\bx J + \boldsymbol{\mu}^T (\bS_t \circ \nabla_\bS J) + \left( C_0(\bx) + r C \right) \fp{J}{C} + \frac{1}{2} \tr \left( \Xi_S \nabla_\bS^2 J \right) }_{\text{Drift terms from } \bx_t, \bS_t, C_t}
= \underbrace{\frac{1}{\beta} \log \sum_{k} \omega_k e^{ - \beta \mathcal{H}_k } (J) }_{\text{Instantaneous optimal cost } \hat{c}}.
\end{align}
This transforms \eqref{dJ} to
\begin{equation} \label{dJ1}
dJ = \left( \frac{1}{\beta} \log \sum_{k} \omega_k e^{ - \beta \mathcal{H}_k}(J) \right) dt + (\bS_t \circ \boldsymbol{\sigma} \circ \nabla_\bS J)^T d\bm{W}_t.
\end{equation}
Since all noise enters through $\bS_t$, we eliminate $d\bm{W}_t$ using
\eqref{SDEs}. From the diffusion part of \eqref{SDEs},
\begin{equation}
d\bm{W}_t = \frac{1}{\boldsymbol{\sigma}} \circ \frac{1}{\bS_t} \circ
\left( d\bS_t - \bS_t \circ \boldsymbol{\mu}\, dt \right),
\end{equation}
where the division is elementwise. Substituting into \eqref{dJ1} and
using $(\bS_t \circ \boldsymbol{\sigma} \circ \nabla_\bS J)^T
(\boldsymbol{\sigma}^{-1} \circ \bS_t^{-1}) = \nabla_\bS J^T$ elementwise,
we obtain
\begin{align} \label{HJfin1}
dJ(t, \bx_t, \bS_t, C_t)
&= \left[ \frac{1}{\beta} \log \sum_{k} \omega_k e^{ - \beta \mathcal{H}_k (J) }
   - \boldsymbol{\mu}(\bx_t, \ba_t)^T (\bS_t \circ \nabla_\bS J)
   \right] dt
   + \nabla_\bS J^T (d\bS_t \circ \bS_t^{-1} \circ \bS_t) \nonumber \\
&= \left[ \frac{1}{\beta} \log \sum_{k} \omega_k e^{ - \beta \mathcal{H}_k (J) }
   - \boldsymbol{\mu}^T (\bS_t \circ \nabla_\bS J)
   \right] dt
   + (\bS_t \circ \nabla_\bS J)^T \frac{d\bS_t}{\bS_t},
\end{align}
where $d\bS_t / \bS_t$ denotes the elementwise (log) return. Note that
$\bx_t$ contributes only through its drift $\ba_t$ to the $dt$ terms, and
no $\nabla_\bx J \cdot d\bx_t$ stochastic term appears since $d\bx_t$ is
purely deterministic.

This is the Hamilton-Jacobi equation in its pathwise form, evaluated along
an optimal trajectory. The term $(\bS_t \circ \nabla_\bS J)^T (d\bS_t / \bS_t)$
represents the instantaneous change in the value function due to the stochastic
change in state $d\bS_t$. The $dt$ term compensates such that the deterministic drift of $dJ$ exactly balances the running cost.

This derivation is fundamentally an application of the \emph{verification
theorem} \cite{bertsekas2012dynamic2}, which states that if a sufficiently
smooth function $J$ satisfies the HJB equation, then integrating the
differential $dJ$ from time $t$ to the terminal time $T$ recovers the total
cost functional $J(t, \bx, \bS, C)$.

Note that \eqref{HJfin1} may be represented in the standard form
\begin{equation} \label{HJfin}
\fp{J}{t} + \mathcal{H}_{HJ} \left(t, \bx, \bS, C, \ba_t,
\frac{d\bS_t}{\bS_t}, \nabla J \right) = 0,
\end{equation}
with the effective Hamiltonian
\begin{align} \label{H_HJ}
\mathcal{H}_{HJ} \left(t, \bx, \bS, C, \ba_t, \frac{d\bS_t}{\bS_t},
\nabla J \right) &=
\frac{1}{\beta} \log \sum_{k} \omega_k e^{ - \beta \mathcal{H}_k (J) }
+ \left( \frac{\dot{\bS}_t}{\bS_t} - \boldsymbol{\mu}_0(\bx_t, \ba_t) \right)^T
  (\bS_t \circ \nabla_\bS J),
\end{align}
where $\dot{\bS}_t / \bS_t$ is the vector of realized log-returns. The gradient term captures the excess log-return of $\bS_t$ above its drift.

The HJ equation \eqref{HJfin} provides a \emph{pathwise} backward recursion. This
formulation enables the computation of the value function $J$ and its partial
derivatives at time $t$ from their values at time $t + dt$ along each individual
trajectory. In contrast to \eqref{HJBh_nonlin_GM}, where the locality and
causality of the dynamics remain \emph{implicit}, the equivalent pathwise
representation \eqref{HJfin} makes both properties \emph{explicit}. This
explicitness may offer stronger training signals for learning algorithms.

Our approach thus enables building a learning method which parameterizes $J(t,
\bx_t, \bS_t, C_t)$ with a neural network $J_{\theta}$ (where $\theta$ can also
include the model (financial) parameters). The network can be trained by
minimizing a loss function that incorporates the residual of equation
\eqref{HJfin}, computed using the realized values $\bx_t, \bS_t, \ba_t$ and
$d\bS_t$. This constitutes a physics-informed learning scheme where the HJ
equation itself guides the estimation from data.

\subsubsection{Boundary conditions for the HJ equation} \label{sect_bc}

At the first glance, \eqref{HJfin} looks like a nonlinear, first-order PDE with independent variables $(t, \bx_t, C_t)$, and as such it requires one boundary condition per each variable (compare with \cref{pideSolver}). However, a more close look reveals that:
\begin{itemize}
\item We treat trajectories $(\bx_t, \bS_t, C_t, \dot{\bx}_t, \ba_t$  as given functions of time $t$ obtained from data or simulation.
\item Our unknown is a vector function $J(t, \bx_t,\bS_t, C_t)$.
\item We will exploit \eqref{HJfin} along these known trajectories.
\end{itemize}

Therefore, \eqref{HJfin} is no longer a PDE to be solved in the full $(t, \bx, \bS, C)$ space. For each observed market trajectory $t \mapsto (\bx(t), \bS(t), C(t))$, the terms $\bx, \dot{\mathbf{x}}, \bS, \dot{\bS}$, and $C$ in the Hamiltonian become known time-dependent functions. Consequently, the PDE simplifies to a scalar ODE in time governing the evolution of $J(t, \bx(t), \bS(t), C(t))$ along that specific trajectory
\begin{equation} \label{ODE}
\frac{dJ(t)}{dt} = \mathcal{H}\big(t, J(t,\by), \nabla_\by J(t,\by)\big) =
\frac{1}{\beta} \log \sum_{k} \omega_k e^{ - \beta \mathcal{H}_k}(J) + \left( \frac{\dot{\bS}_t}{\bS_t} - \boldsymbol{\mu}(\bx_t, \ba_t) \right)^T
  (\bS_t \circ \nabla_\bS J).
\end{equation}
To recap, by using market data (trajectories), we are reducing the PDE to a family of ODEs, parameterized by the different possible market trajectories. For each path, we have one scalar ODE in time.

The complication is that $\mathcal{H}$  depends not only on $J(t, \by)$  but also on $\nabla_\bx J(t,\by), \nabla_C J(t,\by)$, which are the spatial gradients of $J$  evaluated along the known path $(\bx_t, \dot{\bx}_t, C_t$. Therefore this ODE is insufficient to simultaneously determine both $J(t,\by)$  and $\nabla_\bx J(t,\by), \nabla_C J(t,\by)$ without additional information.

Since \eqref{ODE} is an ODE, it now requires only terminal conditions for $J$ and its gradients along each trajectory. They follow from the terminal condition in  \eqref{J_T} and read
\begin{equation} \label{bc}
J(T, \bx_T, \bS_T,C_T) = U(C_T), \qquad J_\bx(T, \bx_T, \bS_T,C_T) = 0, \qquad
J_C(T, \bx_T, \bS_T, C_T) = U'_C(C_T).
\end{equation}

For the three-variable state $(\bx, \bS, C)$, the nature of each variable must be accounted for separately. The holdings $\bx$ and the wealth $C$ are locally deterministic (given the action $\ba$ and the current prices), while the price vector $\bS$ is the sole source of stochasticity. Consequently, the likelihood ratio used below is defined purely over the space of the stochastic variables $\bS$, with the drifts of $\bx$ and $C$ entering the value-function dynamics deterministically; the empirical measure is therefore defined on the observed price paths.

\subsection{Likelihood ratios for paths with different drifts} \label{sect_path_probs}

This work addresses an offline learning problem. In contrast to simulation-based methods that sample Brownian trajectories $\{\mathbf{W}_t \}$, we work directly with a fixed dataset of realized state paths $\{ \bx_t, \bS_t, C_t \}_{t=0}^T$. Given that $d\bx_t = \ba_t dt$ and $dC_t$ are determined by the action and the price changes, the stochasticity in the system is driven entirely by the price vector $\bS \in \mathbb{R}^N$. A key methodological shift is therefore to move from the \emph{Wiener measure} $\mathcal{D} \bm{W}_t$ to an \emph{empirical path-space measure} $\mathcal{D} \bS_t$ constructed from observed data.

To proceed, we look for a short-time approximation of the transition probability density $p(\bS_{t+\Delta t} | \bS_t, \bx_t, C_t)$ for small $\Delta t$. We assume that:
\begin{itemize}
\item $\Delta t$ is small enough that the drift $\boldmu(\bx, \bS)$ and diffusion $\Xi_S(\bx, \bS)$ can be treated as constant over $[t, t+\Delta t]$.

\item The diffusion matrix $\Xi_S$ of the price process is non-degenerate (positive definite).

\item Coefficients are smooth enough to ensure the approximation error is $o(\Delta t)$.
\end{itemize}

The transition density for the stochastic component $\bS$ follows the multidimensional Gaussian form:
\begin{align} \label{tranProb}
p(\bS_{t+\Delta t} | \bS_t) &\approx \frac{1}{(2\pi \Delta t)^{N/2} \sqrt{\det \Xi_S}} \, \exp\Bigg[ -\frac{1}{2\Delta t} \left[\big(\Delta\bS - D_S \boldmu \Delta t \big)^\top \Xi_S^{-1} \big(\Delta\bS - D_S \boldmu \Delta t\big) \right] \Bigg],
\end{align}
where $\Delta\bS = \bS_{t+\Delta t} - \bS_t$ and the coefficients are evaluated at the current state.

We represent \eqref{tranProb} via the action $\calS(\bS_t, \bS_{t+\Delta t}, \boldmu)$:
\begin{equation} \label{prob_trans_path}
P( \bS_{t+\Delta t} | \bS_t) = \frac{1}{\sqrt{ (2 \pi \Delta t)^{N} \det \Xi_S }} e^{ - \calS(\bS_t, \bS_{t+\Delta t}, \boldmu)},
\end{equation}
where the short-term action is related to the Lagrangian $\mathcal{L}$ by $S = \mathcal{L} \Delta t$:
\begin{align} \label{action}
\calS(\bS_t, \bS_{t+\Delta t}, \boldmu) &= \frac{1}{2\Delta t} \left[ (\Delta\bS - D_S \boldmu \Delta t)^\top \Xi_S^{-1} (\Delta\bS - D_S \boldmu \Delta t) \right] \\
&= \frac{\Delta t}{2} \left( \dot{\bS} - D_S \boldmu \right)^\top \Xi_S^{-1} \left( \dot{\bS} - D_S \boldmu \right). \nonumber
\end{align}

The transition probability formula \eqref{prob_trans_path} allows us to obtain the likelihood ratio for an observed transition $\bS_t \rightarrow \bS_{t + \Delta t}$ under two different policy-induced drifts $\boldmu^{(0)}$ and $\boldmu^{(1)}$. Since $\Xi_S$ is invariant under the change of drift, the ratio is:
\begin{align} \label{likelihood_ratio}
\frac{ p^{(\mu^{(1)})} ( \bS_{t+\Delta t} | \bS_t)}{ p^{(\mu^{(0)})} ( \bS_{t+\Delta t} | \bS_t)} &= e^{ - \left( \calS(\bS_t, \bS_{t+\Delta t}, \boldmu^{(1)}) - \calS(\bS_t, \bS_{t+\Delta t}, \boldmu^{(0)}) \right) } = e^{ - \Delta S(\bS_t, \bS_{t+\Delta t}) }.
\end{align}
Expanding the exponent and defining $\dot{\bS} = \Delta \bS / \Delta t$, we obtain:
\begin{equation}
\frac{p^{(\mu^{(1)})}}{p^{(\mu^{(0)})}} = \exp\left[ \Delta t \left( \dot{\bS}^\top \Xi_S^{-1} D_S (\boldmu^{(1)} - \boldmu^{(0)}) - \frac{1}{2} \left( (D_S \boldmu^{(1)})^\top \Xi_S^{-1} D_S \boldmu^{(1)} - (D_S \boldmu^{(0)})^\top \Xi_S^{-1} D_S \boldmu^{(0)} \right) \right) \right].
\end{equation}

\begin{proposition}
The likelihood ratio in \eqref{likelihood_ratio} is the discrete-time equivalent of the Girsanov theorem for the stochastic price process.
\begin{proof}
For the measure $\mathbb{P}^{(k)}$ induced by $d\bS_t^{(k)} = D_S \boldmu^{(k)} dt + \sigma_S d\mathbf{W}_t$, the Radon–Nikodym derivative is:
\begin{equation}
\frac{d\mathbb{P}^{(1)}}{d\mathbb{P}^{(0)}} \bigg|_{\mathcal{F}_t} = \exp\left[ \int_0^t (D_S \Delta \boldmu)^\top \Xi_S^{-1} d\bS_s - \frac{1}{2} \int_0^t \left( (D_S \boldmu^{(1)})^\top \Xi_S^{-1} D_S \boldmu^{(1)} - (D_S \boldmu^{(0)})^\top \Xi_S^{-1} D_S \boldmu^{(0)} \right) ds \right].
\end{equation}
By applying a first-order discretization where $\int_t^{t+\Delta t} d\bS_s \approx \dot{\bS} \Delta t$, the continuous exponent maps exactly to the exponent in \eqref{likelihood_ratio}.
\end{proof}
\end{proposition}

With our price impact model in \cref{appImpact},  $\Xi_S^{-1} = D_S^{-1} \Sigma^{-1} D_S^{-1}$ cancels the $D_S$, so
\begin{equation}
\mathcal S(\boldmu) = \tfrac{\Delta t}{2} (\dot\bS/\bS - \boldmu)^\top \Sigma^{-1}(\dot\bS/\bS - \boldmu),
\end{equation}
so that no blow-up with the price scale occurs. Also, $\boldmu^{(0)} = \bar\boldmu(\pi_0)$, $\boldmu^{(1)} = \bar\boldmu(\pi(J))$ - both mixture-averaged with the plain-square second moment $u_{k,i}^2 + \mathrm{Var}_{k,i}$.

The log-likelihood ratio $L(\boldmu^{(1)})$ is quadratic and concave in $\boldmu^{(1)}$, with a Hessian $\nabla_{\boldmu^{(1)}}^2 L = -\Delta t D_S \Xi_S^{-1} D_S$. Since $\Xi_S$ is positive definite and the diagonal matrix $D_S$ is non-singular (for non-zero prices), the likelihood ratio identifies a unique optimal drift that matches the realized price velocity $\dot{\bS}$. In our setting, $\boldmu^{(1)}$ is a functional of the value function $J$ through the optimal policy $\pi$. Thus, $\Delta S$ serves as a data-driven cost functional for calibrating $J$ against the observed trajectories of the three-variable system.

\subsection{A PINN solver for offline learning from data} \label{sect_deep_DOCTRL}

As mentioned, we use offline RL to learn the optimal policy directly from data, solving the semilinear PDE \eqref{HJBh_nonlin_GM} with a PINN. The solution is encoded into a DNN $J_{\theta}(t, \bx_t, \bS_t, C_t)$, where $\theta$ represents a combined set of trainable parameters, which include a) the PINN trainable parameters (weights) and the model parameters specified at the end of \cref{appImpact}, and b) the model parameters $r, {\bm w}$ in \eqref{r_t_one_more}, $z_{\rm tg}, \Lambda$ in \eqref{quadratic_U}, $\sigma_k, \omega_k, \, k\in[1,K]$ in \eqref{pi_0_GM}.

We begin by discretizing the time interval $[0, T]$ into $N$ steps of size $\Delta t$, indexed by $n = 0, 1, \ldots, N-1$, such that $t_n = n \Delta t$, with $t_0 = 0$ and $t_{N-1} = T$. For a parameterized function $J_{\theta}(t, \bx_t, \bS_t, C_t)$, the time-discretized form of the HJ equation in \eqref{HJBh_nonlin_GM} is interpreted as a regression problem for given states $\by_t = (\bx_t, \bS_t, C_t)$ and $\by_{t+1} = (\bx_{t+1}, \bS_{t+1}, C_{t+1})$:
\begin{equation} \label{backward_X_2_0}
\Delta J_{\theta}(t, \bx_t, \bS_t, C_t) = \mathcal{H}_{HJ} \left( \bx_t, \bS_t, C_t, \bx_{t+1}, \bS_{t+1}, J_{\theta} \right) \Delta t + v\varepsilon_t,
\end{equation}
where $\varepsilon_t \sim \mathcal{N}(0,1)$ and $v^2$ is the noise variance. Importantly, $\varepsilon_t$ models errors in the function approximation $J_{\theta}(t, \bx_t, \bS_t, C_t)$, not the stochasticity of the underlying dynamics. In the backward-in-time learning scheme of \eqref{backward_X_2_0}, which conditions on the future state $(\bx_{t+1}, \bS_{t+1}, C_{t+1})$, the dynamical stochasticity (which in this model is driven by the prices $\bS$) is effectively "frozen" or marginalized out.

Assuming that the function $J_{\theta}(t, \bx_t, \bS_t, C_t)$ is known, \eqref{backward_X_2_0} suggests that the joint probability of observing the transition $\by_t \rightarrow \by_{t'}, \, t'= t+\Delta t$ along with the values $J_{\theta}(t, \by_t)$ and $J_{\theta}(t', \by_{t'})$ is given by the product of two factors: (i) the probability of the transition $\bS_t \rightarrow \bS_{t'}$ (the stochastic component), and (ii) the probability of observing the change $J_{\theta}(t', \by_{t'}) - J_{\theta}(t, \by_t)$ for the given values of $\by_t$ and $\by_{t'}$. We stress that these two factors model \emph{distinct and independent} sources of randomness: factor (i) is the genuine dynamical (Girsanov) likelihood of the price increment under a given drift, whereas factor (ii) is a Gaussian model of the residual of the HJ relation \eqref{backward_X_2_0} arising purely from the imperfection of the approximation $J_\theta$. The objective below is the negative logarithm of this joint density; minimizing it, therefore simultaneously (i) drives the HJ residual of $J_\theta$ to zero, fixing the value-function dynamics, and (ii) selects, through the likelihood ratio, the price drift consistent with the realized increments. Because that drift is itself a functional of $J_\theta$ via the optimal policy, the minimizer is the value function whose induced optimal policy is consistent with the observed data.

In order to learn the optimal policy $\pi$ using a generally sub-optimal behavioral policy $\pi_0$, the SciPhyRL approach of \cite{Halperin2023} relies on the exact likelihood ratio \eqref{likelihood_ratio} with price drifts $\boldmu^{(0)}$ and $\boldmu^{(1)}$ corresponding, respectively, to the behavioral policy $\pi_0$ and the optimal policy $\pi$. Using \cref{pi_GM_params,pi_0_GM}, the action difference $\Delta S$ in \eqref{likelihood_ratio}, now defined over the price trajectories, can be expressed as an explicit function of $J_{\theta}$ and its partial derivatives $\partial J/\partial C$ and $\partial J/\partial \bS$. Further, flipping the sign and rescaling by $v^2$, we obtain an empirical negative log-likelihood $-\calL(\theta)$ (the loss function) along trajectories $\{\bx_{t}^{(n)}, \bS_{t}^{(n)}, C_{t}^{(n)}\}$, $n=1,\ldots,N$ corresponding to the behavioral policy $\pi_0$.

However, we also need to account for the terminal conditions discussed in \cref{sect_bc}. One approach is to parameterize $J(t, \bx, \bS, C)$ using a neural network $J_\theta(t, \bx, \bS, C)$. The gradients $\nabla_\bx J_\theta$, $\nabla_{\bS} J_\theta$, and $\nabla_C J_\theta$ are then obtained via automatic differentiation of this parameterization, while the HJ equation is satisfied along each trajectory. The terminal conditions are enforced by incorporating four penalty terms into the loss function corresponding to: (a) the terminal condition on the value function, $\|J_\theta(T, \by_T) - U(C_T)\|^2$; (b) the gradient terminal condition with respect to $\bx$: $\|\nabla_\bx J_\theta(T, \by_T)\|^2$; (c) the gradient terminal condition with respect to $\bS$: $\|\nabla_{\bS} J_\theta(T, \by_T)\|^2$; and (d) the gradient terminal condition with respect to $C$: $\|\nabla_C J_\theta(T, \by_T) - U'_C(C_T)\|^2$. Note that for $\bx$ and $\bS$, the optimal terminal gradients are zero as they do not enter the terminal utility directly. Thus, the final loss function is
\begin{align} \label{NLL_PDE}
-\calL(\theta) &= \frac{1}{N} \sum_{n=1}^{N} \sum_{t=0}^{T-1}
\Biggl\{ \frac{1}{2} \left[ \Delta J_\theta (t, \by_t^{(n)}) - \mathcal{H}_{HJ}
\left( \by_t^{(n)}, \by_{t+\Delta t}^{(n)}, J_\theta \right) \Delta t \right]^2
+ v^2 \Delta S(\bS_t^{(n)}, \bS_{t+\Delta t}^{(n)}, J_{\theta}) \Biggr\} \\
&+ \lambda_{\TC} \Biggl\{ \left[J_\theta(T, \by_T^{(n)}) - U(C_T^{(n)})\right]^2 + |\nabla_\bx J_\theta(T, \by_T^{(n)})|^2 + |\nabla_{\bS} J_\theta(T, \by_T^{(n)})|^2 \nonumber \\
&+ \left[\nabla_C J_\theta(T, \by_T^{(n)}) - U'_C(C_T^{(n)})\right]^2 \Biggr\}. \nonumber
\end{align}
Here, $\lambda_{\TC} \ge 0$ is a constant regularization parameter that should be chosen such that the terminal condition terms in \eqref{NLL_PDE} are of the same magnitude as the others. Since all terms except the second one in \eqref{NLL_PDE} are non-negative while the second term has the lower bound (see the end of \cref{sect_path_probs}), $-\calL(\theta)$ is bounded from below.

This loss function is intuitively appealing as it balances two primary components: the model loss for fixed trajectories (first term) and the cost of mismatches between observed and expected dynamics (second term), with $v^2$ as the sole hyperparameter. More specifically, the second term, $\Delta S(\bS_t^{(n)}, \bS_{t+\Delta t}^{(n)}, J_{\theta})$ in \eqref{NLL_PDE}, enforces consistency between the model $J_{\theta}(t, \by_t)$ and the price dynamics jointly implied by the observations and the model itself as defined in \eqref{likelihood_ratio}.

Therefore, by interpreting behavioral data as optimal data modified by the likelihood ratio in \eqref{likelihood_ratio}, the loss function \eqref{NLL_PDE} enables the direct learning of an optimal policy. This converts the offline RL problem into a straightforward supervised learning (inference) problem. Unlike other offline RL methods that typically rely on value or policy iteration, this approach solves the optimal control problem in a single pass by leveraging the exact likelihood ratio as a functional of the optimal value function $J_{\theta}$ and its derivatives.

Note that the joint learning from both the data and model errors, implemented via the loss function \eqref{NLL_PDE}, is largely driven by the derivatives $\partial J_{\theta}/\partial \bx_t, \partial J_{\theta}/\partial \bS_t$ and $\partial J_{\theta}/\partial C_t$. Indeed, if we consider a limit where these derivatives are negligible, the first term in \eqref{NLL_PDE} vanishes. In this limit, the minimum of the negative log-likelihood is attained at the minimum of the second term, $\Delta S(\bS_t^{(n)}, \bS_{t+\Delta t}^{(n)}, J_{\theta})$, which occurs at the realized price velocity $\dot{\bS}$.

Any policy optimization would be lost in this limit, as can be seen from \cref{pi_GM_params}, implying that $\boldmu^{(1)} = \boldmu^{(0)}$ holds in this scenario. Therefore, vanishing derivatives would simply enforce a match between the model drift $\boldmu^{(0)}(\bx_t, \bS_t, \pi_0)$ (which corresponds to the behavioral policy $\pi_0$) and the observed velocities $\dot{\bS}_t$, while the policy optimization component disappears. In other words, without these derivatives, the approach could at best recover a form of behavioral cloning, lacking any optimization of the policy. This highlights that the "physics" (that is, the propagation of information through the function derivatives across the wealth, holding, and price dimensions) is critical for ensuring consistency between the neural network model for the cost function $J_{\theta}$ and the dynamics jointly implied by the observations and the learned function $J_{\theta}$.

As the loss function \eqref{NLL_PDE} involves only the observable data along with the unknown function $J_{\theta} (t, \bx_t, \bS_t, C_t)$ and its derivatives, a single neural network can be used to encode $J_{\theta}$ together with the model parameters. The derivatives of this function can be computed by using automatic differentiation available via software such as TensorFlow or PyTorch. In more detail, see \cite{Halperin2023}.

\section{Experiments} \label{sect_Experiments}

For our numerical experiments, we selected a diverse universe of 14 exchange-traded funds (ETFs) representing various asset classes: SPY, IWM, QQQ, DIA, EEM, EFA, FXE, FXI, TLT, HYG, SLV, GLD, USO, and VNQ. The portfolio includes US equity trackers (SPY, IWM, QQQ, DIA), international equity ETFs (EFA for developed markets ex-North America, EEM for emerging markets), a currency ETF (FXE for the Euro), a Chinese large-cap equity ETF (FXI), fixed-income ETFs (TLT for 20-year US Treasury bonds, HYG for high-yield corporate bonds), and commodities (SLV for silver, GLD for gold, USO for oil). The selection concludes with VNQ, which tracks the MSCI US REIT Index, covering US real estate.

All daily market data for this $N=14$-dimensional asset universe span the period from Jan.~1, 2019, to Dec.~31, 2025, so full 7 years. We employ actual historical data from Yahoo Finance on \emph{Open\slash Close\slash High\slash Low\slash Volume} data and also the data for ADV and Market Cap. Note that historical bid-ask spread is not available there, therefore we obtain it by using Corwin-Schultz approximation, \cite{Corwin2012}. All these data are plugged into the price impact model in \eqref{appImpact}.

\subsection{Construction of the experimental framework}

\myparagraph{Behavioral policy ${\bm \pi_0}$.}
The construction of the behavioral policy $\pi_0$ draws inspiration from an equally-weighted long portfolio. At inception the holdings are set to equal dollar weights, $x_{0,i} = (\Not/N)/p_{0,i}$, $i \in [1,N]$, where $x_{t,i}$ denotes the holding of asset $i$ in shares and the action is the trade rate, $a_{t,i} \, \Delta t = x_{t,i} - x_{t-\Delta t,i}$. Since prices move, a portfolio with frozen holdings drifts away from equal weights. Therefore, the baseline action trades back toward the current equal-weight target at a rebalancing rate $\kappa$,
\begin{equation} \label{a_equally_w}
a^{\mathrm{EW}}_{t,i}(\bx_t) = \kappa \left( \frac{\Pi_t}{N \, p_{t,i}} - x_{t,i} \right), \qquad \Pi_t = \sum_{i=1}^N x_{t,i} \, p_{t,i},
\end{equation}
where $\Pi_t$ is the portfolio notional marked at current prices. The parameter $\kappa > 0$ is the rebalancing speed, so deviations from equal weight decay with characteristic time $1/\kappa$; the limits $\kappa \to 0$ and $\kappa \Delta t = 1$ correspond to buy-and-hold and to full per-step rebalancing, respectively. This trade is dollar-neutral by construction, $\sum_i p_{t,i} \, a^{\mathrm{EW}}_{t,i} = \kappa (\Pi_t - \Pi_t) = 0$, so the notional changes only through price moves, and, under the assumptions of no corporate actions (plausible for ETFs, where such events are rare) and no dividend payments, the resulting portfolio is self-financing.

To capture potential deviations from the baseline action in \eqref{a_equally_w}, we specify a stochastic Gaussian policy. This policy has mean $\mathbf{a}^{\mathrm{EW}}_t$ and a variance $\mathbf{\Omega}_1^{(0)}$ drawn from a uniform distribution on $[0.2, 0.7]$. It serves as an exploitation policy, capitalizing on the robust historical performance of equally weighted portfolios relative to more complex strategies.

On the other hand, to create a dataset suitable for learning an optimal policy, we must also incorporate \emph{exploration} into the allocation decisions. To this end, we generate a random vector $(\alpha_1, \ldots, \alpha_N)$ whose elements $\alpha_i$ are uniformly sampled from the interval $[0.5, 2]$, once per episode. These scaling factors adjust the rebalancing trades up or down relative to the equal-weighted baseline and are de-meaned to preserve neutrality in share terms. The exploration policy is then defined as a Gaussian distribution $\mathcal{N}(\mathbf{a}_{t}^{\mathrm{EX}}, \mathbf{\Omega}_2^{(0)})$ with variance $\mathbf{\Omega}_2^{(0)}$ drawn from the same distribution as $\mathbf{\Omega}_1^{(0)}$, and mean
\begin{equation} \label{a_ex}
a_{t,i}^{\mathrm{EX}} (\bx_t) =  \alpha_i \, a_{t,i}^{\mathrm{EW}}( \bx_t) - \frac{1}{N} \sum_{j=1}^{N} \alpha_j \, a_{t,j}^{\mathrm{EW}}( \bx_t).
\end{equation}
The behavioral policy is given by a two-component Gaussian mixture combining the exploitation and exploration policies
\begin{equation} \label{pi_0_GM_2}
\pi_0 (\mathbf{a}_t | \mathbf{x}_t) = (1 - \omega_{E}) \mathcal{N} \left( \mathbf{a}_{t}^{\mathrm{EW}}(\mathbf{x}_t), \mathbf{\Omega}_1^{(0)}\right)
+ \omega_{E} \mathcal{N} \left( \mathbf{a}_{t}^{\mathrm{EX}}(\mathbf{x}_t), \mathbf{\Omega}_2^{(0)} \right),
\end{equation}
with $0 \leq \omega_E \leq 1$ being a parameter that controls the relative weight of exploration versus exploitation in the behavioral policy. Its means, variances, and weights are exactly the prior Gibbs components $\{u_k, \Omega^{(0)}_k, \omega_k\}$ read by the Hamiltonian.

Note that when Gaussian stochastic policies are used as components of the mixture, the self-financing condition holds only on average, not for each individual sample drawn from the distribution. The extent of possible violations in individual samples is governed by the covariance matrices $\mathbf{\Omega}_1^{(0)}$ and $\mathbf{\Omega}_2^{(0)}$ and by the mixture weight $\omega_{E}$. As long as such deviations remain small, they can be retained in the generated dataset, provided that the net allocation sum $\mathbf{1}^T \mathbf{a}_t$ is treated as a cash inflow (or outflow, depending on sign) to the fund and is deducted accordingly when computing the fund's return. Note that under the normalization $p_{0,i} = 1$ the share-space sum $\mathbf{1}^T \mathbf{a}_t$ approximates the dollar imbalance $\mathbf{p}_t^T \mathbf{a}_t$.

\myparagraph{Weight-based control and state transitions.}
In the classical continuous-time formulation, the action $\ba_t$ represents the trading rate, leading to the state transition $d\bx_t = \ba_t\,dt$. However, within a short discrete holding period, e.g., one month, this rate-based integration introduces a structural lag: the position $\bx_t$ becomes a mean-reverting integral of small tilts and cannot reach the signal-implied target holdings before the rolling window terminates.

To enable cross-sectional alpha capture over short horizons, we reparameterize the control in terms of the \emph{target holding} $\bh_t$, which carries the same units as $\bx_t$, while $\ba_t$ retains its meaning of a trading rate. The two are related by the affine bijection
\begin{equation} \label{eq:induced_rate}
\ba_t = \frac{\bh_t - \bx_t}{\Delta t},
\qquad \text{equivalently} \qquad
\bh_t = \bx_t + \ba_t\,\Delta t,
\end{equation}
so that the Euler transition of the rate dynamics \eqref{eq:post_rebalance_values}
lands exactly on the target. In the notation of \eqref{eq:post_rebalance_values},
the target coincides with the post-rebalance holding, $\bh_t = \bx^+_t$. Since
rebalancing occurs at the beginning of the interval, the position actually carried
over $[t, t+\Delta t]$ is $\bh_t$, and the agent earns the return and bears the
risk of that position. Accordingly, in the discrete-time cost the holding-return
and risk terms of \eqref{cost_quadratic} are evaluated at the post-jump holding
$\bh_t$ rather than at $\bx_t$. In the rate-based formulation this distinction is
immaterial. There the difference $\bh_t - \bx_t = \ba_t \Delta t$ is $O(\Delta t)$,
and the cost itself carries a factor of $\Delta t$, so the two evaluations differ
only at $O(\Delta t^2)$. Under the jump transition the trade size is $O(1)$, and
the two evaluations differ at leading order. Evaluating the cost at $\bx_t$ would
remove the target from the return and risk terms altogether, so the signal and the
risk penalty would not enter the minimization over $\bh_t$. With the cost written
at $\bh_t$, the coefficients $C_0$ and $C_2$ acquire an explicit action dependence,
which the analytic Gibbs step absorbs into the couplings of
\eqref{eq:couplings_jump}.

We write $\delta\bh_t = \bh_t - \bx_t = \ba_t \Delta t$ for the per-step trade
size. The reformulation removes the transient accumulation entirely, since
\eqref{eq:post_rebalance_values} is a change of control variable rather than a
change of dynamics. All transaction cost and price impact terms in
\eqref{cost_quadratic} and \eqref{quadratic_impact_vec} retain their original
definitions and are evaluated at the induced rate $\ba_t$ of
\eqref{eq:induced_rate}, so the agent is charged the full execution cost of the
transition. Because the per-step trade size $\delta\bh_t$ is $O(1)$ rather than
$O(\Delta t)$, the per-step quadratic cost is $O(1/\Delta t)$. The formulation is
inherently discrete-time, with $\Delta t$ fixed at one trading day, and no
$\Delta t \to 0$ limit is taken. The discrete-time derivation, the relation to
the continuous-time couplings, and the budget constraint are given in
\cref{app:jump_control}.

\myparagraph{Predictive signals.}
For model training, we replace proprietary predictive signals with a \emph{synthesized benchmark} signal, an "engineered oracle". This approach allows us to precisely control the predictive quality of the signal and perform systematic sensitivity analysis.

We begin with the true future realized log-return of asset $i$ at time $t$, denoted $r_{t, i}$, which represents perfect foresight. We standardize this return by the asset's step volatility $\sigma_i \sqrt{\Delta t}$ and degrade it by injecting calibrated Gaussian noise to construct the raw signal $z_{t, i}$:
\begin{equation} \label{eq:raw_oracle}
    z_{t, i} = \sqrt{q} \left( \frac{r_{t, i}}{\sigma_i \sqrt{\Delta t}} \right) + \sqrt{1 - q} \, \epsilon_{t, i}, \qquad \epsilon_{t, i} \sim \mathcal{N}(0, 1)
\end{equation}
where $q \in [0, 1]$ is the target predictive power, that is, the $R^2$ of a regression of next-step returns on the signal. On daily returns, empirical models rarely reach an out-of-sample $R^2$ of one percent, so the values of $q$ used below lie well above what a real forecast delivers. This is deliberate. The construction is an oracle by design, and the experiments measure how much of a signal of \emph{known} quality the framework converts into portfolio performance, not whether such a signal exists. By varying $q$ we explore how signal quality influences the learned policy, and the $R^2$ actually achieved on the model's inputs is reported in \cref{tab:predictive_metrics}.


\myparagraph{Implicit turnover control.}
Note that despite we don't explicitly impose constraints on turnover (which is a popular way in practice), two existing parameters are already quadratic turnover penalties in all but name:
\begin{enumerate}
\item $\eta_{\text{tc}}$: the transaction cost is $\frac{\eta}{\Delta t}\sum_i S_i\,\delta h_i^2$ under the jump. A turnover penalty $\lambda_{\mathrm{turn}}\sum S_i \delta h_i^2$ is mathematically indistinguishable from raising $\eta_{\text{tc}}$ by $\lambda_{\mathrm{turn}}\Delta t$. The only reason to keep them separate is interpretation: $\eta_{\text{tc}}$ is calibrated to real slippage and belongs to the market model, while a turnover "preference" is a mandate constraint.

\item $\sigma_{h,k}$ - the target-space prior widths. The KL term already penalizes deviation from the behavioral prior, whose means are near-current-holding rebalancing targets. Shrinking $\sigma_{h,k}$ tightens the policy around small trades. This is the turnover control intrinsic to the G-learning formulation, and it requires no modification of the cost model.
\end{enumerate}

\myparagraph{Behavioral priors and risk penalties.}
The Gibbs mixture in \eqref{pi_0_GM_2} relies on prior means $\mathbf{u}_k(\mathbf{x}_t, \mathbf{S}_t)$. Since the action space is now the target holding, the behavioral priors $a^{EW}$ and $a^{EX}$ must be converted from trade rates to target holdings before being passed to the HJB solver. We achieve this by integrating the rate over the time step:
\begin{equation}
    \mathbf{u}_k^{\text{target}} = \mathbf{x}_t + \mathbf{a}_t^{k, \text{rate}} \Delta t.
\end{equation}
This maintains the scale compatibility required for the analytic Gibbs minimizer.

Furthermore, to prevent the agent from jumping into excessively concentrated or high-volatility target positions, the risk penalty in the running cost \eqref{running_cost} is modified to penalize the \emph{tracking error} of the target holding $\mathbf{h}_t$ against the equal-dollar-weight allocation implied by the same target,
\begin{equation} \label{eq:risk_target}
    \text{Risk}(\mathbf{h}_t) = \Lambda \left[\mathbf{S}_t \circ \left(\mathbf{h}_t - \mathbf{x}^{\rm ew}(\mathbf{h}_t)\right)\right]^T \Sigma \left[\mathbf{S}_t \circ \left(\mathbf{h}_t - \mathbf{x}^{\rm ew}(\mathbf{h}_t)\right)\right],
    \qquad
    \mathbf{x}^{\rm ew}(\mathbf{h}_t) = \frac{\bm{1}^T (\mathbf{S}_t \circ \mathbf{h}_t)}{N}\, \mathbf{S}_t^{\circ (-1)},
\end{equation}
where $\mathbf{S}_t^{\circ (-1)}$ denotes the elementwise inverse. In dollar terms, with $\mathbf{y} = \mathbf{S}_t \circ \mathbf{h}_t$ and the centering projector $P = I - \bm{1}\bm{1}^T/N$, the penalty is the quadratic form $\Lambda\, \mathbf{y}^T K \mathbf{y}$ with $K = P^T \Sigma P$. Penalizing the centered allocation rather than the raw positions leaves the total notional unconstrained (that role belongs to the budget penalty of \cref{sect_self_financing}) and vanishes on the equal-weight portfolio, so the term prices only the cross-sectional concentration of the tilt.

The full matrix $K$ is not diagonal, whereas the analytic Gibbs solver requires a diagonal quadratic-in-$\mathbf{h}$ exponent. We therefore split the penalty at the current holding $\mathbf{x}_t$: the quadratic part enters the coupling $\mathbf{C}$ in \eqref{pi_GM_params} through $\text{diag}(K)$ only, while the linear part, which pulls the target toward equal weight, retains the full matrix,
\begin{equation} \label{eq:risk_couplings}
    \Delta \mathbf{C} = 2\beta \Lambda\, \Delta t\, \mathbf{S}_t^{\circ 2} \circ \text{diag}(K),
    \qquad
    \Delta \mathbf{D} = 2\beta \Lambda\, \Delta t\, \mathbf{S}_t \circ \left[K (\mathbf{S}_t \circ \mathbf{x}_t)\right] - \Delta \mathbf{C} \circ \mathbf{x}_t,
\end{equation}
so that the gradient of the Gibbs exponent at $\mathbf{h}_t = \mathbf{x}_t$ matches that of the exact penalty, and only the off-diagonal curvature of $K$ is approximated. This injects the required diagonal quadratic term into the Hamiltonian coupling $\mathbf{C}$, explicitly constraining the optimal target $\mathbf{h}_t$ within the HJB optimization step, while the linear coupling preserves the pull toward the passive benchmark under the full covariance.

\myparagraph{Optimal policy ${\bm \pi}$.}
The HJB equation is solved using a PINN that parameterizes the value function $J_\theta(t, \mathbf{x}, \mathbf{S}, C)$. The network is trained by minimizing the loss function \eqref{NLL_PDE}, which enforces the pathwise HJ equation along observed trajectories.

The control gradients $\nabla_{\mathbf{x}} J$, $\nabla_{\mathbf{S}} J$, and $\nabla_C J$ in our PINN-based reinforcement learning framework can be computed twofold. The first one is via automatic differentiation of the trained PINN (\texttt{control\_mode=\allowbreak 'network'}). In the second mode (\texttt{control\_mode=\allowbreak 'analytic'}) the gradient $\nabla_C J$, which governs the policy updates, is derived analytically as the leading-order sensitivity of the cost function $J$ with respect to the portfolio weights. This analytical expression comprises two primary, signed terms: the signal-following term, $-S \cdot (w \zeta)$, which drives the portfolio toward the predictive alpha $\zeta$; and the tracking-error regularization term, $+2 \Lambda \Sigma S x_{\text{diff}}$, which anchors holdings toward the equal-weight benchmark. This gradient is scaled by the time envelope function $g(\tau)$ and is computed using pure Python \emph{numpy} operations.

Because this expression is treated as a constant with respect to the neural network parameters $\theta$, it remains differentiable via the framework's automatic differentiation engine during the PINN loss minimization. This approach contrasts with relying solely on the automatic differentiation of the PINN, as it embeds structural financial domain knowledge (specifically the known cost-gradient sensitivity) directly into the supervision target, ensuring the model respects the underlying optimal control objective throughout training.

We found that \texttt{control\_mode='analytic'} provides better performance of our method, so in what follows we use it everywhere. To highlight, these gradients drive the couplings $\mathbf{C}(J)$ and $\mathbf{D}(J)$ in \eqref{pi_GM_params}, producing the Gibbs mixture-mean target holding $\mathbf{a}^* = \sum_{k=1}^K \omega_k(J) \mathbf{u}_k(J)$. At each step $t$, the portfolio instantaneously jumps to $\mathbf{x}_{t+\Delta t} = \mathbf{h}^*$, and the realized transaction cost is deducted based on the jump size $(\mathbf{h}^* - \mathbf{x}_t)$.

\myparagraph{Price impact function.}
To assess the impact of investment decisions on portfolio returns, we must specify parameters for the price impact model in \eqref{quadratic_impact_vec}. We adopt the reduced five-parameter reparameterization $\mathbf{\Theta} = \{ \nu, \lambda, \theta, \kappa_3, \phi \}$ described at the end of \cref{appImpact}. For our baseline experiments, we use values calibrated to the observed microstructure of our ETF universe: $\nu = 0.001$, $\lambda = 0.001$, $\theta = 0.001$, $\kappa_3 = 0.01$, and $\phi = 0.5$.

Under the weight-based control formulation, the generation of offline trajectories proceeds as follows. We utilize rolling historical price windows for the asset prices $\bS_t$. At each time step $t$, a target holding $\bh_t$ is sampled from the behavioral policy \eqref{pi_0_GM_2}, specified in target space as described in \cref{app:jump_control}. The portfolio state updates instantaneously via the jump transition \eqref{eq:induced_rate}, and the trading rate executed over the interval $\Delta t$ is the induced rate $\ba_t = (\bh_t - \bx_t)/\Delta t$ of \eqref{eq:induced_rate}.

This rate $\ba_t$ enters the price impact function $\mathbf{f}(\ba_t)$, which provides a model-based ``counterfactual'' correction to the historical drift. The action-independent component of the drift, $\bS_t \circ \left( r {\bf 1} + {\bf w} \tilde{\bm{\zeta}}_t \right)$ as defined in \eqref{mu_sigma_SDE}, is computed directly from the observed market returns and is thus data-driven. The remaining term arises from portfolio trading and is model-dependent, scaling with the per-step trade size $\delta\bh_t = \bh_t - \bx_t$.

When the impact parameters $\nu, \lambda, \theta$ are sufficiently small, the trading-dependent part offers a straightforward way to enrich a single historical return series with realistic, trading-driven variation without completely obscuring the baseline market dynamics. We note that if several long historical return sequences are available (for instance, 30 years of data can be segmented into multiple non-overlapping windows), an even richer set of scenarios can be generated by using these sequences as different baseline realizations of the drift. Alternatively, synthetic time series for the drift can be constructed to resemble stress scenarios by deterministically scaling historical returns upward or downward.

\myparagraph{Feed-forward neural network.}
The optimal value function $J_{\theta}({\bf y}_t, t)$ is approximated by a feedforward neural network with 3 hidden layers, each having 150 neurons with a softplus activation function. Training is performed using the Adam optimizer with a learning rate of $5 \times 10^{-4}$ and mini-batches of size 512. The terminal boundary conditions are enforced via the hard-terminal ansatz 
\begin{equation} \label{ansatz_term} 
J(\tau, \mathbf{x}, \mathbf{S}, C) = U(C) + g(\tau) h_\theta(\mathbf{x}, \mathbf{S}, C),
\end{equation}
where $g(\tau) = \tau_{\text{sat}} \tanh(\tau / \tau_{\text{sat}})$ is a saturating time envelope that bounds the policy gradient at long horizons while preserving the exact terminal condition.

\subsection{Results} \label{results}

We conducted several experiments using a Python implementation of this framework. The baseline model parameters are detailed in \cref{tab:exp_params}.
\begin{table}[!htb]
    \centering
    \setlength{\tabcolsep}{4pt}
    \begin{tabular}{llc | llc}
        \toprule
        \textbf{Network} & \textbf{Symbol} & \textbf{Value} & \textbf{Simulation} & \textbf{Symbol} & \textbf{Value} \\
        \hline
        Hidden Layers & - & (150, 150, 150) & Inv. temperature & $\beta$ & 15.0 \\
        Training episodes & $n_{\text{train}}$ & 1008 & Terminal BC Penalty & $\lambda_{tc}$ & 1.0 \\
        Testing episodes & $n_{\text{test}}$ & 120 & Terminal Utility & - & \eqref{quadratic_U} \\
        Random Seed & - & 42 & Time Envelope Sat. & $\tau_{\text{sat}}$ & None \\
        & & & Notional & $\Not$ & 10.0 USD \\
        & & & Notional penalty & $\lambda_{\text{not}}$ & 0.1 \\
        & & & Risk Aversion & $\Lambda$ & 10.0 \\
        & & & Exploration weight & $\omega_E$ & 0.5 \\
        & & & Raw Target $R^2$ & $q$ & 0.1 \\
        & & & Target CML Cost & $z_{\text{target}}$ & $\Not(1.0 - e^{0.1 T})$ \\
        & & & Rebalancing rate & $\kappa$ & 2 \\
        \hline
        \textbf{Optimization} & \textbf{Symbol} & \textbf{Value} & \textbf{Market Impact} & \textbf{Symbol} & \textbf{Value} \\
        \hline
        Epochs & - & 3000 & Trans.Cost/Slippage & $\eta_{tc}$ & 0.0001 \\
        Learning Rate & $\alpha$ & $5 \cdot 10^{-4}$ & Impact Parameter 1 & $\nu$ & 0.001 \\
        Control Mode & - & \texttt{analytic} & Impact Parameter 2 & $\lambda$ & 0.001 \\
        Batch Size & - & 512 & Impact Parameter 3 & $\theta$ & 0.001 \\
        Terminal Batch Size & - & 512 & Impact Parameter 4 & $\kappa_3$ & 0.01 \\
        Optimmizer & - & \texttt{Adams} & Impact Parameter 5 & $\phi$ & 0.5 \\
        \bottomrule
    \end{tabular}
    \caption{Experimental parameters and hyperparameters.}
    \label{tab:exp_params}
\end{table}

Here, $n_{\rm train}$ and $n_{\rm test}$ are the numbers of overlapping rolling
windows (episodes) used for training and for testing, not counts of calendar
days. Each episode is a price path of $T+1$ days, and consecutive episodes are
shifted by one day, so the $120$ test episodes at $T = 63$ occupy a contiguous
block of $183$ trading days. Parameters in \cref{tab:exp_params} are chosen just to demonstrate the robustness of the approach. Ideally, they have to be found either by calibration to the available data, or, at least, based on some selection procedure (or sensitivity) which we leave for the future research.

Note that the oracle experiment measures the conversion of predictive information into portfolio performance, not forecasting ability, so the achieved signal quality has to be checked. Since the signal is constructed from the realized future return with a nominal target $R^2 = q$, the predictive $R^2$ evaluated on the model's inputs differs from the stated target. It overshoots the target in sample and falls short of it out of sample.

This mechanism is identified as follows. Because the target $R^2$ is applied to raw, variance-weighted returns, measuring the achieved $R^2$ on cross-sectionally standardized returns alters the correlation structure. This standardization removes variance-weighting across assets and inadvertently captures cross-sectional mean differences, which shifts the achieved metric away from the nominal target.

The divergent behavior of the achieved $R^2$ reveals the source of this discrepancy. A structural artifact of the estimator would shift the metric in both samples uniformly. However, because the achieved $R^2$ moves in opposite directions, the underlying driver must be a factor that varies between the in-sample and out-of-sample blocks. That factor is the realized volatility. Specifically, the standardization in \eqref{eq:raw_oracle} divides the realized return by the static, full-sample volatility $\sigma_i$, rather than the local realized volatility of the specific block where the signal is evaluated.

Let $v$ be the ratio of the realized return variance of a block to the full-sample
variance used in the standardization. The informative part of the signal then  carries variance $q v$, while the noise floor stays at $1-q$. Hence
\begin{equation} \label{eq:r2_regime}
\text{Var}(z_t) = q v + (1 - q), \qquad
R^2_{\rm achieved} = \frac{q v}{q v + 1 - q}.
\end{equation}
The achieved $R^2$ equals the nominal target only when $v = 1$. The gap is therefore a property of the static standardization and not of the method. Standardizing each window by its own realized volatility would set $v =
1$ and recover $R^2 = q$ in both blocks. We keep the static form, since
$\sigma_i$ also sets the signal scale ${\bf w}$ and the covariance $\Sigma$, and
we report the achieved values as measured.

To measure the realized OOS $R^2$ we choose true OOS predictive $R^2$ (Gu–Kelly–Xiu convention (GKX), \cite{gu2020empirical}) and fit the linear map IS, then apply it frozen OOS:
\begin{equation}
\hat b = \frac{\mathrm{cov}_{\text{IS}}(\tilde\zeta, y)}{\mathrm{var}_{\text{IS}}(\tilde\zeta)}, \qquad R^2_{\text{OOS}} = 1 - \frac{\sum_{\text{OOS}} (y_{t,i} - \hat b\,\tilde\zeta_{t,i})^2}{\sum_{\text{OOS}} y_{t,i}^2}.
\end{equation}
This is the strict version, meaning it can go negative if the IS-fitted scale doesn't transfer.

Also, we compute IC (cross-sectional information coefficient. For each OOS date $t$ (and holding period), the cross-sectional correlation across the $N=14$ assets,
$t_{\rm IC} = \overline{\mathrm{IC}}/[\mathrm{std}(\mathrm{IC}_t)/\sqrt{T_{\rm oos}}]$, reported as mean IC with a t-stat $= \overline{\mathrm{IC}} \sqrt{T_{\text{oos}}} / \mathrm{sd}(\mathrm{IC}_t)$. This measures the ranking ability the portfolio actually exploits, and is robust to the pooled estimator being inflated by common time-series variation. These values computed for our dataset at $T = 31$ are presented in \cref{tab:predictive_metrics}.
\begin{table}[!htb]
\centering
\begin{tabular}{c c c c c c c}
\toprule
& \multicolumn{3}{c}{\textbf{Horizon T=31}} & \multicolumn{3}{c}{\textbf{Horizon T=63}} \\
\cmidrule(lr){2-4} \cmidrule(lr){5-7}
\textbf{$q$} & \textbf{$R^2_{\text{OOS}}$ (GKX)} & \textbf{Mean IC} & \textbf{t-stat} & \textbf{$R^2_{\text{OOS}}$ (GKX)} & \textbf{Mean IC} & \textbf{t-stat} \\
\hline
0.05 & 0.0280 & 0.129 & 28.5 & 0.0256 & 0.126 & 40.5 \\
0.10 & 0.0501 & 0.186 & 41.5 & 0.0610 & 0.180 & 58.3 \\
0.20 & 0.1147 & 0.273 & 61.7 & 0.1166 & 0.263 & 86.6 \\
\bottomrule
\end{tabular}
\caption{Predictive Performance Metrics across Horizons}
\label{tab:predictive_metrics}
\end{table}

\myparagraph{Terminal utility and the profit target.}
The terminal utility \eqref{quadratic_U} is quadratic in the deviation of the cumulative cost from a target, $U(C_T) = (C_T - z_{\mathrm{\rm tg}})^2$. This is a tracking objective in the spirit of a capital-market-line benchmark: the policy is rewarded for landing near a stated performance level with low dispersion, rather than for unbounded profit. For an institutional mandate this is often the relevant criterion, and the quadratic form keeps $U'(C)$ linear in $C$, which preserves the tractability of the terminal conditions in \eqref{NLL_PDE} and of the analytic Gibbs step.

The target level is set as $z_{\mathrm{\rm tg}} = \Not\,(1 - e^{0.1 T})$, matching \eqref{z_tg_CML} with $\Pi_0 = \Not$ and $r_{\rm tg} = 0.1$, which for horizons of one to three months is numerically indistinguishable from $-0.1\,\Not\,T$. Since negative cumulative cost corresponds to profit, this amounts to a profit target of roughly ten percent annualized, a level consistent with the predictive quality of the engineered signal. The choice involves a trade-off. A symmetric quadratic penalizes overshooting the target as much as missing it, so it caps the upside by construction, and a target set well beyond what the signal supports pushes the policy toward higher turnover without commensurate return. A linear utility, or an asymmetric shortfall penalty, would remove the cap at the price of losing the closed-form terminal gradient; we regard the target level and the utility shape as modeling inputs to be set by the mandate.

\myparagraph{Choice of the saturation scale $\tau_{\mathrm{sat}}$.}
The envelope $g(\tau)$ interpolates between two regimes. For $\tau \ll \tau_{\mathrm{sat}}$ it reduces to $g(\tau) \approx \tau$, reproducing the standard near-maturity expansion of the value function, so at horizons up to the saturation scale the ansatz coincides with the plain linear-in-$\tau$ form. For $\tau \gg \tau_{\mathrm{sat}}$ the state-gradients of $J_\theta$ saturate to a horizon-independent level, consistent with the turnpike intuition that far from maturity the marginal value of a state perturbation approaches a stationary shadow price rather than growing linearly with time-to-go.

The reported experiments use the unsaturated envelope $g(\tau) = \tau$, corresponding to $\tau_{\mathrm{sat}} = \text{None}$ in \cref{tab:exp_params}; the saturation is therefore described but not exercised in the results below. A natural scale for longer horizons is one trading month, $\tau_{\mathrm{sat}} = 21/252$: beyond the horizon over which the signal retains predictive content, the value of additional time-to-go should not accumulate, and $\tau_{\mathrm{sat}}$ then directly controls the growth of the policy tilt and hence of turnover. A one-figure sensitivity study, reporting out-of-sample Sharpe ratio and turnover at $T = 63$ over a grid such as $\tau_{\mathrm{sat}} \in \{10, 21, 42, 63\}/252$ together with the unsaturated limit $g(\tau) = \tau$, quantifies this dependence and is a natural robustness check for the device.

\myparagraph{Turnover.}
Throughout the results, \emph{turnover} denotes the total one-way fraction of the
portfolio traded over the holding period (episode),
\begin{equation} \label{eq:turnover_def}
\mathrm{Turnover} = \sum_{t=0}^{T-1} \frac{1}{\No{0}} \sum_{i=1}^{N} S_{t,i}\, \big| x_{t+\Delta t, i} - x_{t,i} \big|,
\end{equation}
i.e., the dollar value traded per step, normalized by the portfolio notional and
summed over the holding period. A value of $1.0$ means the strategy trades its
entire book once over the episode.

\paragraph{Test $\bm{T = 31}$ days.}

The results for $T = 31$ days with $q = 0.1$ are presented in \cref{tab:performance_metrics_31} and \cref{fig:is_metrics_31,fig:os_metrics_31} (\Cref{panel1}). These computations were performed using a Python script on an Intel Quad-Core i7-4790 CPU (3.80 GHz), with a total execution time of 260 seconds.

\begin{panel}{$T=31$ days, $q = 0.1, R^2 = 0.0501$, other parameters as in \cref{tab:exp_params}.}[panel1]
\begin{tabular}{l c c c c}
    \toprule
    \textbf{Policy} & \textbf{Sharpe} & \textbf{Ann. Ret.} & \textbf{Ann. Vol.} & \textbf{Turnover} \\
    \hline
    \multicolumn{5}{c}{\textit{Panel A: In-Sample (1008 episodes, 14 assets, $T=31$)}} \\
    \hline
    Gibbs      & 0.611 & 0.074 & 0.120 & 0.5475 \\
    Equal      & 0.502 & 0.078 & 0.155 & 0.2349 \\
    Behavioral & 0.441 & 0.067 & 0.153 & 1.2428 \\
    \hline
    \multicolumn{5}{c}{\textit{Panel B: Out-of-Sample (120 episodes, 14 assets, $T=31$)}} \\
    \hline
    Gibbs      & 1.120 & 0.090 & 0.080 & 0.5414 \\
    Equal      & 1.074 & 0.109 & 0.102 & 0.1971 \\
    Behavioral & 0.937 & 0.096 & 0.102 & 1.2421 \\
    \bottomrule
\end{tabular}
\captionof{table}{Performance Metrics: In-Sample vs. Out-of-Sample ($T=31$ days, $q = 0.1, R^2 = 0.0501$).}
\label{tab:performance_metrics_31}
\vspace{2em}

\includegraphics[width=\textwidth]{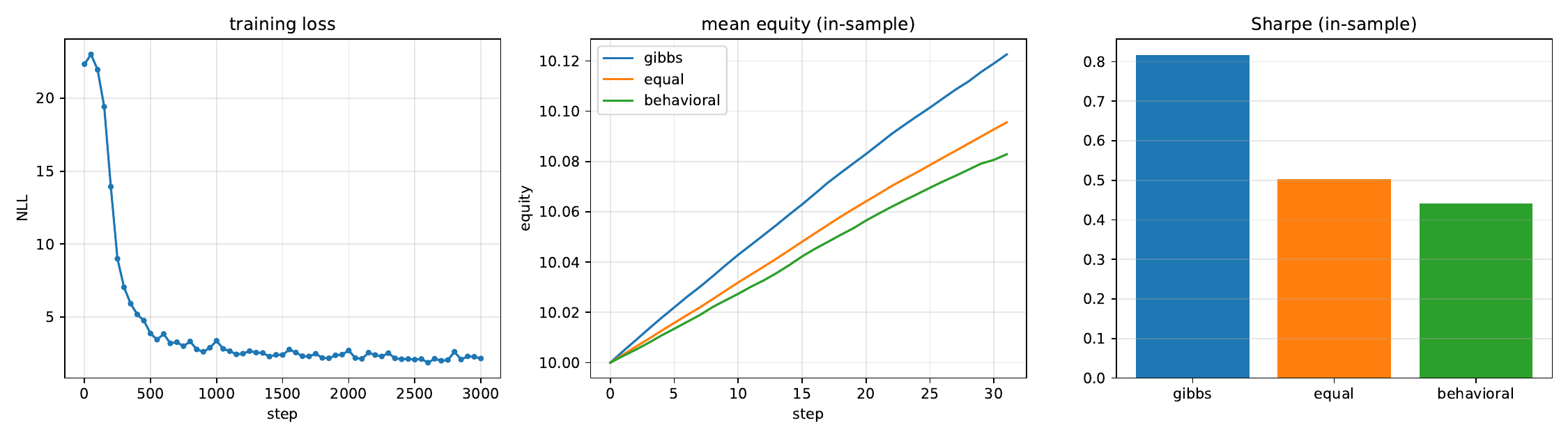}
\captionof{figure}{In-sample training and performance metrics for $T=31$ days, $q = 0.1, R^2 = 0.0501$. \textbf{(Left)} The negative log-likelihood (NLL) training loss converging over 3,000 optimization steps. \textbf{(Middle)} Mean equity trajectories over the in-sample period, comparing the trained Gibbs policy against the equal-weight and behavioral baselines. \textbf{(Right)} In-sample Sharpe ratios for the three policies, illustrating the superior risk-adjusted performance of the Gibbs strategy.}
\label{fig:is_metrics_31}
\vspace{2em}

\includegraphics[width=\textwidth]{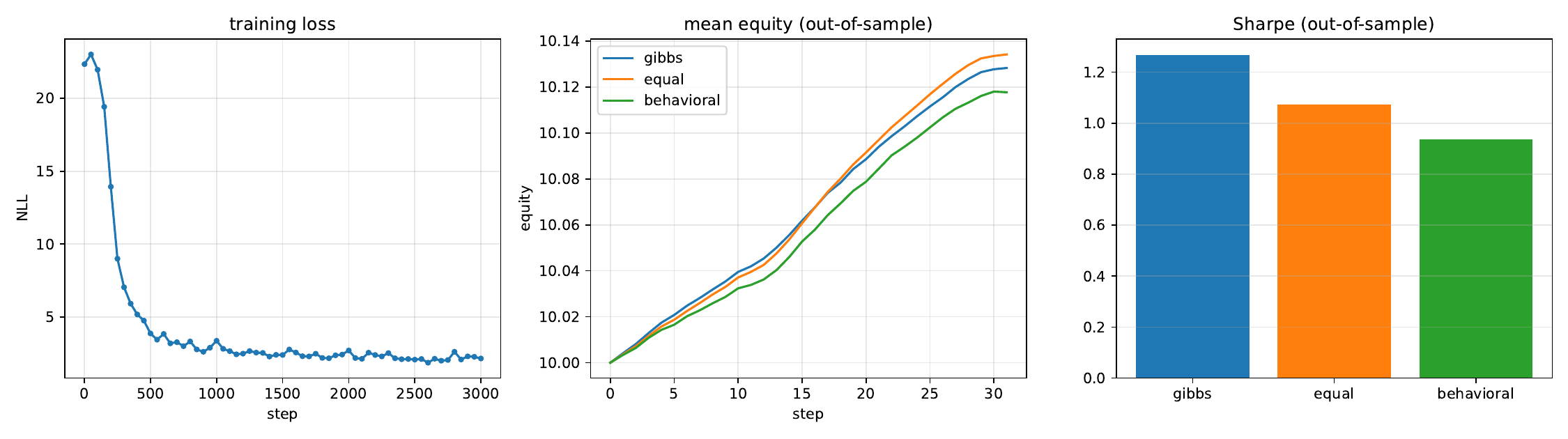}
\captionof{figure}{Same test as in \cref{fig:is_metrics_31} but out-of-sample training and performance metrics.}
\label{fig:os_metrics_31}

\end{panel}

Similar results for $q = 0.2$ are presented in \cref{tab:performance_metrics_31_02} and  \cref{fig:is_metrics_31_02,fig:os_metrics_31_02} (\Cref{panel2}).

\begin{panel}{$T=31$ days, $q = 0.2, R^2 = 0.1147$, other parameters as in \cref{tab:exp_params}.}[panel2]
\begin{tabular}{l c c c c}
\toprule
\textbf{Policy} & \textbf{Sharpe} & \textbf{Ann. Ret.} & \textbf{Ann. Vol.} & \textbf{Turnover} \\
\hline
\multicolumn{5}{c}{\textit{Panel A: In-Sample (1008 episodes, 14 assets, $T=31$)}} \\
\hline
Gibbs      & 1.180 & 0.140 & 0.118 & 1.1987 \\
Equal      & 0.502 & 0.078 & 0.155 & 0.2349 \\
Behavioral & 0.441 & 0.067 & 0.153 & 1.2428 \\
\hline
\multicolumn{5}{c}{\textit{Panel B: Out-of-Sample (120 episodes, 14 assets, $T=31$)}} \\
\hline
Gibbs      & 1.489 & 0.123 & 0.082 & 1.1485 \\
Equal      & 1.074 & 0.109 & 0.102 & 0.1971 \\
Behavioral & 0.937 & 0.096 & 0.102 & 1.2421 \\
\bottomrule
\end{tabular}
\captionof{table}{Performance Metrics: In-Sample vs. Out-of-Sample ($T=31, q = 0.2, R^2 = 0.1147$).}
\label{tab:performance_metrics_31_02}

\vspace{2em}

\includegraphics[width=\textwidth]{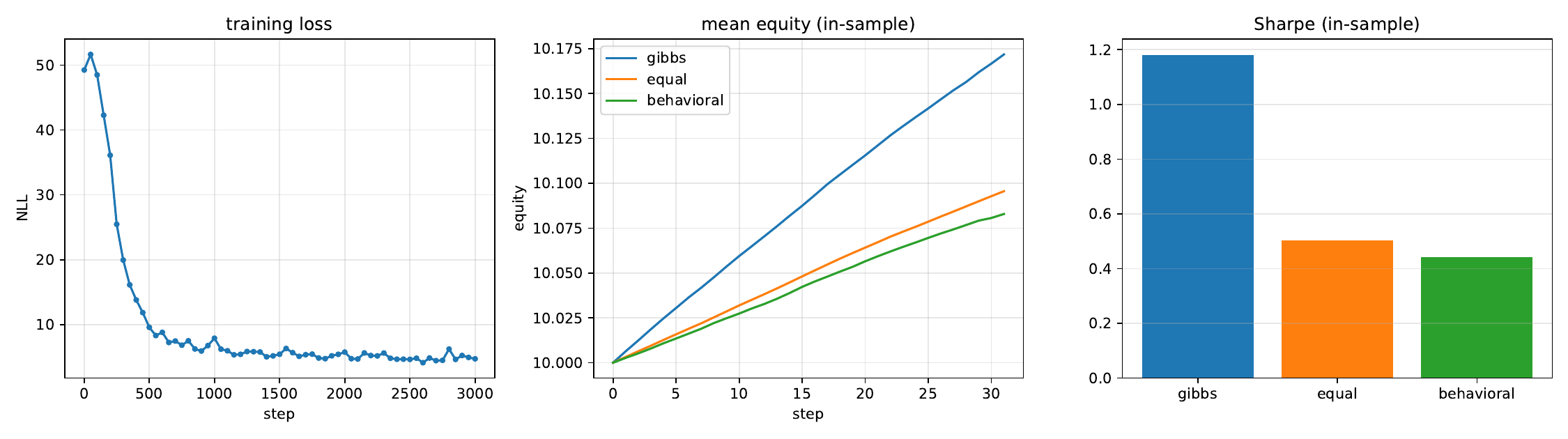}
\captionof{figure}{In-sample training and performance metrics for $T=31$ days and $q = 0.2, R^2 = 0.1147$. \textbf{(Left)} The negative log-likelihood (NLL) training loss converging over 3,000 optimization steps. \textbf{(Middle)} Mean equity trajectories over the in-sample period, comparing the trained Gibbs policy against the equal-weight and behavioral baselines. \textbf{(Right)} In-sample Sharpe ratios for the three policies, illustrating the superior risk-adjusted performance of the Gibbs strategy.}
\label{fig:is_metrics_31_02}

\vspace{2em}

\includegraphics[width=\textwidth]{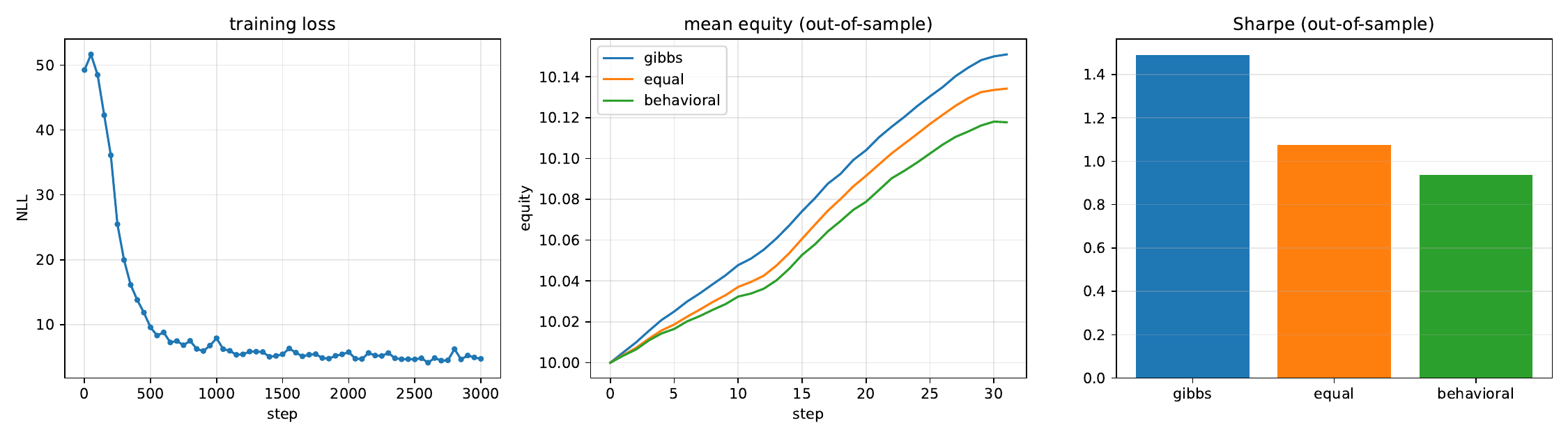}
\captionof{figure}{Same test as in \cref{fig:is_metrics_31_02} but out-of-sample training.}
\label{fig:os_metrics_31_02}

\end{panel}

\paragraph{Test $\bm{T = 63}$ days.}

The results for $T = 63$ days with $q = 0.2$ are presented in \cref{tab:performance_metrics_63} as well as   \cref{fig:is_metrics_63,fig:os_metrics_63} (\Cref{panel3}).

\begin{panel}{$T=63$ days, $q = 0.2, R^2 = 0.1166$, other parameters as in
\cref{tab:exp_params}.}[panel3]

\begin{tabular}{l c c c c}
\toprule
\textbf{Policy} & \textbf{Sharpe} & \textbf{Ann. Ret.} & \textbf{Ann. Vol.} & \textbf{Turnover} \\
\hline
\multicolumn{5}{c}{\textit{Panel A: In-Sample (1008 episodes, 14 assets, $T=63$)}} \\
\hline
Gibbs      & 1.139 & 0.127 & 0.112 & 2.4322 \\
Equal      & 0.447 & 0.069 & 0.154 & 0.4878 \\
Behavioral & 0.390 & 0.059 & 0.151 & 2.5388 \\
\hline
\multicolumn{5}{c}{\textit{Panel B: Out-of-Sample (120 episodes, 14 assets, $T=63$)}} \\
\hline
Gibbs      & 0.526 & 0.043 & 0.081 & 2.2698 \\
Equal      & 0.126 & 0.013 & 0.104 & 0.3837 \\
Behavioral & 0.025 & 0.003 & 0.105 & 2.5090 \\
\bottomrule
\end{tabular}
\captionof{table}{Performance Metrics: In-Sample vs. Out-of-Sample ($T=63, q = 0.2, R^2 = 0.1166, \beta = 15$).}
\label{tab:performance_metrics_63}
\vspace{2em}

\includegraphics[width=\textwidth]{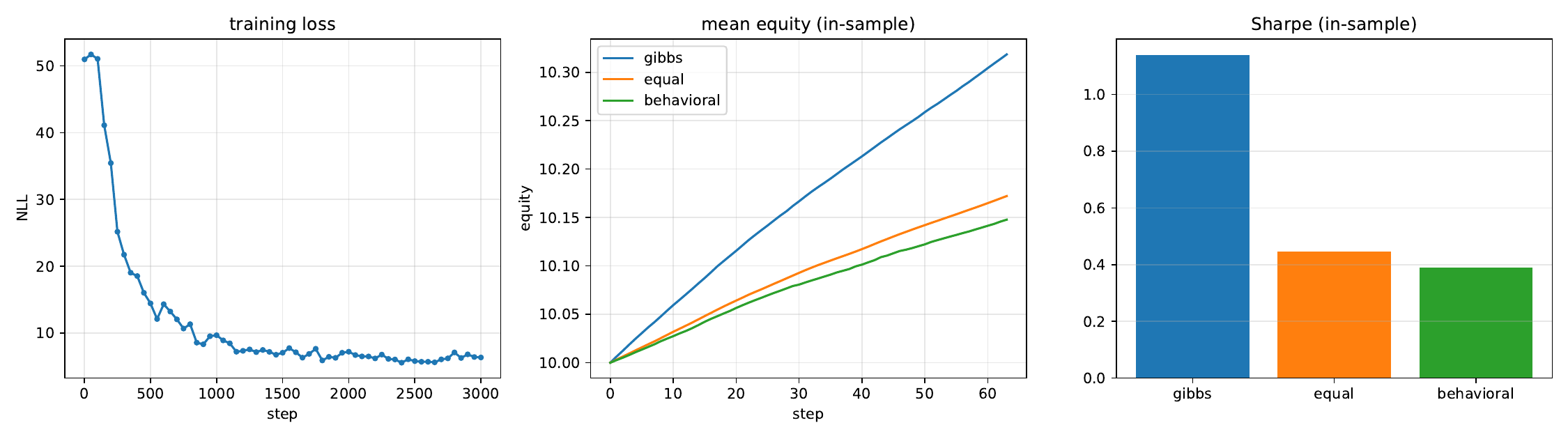}
\captionof{figure}{In-sample training and performance metrics, $T = 63$ days and $q = 0.2, R^2 = 0.1166, \beta=15$. \textbf{(Left)} The negative log-likelihood (NLL) training loss converging over 3,000 optimization steps. \textbf{(Middle)} Mean equity trajectories over the in-sample period, comparing the trained Gibbs policy against the equal-weight and behavioral baselines. \textbf{(Right)} In-sample Sharpe ratios for the three policies, illustrating the superior risk-adjusted performance of the Gibbs strategy.}
\label{fig:is_metrics_63}
\vspace{2em}

\includegraphics[width=\textwidth]{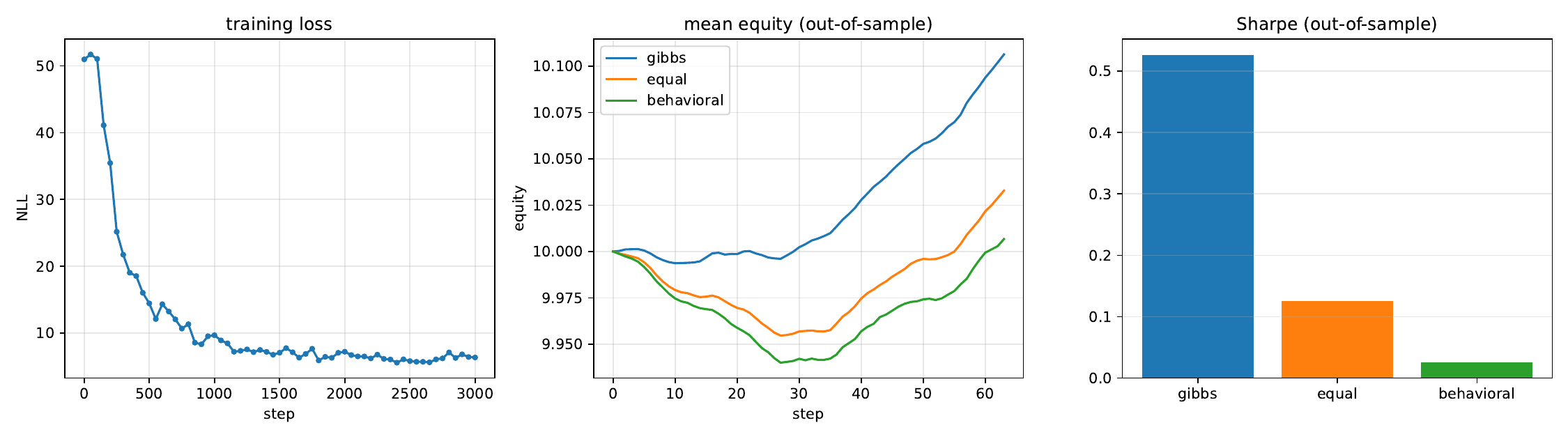}
\captionof{figure}{Same test as in \cref{fig:is_metrics_63} but out-of-sample training and performance metrics, $T = 63$ days.}
\label{fig:os_metrics_63}
\end{panel}

Similar results utilizing $\beta = 5$ and a target CML cost of $z_{\rm tg} = \Not(1.0 - e^{0.2 T})$ are presented in \cref{tab:performance_metrics_63_B5} and \cref{fig:is_metrics_63_B5,fig:os_metrics_63_B5} (\Cref{panel4}). While adjusting $\beta$ does not alter the intrinsic characteristics of the signals, the results clearly demonstrate that both $\beta$ and $z_{\mathrm{\rm tg}}$ significantly influence the resulting Sharpe ratio and annual return. Consequently, these can be treated as free parameters, allowing the model's performance to be optimized through empirical calibration.

\begin{panel}{$T=63$ days, $q = 0.2, R^2 = 0.1166, \beta = 25$, $z_{\rm tg} = \Not(1.0 - e^{0.2 T})$, other parameters as in \cref{tab:exp_params}.}[panel4]
\begin{tabular}{l c c c c}
\toprule
\textbf{Policy} & \textbf{Sharpe} & \textbf{Ann. Ret.} & \textbf{Ann. Vol.} & \textbf{Turnover} \\
\hline
\multicolumn{5}{c}{\textit{Panel A: In-Sample (1008 episodes, 14 assets, $T=63$)}} \\
\hline
Gibbs      & 1.618 & 0.137 & 0.085 & 4.1490 \\
Equal      & 0.447 & 0.069 & 0.154 & 0.4878 \\
Behavioral & 0.390 & 0.059 & 0.151 & 2.5388 \\
\hline
\multicolumn{5}{c}{\textit{Panel B: Out-of-Sample (120 episodes, 14 assets, $T=63$)}} \\
\hline
Gibbs      & 0.706 & 0.044 & 0.062 & 3.8942 \\
Equal      & 0.126 & 0.013 & 0.104 & 0.3837 \\
Behavioral & 0.025 & 0.003 & 0.105 & 2.5090 \\
\bottomrule
\end{tabular}
\caption{Performance Metrics: In-Sample vs. Out-of-Sample ($T=63, q = 0.2, \beta = 25$).}
\label{tab:performance_metrics_63_B5}
\vspace{2em}

\includegraphics[width=\textwidth]{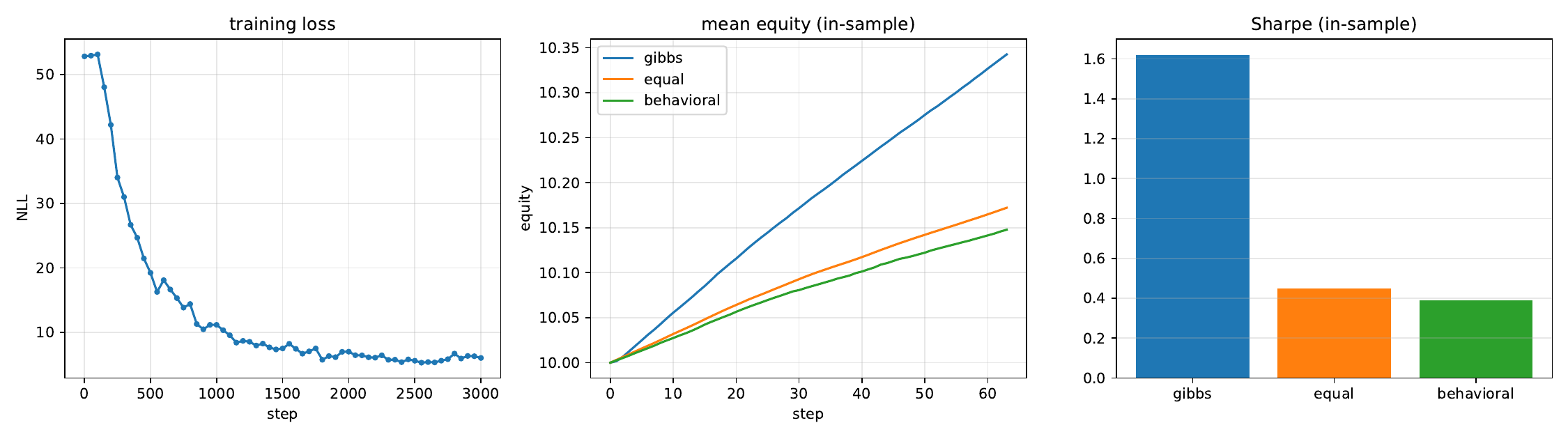}
\captionof{figure}{In-sample training and performance metrics, $T = 63$ days and $q = 0.2, \beta=25$ and target CML cost $\Not(1.0 - e^{0.2 T})$. \textbf{(Left)} The negative log-likelihood (NLL) training loss converging over 3,000 optimization steps. \textbf{(Middle)} Mean equity trajectories over the in-sample period, comparing the trained Gibbs policy against the equal-weight and behavioral baselines. \textbf{(Right)} In-sample Sharpe ratios for the three policies, illustrating the superior risk-adjusted performance of the Gibbs strategy.}
\label{fig:is_metrics_63_B5}
\vspace{2em}

\includegraphics[width=\textwidth]{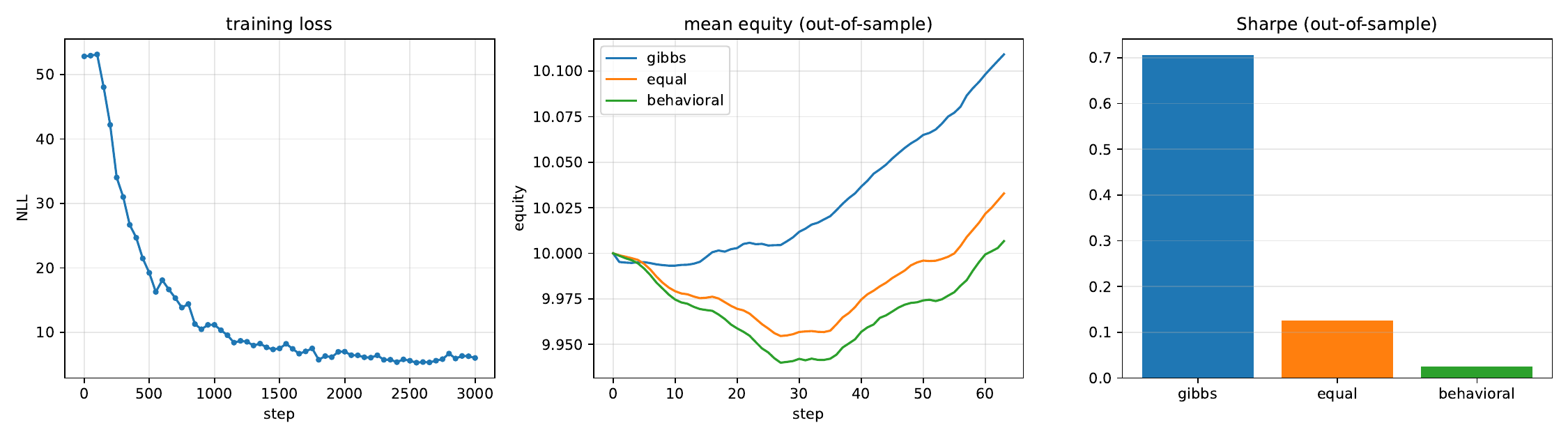}
\captionof{figure}{Same test as in \cref{fig:is_metrics_63_B5} but out-of-sample training and performance metrics, $T = 63$ days.}
\label{fig:os_metrics_63_B5}
\end{panel}

Comparing \cref{tab:performance_metrics_31_02} with \cref{tab:performance_metrics_63}, the OOS Sharpe ratio exceeds the IS one at $T=31$ but falls below it at $T=63$. This reversal is a property of the test window, not of the learned policy. The signal-free equal-weight benchmark exhibits the same flip: 0.502 IS versus 1.074 OOS at $T=31$, and 0.447 versus 0.126 at $T=63$. The reason is that the two test blocks occupy different stretches of 2023. The one-month episodes run from mid-February to late September, a calm rising period. The three-month episodes run from early April to late December and straddle the autumn drawdown, cf. Fig.~\ref{fig:os_metrics_31} and Fig.~\ref{fig:os_metrics_63}.

The position of that drawdown is fixed by the chronological split, and it can be
dated. The sample carries about $1762$ daily observations, and at $T = 63$ the
window stride resolves to a single day, so window $i$ begins at observation $i$.
Training occupies windows $0$ to $1007$, whose start dates cover almost exactly
the four calendar years 2019 to 2022. A purge gap of $63$ windows follows. The
test block then occupies windows $1071$ to $1190$, whose price paths run from
early April 2023 to late December 2023. The market dislocation of February and
March 2020 therefore lies inside the training sample, and it cannot account for
the out-of-sample trough.

The trough is the joint drawdown of August to October 2023, when equities,
Treasuries, credit and real estate fell together as the ten-year yield approached
$5\%$, followed by the broad recovery of November and December. All fourteen
assets moved in the same direction, so the decline is present in the passive
equal-weight curve as well as in the learned one, which places it in the price
paths rather than in the policy. Two checks support the identification. The
trough lies about $0.5\%$ below the initial notional, whereas a window of $63$
days containing the crash of February and March 2020 would leave an
equal-weighted multi-asset book $15$ to $20\%$ below its starting value. The
in-sample panels do contain that crash and show no trough at all, since only
about $8\%$ of the $1008$ training windows overlap it, and the average over
windows dilutes any single episode.

In both cases transaction costs are around 10 bps. To verify that the reward is well scaled, we monitor two diagnostics along simulated trajectories. The first is the per-step tracking error, defined as the standard deviation of the active return, $\sqrt{\Delta t \, \mathbb{E}[x_t^\top \Sigma x_t]}$ with $x_t = S_t \Delta w_t$, which quantifies the magnitude of active risk taken by the policy per rebalancing interval. In these experiments it is typically $4\cdot 10^{-3}$. The second is the risk-to-signal ratio, $\lambda \, \mathbb{E}[x_t^\top \Sigma x_t] / |\mathbb{E}[x_t^\top (w_t \circ \zeta_t)]|$, the ratio of the mean quadratic risk penalty to the mean signal contribution in the reward. A ratio of order one indicates that the risk aversion parameter $\Lambda$ is calibrated so that neither term dominates, whereas a large ratio signals that positions are dictated by the penalty rather than by the predictive signal. In our experiments this ratio is around 7-15.

Measured relative to the benchmark, the OOS Sharpe advantage of the Gibbs policy is nearly identical at both horizons, 0.415 at $T=31$ and 0.400 at $T=63$. The edge attributable to the method is thus horizon-stable, while its absolute level moves with the market regime. We also note that at $T=63$ the 120 rolling test episodes overlap in 62 of 63 days, so a single adverse stretch affects nearly all of them. The OOS levels at this horizon should be read with this dependence in mind, as discussed further in \cref{sect_Summary}.

\Cref{tab:rl_performance} reports Interquartile Means (IQM) with 95\% bootstrap confidence intervals across 10 random seeds for the test in \cref{tab:performance_metrics_63}.
\begin{table}[!htb]
\centering
\begin{tabular}{lcc}
\toprule
\textbf{Metric} & \textbf{In-Sample} & \textbf{Out-of-Sample} \\
\hline
Sharpe Ratio & 1.13867 [1.13859, 1.13876] & 0.52564 [0.52544, 0.52580] \\
Annual Return & 0.12737 [0.12736, 0.12737] & 0.04251 [0.04250, 0.04253] \\
Annual Volatility & 0.11185 [0.11185, 0.11186] & 0.08088 [0.08087, 0.08088] \\
Turnover & 2.43220 [2.43216, 2.43223] & 2.26964 [2.26958, 2.26970] \\
\bottomrule
\end{tabular}
\caption{Performance Metrics with Interquartile Mean (95\% CI) from bootstrap resampling over 10 random seeds.}
\label{tab:rl_performance}
\end{table}
The narrow width of the bootstrap confidence intervals across all metrics, both in- and out-of-sample, indicates that the results exhibit only a weak dependence on the random seed, so the reported performance is not an artifact of stochastic initialization. These intervals quantify sensitivity to the seed alone. They are not a test of the significance of the Sharpe ratios themselves, which would have to account for the overlap of the rolling windows, and which we leave for future work as discussed in \cref{sect_Summary}.

Another natural question is whether the OOS Sharpe ratio of 1.49 at $T=31$ exhausts the information content of the signal. It does not, and the gap is by design. By the fundamental law of active management, a frictionless strategy exploiting the realized cross-sectional IC of 0.27 from \cref{tab:predictive_metrics} over $N=14$ assets rebalanced daily has a gross annualized information ratio of order $\mathrm{IC}\sqrt{N \cdot 252} \approx 16$. The realized Sharpe ratio thus corresponds to a transfer coefficient of order 0.1, and the tables identify what consumes the remainder.

The first consumer is the cost structure. Under the jump transition the per-step quadratic cost is $O(1/\Delta t)$ in the trade size, so full daily expression of the signal is prohibitively expensive. The signal-tilted baseline of \cref{sect_baselines} shows this directly: it trades 3.55 times the book and reaches only 1.23. The second consumer is the objective itself. The quadratic tracking utility targets a CML cost with $r_{\rm tg}=0.1$, the realized annualized return of 0.123 lands essentially on that target, and the symmetric penalty caps the upside by construction. \cref{tab:performance_metrics_63,tab:performance_metrics_63_B5} quantify the sensitivity: moving $(\beta, z_{\rm tg})$ from $(15,\, r_{\rm tg}=0.1)$ to $(5,\, r_{\rm tg}=0.2)$ halves the OOS Sharpe ratio at $T=63$. The third consumer is the set of shrinkage terms, namely the KL penalty toward the near-equal-weight prior, the risk penalty (realized volatility 0.080 versus 0.102 for the benchmark), and the notional penalty, each of which pulls the policy toward the passive benchmark.

Higher Sharpe ratios are therefore attainable within the same framework by raising $\beta$ and $r_{\rm tg}$, relaxing $\Lambda$, or expanding breadth, since the fundamental law scales as $\sqrt{N}$ and 14 assets is a small universe. The trade-off is that these levers move the policy away from the behavioral support, so extrapolation risk grows exactly as the shrinkage that guards against it is removed. The reported numbers should thus be read as the performance of the mechanism under a conservative institutional mandate, not as the ceiling of the signal.

\myparagraph{A composed memory-aware signal.}
A further remark is warranted regarding the temporal structure of the oracle signal. The formulation in \eqref{eq:raw_oracle} represents a memoryless, contemporaneous process, constructed strictly as a linear combination of the instantaneous normalized log-return and independent Gaussian noise at time $t$. A signal of this kind changes almost entirely from one step to the next, which translates into an unrealistically high portfolio trading turnover. Indeed, for a unit-variance stationary signal $\zeta_t$ the variance of its increments satisfies $\text{Var}(\zeta_t - \zeta_{t-1}) = 2(1 - \rho_{\zeta})$, where $\rho_{\zeta}$ denotes the lag-1 autocorrelation of the signal. This quantity serves as a direct proxy for expected turnover, and for a memoryless signal $\rho_{\zeta} \approx 0$, so the turnover proxy attains nearly its maximal value. To obtain a signal with a realistic persistence profile we impose an endogenous financial constraint: the lag-1 autocorrelation of the signal must match the empirical lag-1 autocorrelation of the underlying asset returns,
\begin{equation}
\rho_{\zeta} = \rho_{\text{asset}}, \qquad \rho_{\text{asset}} \equiv \text{Corr}\!\left(\tilde{r}_t, \tilde{r}_{t-1}\right),
\label{eq:turnover_constraint}
\end{equation}
where $\tilde{r}_t = \dot{\bS}_t / (\sigma \bS_t r_t)$ is the normalized log-return. This choice avoids introducing an arbitrary exogenous persistence parameter and ties the turnover of the synthetic signal to the temporal regime of the asset itself.

To satisfy this constraint the signal must carry memory, which we embed twofolds: a historical inertia, represented by an exponentially weighted moving average (EWMA) of past returns, and a persistence in the unobservable noise component. The composite signal is defined as
\begin{equation} \label{eq:composite_signal}
\zeta_t = \alpha_1 \tilde{r}_{t+1} + \beta_1 \, \text{EWMA}_t(\tilde{r}) + \sqrt{1 - q} \, \eta_t, \qquad
\eta_t = \varphi \, \eta_{t-1} + \sqrt{1 - \varphi^2} \, u_t,
\end{equation}
where $u_t \sim \mathcal{N}(0,1)$ is independent Gaussian noise, $\eta_t$ is a stationary unit-variance AR(1) process with persistence $\varphi \in (-1,1)$.

The autocorrelated form of the noise is not a decorative choice. Suppose instead that $\eta_t$ were white noise. Since white noise contributes nothing to the lag-1 autocovariance of $\zeta_t$, the constraint \eqref{eq:turnover_constraint} would have to be met entirely by the informative component $\alpha \tilde{r}_{t+1} + \beta_1 \, \text{EWMA}_t(\tilde{r})$, which carries only a fraction $q$ of the total variance. Its internal autocorrelation would then need to equal $\rho_{\text{asset}}/q$, a quantity amplified by the factor $1/q$ beyond what any admissible combination of a near-white oracle and a persistent EWMA can produce. For realistic values $q \ll 1$ the resulting system of equations for $(\alpha, \beta_1)$ is infeasible. Allowing the noise to carry persistence removes this obstruction and is also economically natural, since the idiosyncratic component of real alpha signals is itself serially correlated, and for small $q$ it is precisely this component that dominates realized turnover.

The parameters $(\alpha, \beta_1, \varphi)$ are determined sequentially by three conditions. First, a memory share $m \in [0,1)$ fixes the fraction of informative variance attributed to the historical returns (the forward $R^2$ equals $(1-m) q$ up to small cross terms)
\begin{equation} \label{eq:beta_memory_share}
\beta_1^2 \, \text{Var}\!\left(\text{EWMA}_t\right) = m \, q.
\end{equation}
Second, the total informative variance is pinned to $q$ through the quadratic constraint
\begin{equation}
\alpha^2 \, \text{Var}(\tilde{r}_{t+1}) + \beta_1^2 \, \text{Var}\!\left(\text{EWMA}_t\right) + 2 \alpha \beta_1 \, \text{Cov}\!\left(\tilde{r}_{t+1}, \text{EWMA}_t\right) = q,
\label{eq:variance_constraint}
\end{equation}
whose free term equals $q(m-1) < 0$, so that a unique positive root $\alpha > 0$ always exists. Third, the constraint \eqref{eq:turnover_constraint} determines $\varphi$. Since the AR(1) noise contributes $(1-q)\varphi$ to the lag-1 autocovariance of $\zeta_t$, the condition yields $\varphi$ in closed form,
\begin{gather} \label{eq:phi_closure}
\varphi = \frac{\rho_{\text{asset}} - \Sigma_1}{1 - q}, \\
\begin{align*}
\Sigma_1 &= \alpha^2 \, \text{Cov}\!\left(\tilde{r}_{t+1}, \tilde{r}_t\right) + \beta_1^2 \, \text{Cov}\!\left(\text{EWMA}_t, \text{EWMA}_{t-1}\right)
+ \alpha \beta_1 \left[ \text{Cov}\!\left(\tilde{r}_{t+1}, \text{EWMA}_{t-1}\right) + \text{Cov}\!\left(\text{EWMA}_t, \tilde{r}_t\right) \right],
\end{align*}
\end{gather}
where all variances and covariances are computed empirically from the simulated time-series arrays. Because $\Sigma_1 = O(q)$ and $|\rho_{\text{asset}}|$ is small for daily returns, $\varphi$ falls strictly inside the stationarity region $(-1,1)$ for all realistic configurations, and no feasibility projection is required. This framework thus guarantees strict unit variance of the synthetic alpha signal, the designated informational predictive power $q$, and an asset-conforming turnover profile, with each parameter determined by a single interpretable condition.

The results of such a test with $T=63, q = 0.1, \beta = 25, m = 0.3$ and ${\rm span} = 3$ are presented in \cref{tab:ewma}. The computed coefficients in \eqref{eq:composite_signal} are $\alpha_1 = 0.2672, \beta_1 = 0.2943, \phi = -0.0831, \rho_{\text{asset}} = -0.0229$.
\begin{table}[!htb]
\centering
\begin{tabular}{l c c c c}
\toprule
\textbf{Policy} & \textbf{Sharpe} & \textbf{Ann. Ret.} & \textbf{Ann. Vol.} & \textbf{Turnover} \\
\hline
\multicolumn{5}{c}{\textit{Panel A: In-Sample (1008 episodes, 14 assets, $T=63$)}} \\
\hline
Gibbs      & 0.753 & 0.068 & 0.090 & 2.8828 \\
Equal      & 0.447 & 0.069 & 0.154 & 0.4878 \\
Behavioral & 0.390 & 0.059 & 0.151 & 2.5388 \\
\hline
\multicolumn{5}{c}{\textit{Panel B: Out-of-Sample (120 episodes, 14 assets, $T=63$)}} \\
\hline
Gibbs      & 0.414 & 0.025 & 0.061 & 2.8115 \\
Equal      & 0.126 & 0.013 & 0.104 & 0.3837 \\
Behavioral & 0.025 & 0.003 & 0.105 & 2.5090 \\
\bottomrule
\end{tabular}
\caption{Performance Metrics: In-Sample vs. Out-of-Sample ($T=63, q=0.1, \beta=25, {\rm span} = 3, R^2_{OOS} = 0.07$.}
\label{tab:ewma}
\end{table}
The performance hierarchy remains robust, with the Gibbs policy consistently outperforming both the equal-weighted and behavioral benchmarks.

\section{Discussion} \label{sect_Discussion}

\myparagraph{Weight-based control.}
The most consequential choice in the empirical framework is the reformulation of the control from a trading rate to a target holding, \eqref{eq:induced_rate}. In the continuous-time formulation $d\bx_t = \ba_t\,dt$ the position is an integral of the controlled rate. Over a short episode of daily steps this integral behaves as a slowly mean-reverting average of small tilts, so the realized holding cannot reach the signal-implied target before the episode terminates.

The consequence is that the predictive signal moves the instantaneous trade but not the carried position, and the strategy relaxes onto the equal-weight benchmark regardless of signal quality. Replacing the rate dynamics with the instantaneous jump $\bx_{t+\Delta t} = \bh_t$ removes this transient. The target holding is realized in a single step, while the execution cost is charged on the induced rate $\ba_t = (\bh_t - \bx_t)/\Delta t$ through the same impact and transaction-cost terms, \eqref{cost_quadratic}. Cost-awareness is preserved, and cross-sectional information can now be expressed in the portfolio within the horizon.

\myparagraph{Signal persistence and turnover.}
Because the weight-based action sets the target holding directly, an i.i.d.\ signal would drive the target to fluctuate at every step and incur turnover that the transaction-cost term penalizes heavily. A monetizable signal must therefore be persistent relative to the rebalancing frequency. Our signal construction imposes this persistence, so that the optimal target holding stays stable over several steps.

The reported turnover reflects this trade-off. In \cref{tab:performance_metrics_31,tab:performance_metrics_63} the Gibbs policy trades substantially more than the static equal-weight portfolio, yet a fraction of what the exploratory behavioral policy trades, and it does so while improving risk-adjusted return.

\myparagraph{The PINN solver.}
The value function is obtained with a physics-informed neural network that parameterizes $J_\theta(t, \bx, \bS, C)$ and learns it directly from offline data. The object it solves is the Hamilton-Jacobi equation \eqref{HJBh_nonlin_GM}, which is derived analytically from the distributional RL formulation of \cref{Soft_HJB_control} rather than posited. The network enters only as the approximation of that solution, and the oracle signal of \cref{sect_Experiments} is a choice of input that orchestrates the framework, not a separate solver.

Optimality is supplied by the loss \eqref{NLL_PDE}. It enforces the HJ relation pathwise along the observed price trajectories and reweights the behavioral data through the exact path-likelihood ratio, recasting it as draws from the optimal policy. The offline RL problem thereby becomes a single-sweep inference problem.

\myparagraph{Role of the value-function derivatives.}
The policy improvement is carried entirely by the derivatives $\partial J_\theta/\partial\bx$, $\partial J_\theta/\partial\bS$, and $\partial J_\theta/\partial C$. In the limit of vanishing derivatives the objective reduces to matching the behavioral drift to the observed price velocities, that is, to behavioral cloning with no optimization. The outperformance relative to the behavioral policy in our experiments is thus attributable to this derivative propagation across the wealth, holding, and price dimensions.

The hard-terminal ansatz $J(\tau, \bx, \bS, C) = U(C) + g(\tau) h_\theta(\bx, \bS, C)$ enforces the terminal conditions of \cref{sect_bc} exactly. The saturating envelope $g(\tau) = \tau_{\mathrm{sat}} \tanh(\tau/\tau_{\mathrm{sat}})$ additionally bounds the state-gradients of $J_\theta$ at long time-to-go. This keeps the induced policy tilt, and therefore turnover, from growing without bound as the horizon lengthens, while leaving the terminal behavior unchanged.

\myparagraph{Price impact model.}
The impact model of \cref{appImpact} is written as a quadratic form in the trading action. This keeps the running cost and the induced drift tractable for the analytic Gibbs step while retaining the convexity of impact. Its five effective parameters are tied to observable market quantities, in the spirit of Kyle's $\lambda$ and the square-root law, and the model carries cross-impact and a memory component, so that spillover and path-dependence are represented rather than discarded.

This interpretability rests on the modeling assumptions of \cref{appImpact}, and the sensitivity of the learned policy to the parameter reduction described there is not quantified separately.

\myparagraph{Limitations: signal and objective.}
The predictive signal is an engineered oracle, \eqref{eq:raw_oracle}, with a controlled out-of-sample $R^2$. The attainable outperformance depends on the persistence of the signal relative to the cost structure, so a real signal with different decay would produce different turnover and Sharpe. The tracking form of the terminal utility and the level of $z_{\mathrm{\rm tg}}$, discussed above, are likewise modeling inputs rather than outputs of the method. How the reported numbers should be read in consequence is taken up in \cref{sect_Summary}.

\myparagraph{Limitations: data and approximations.}
Further limitations concern the setup. Each rolling window supplies a single realized price path as the drift baseline, whereas the richer scenario generation noted in \cref{sect_Experiments}, from multiple non-overlapping windows and stress-scaled paths, is left to future work.

The analytic Gibbs minimizer also requires a diagonal quadratic exponent, which is
why only the diagonal of the centered covariance $K = P^T \Sigma P$ enters the
precision in \eqref{eq:risk_couplings} while the linear pull toward equal weight
retains the full matrix; this off-diagonal truncation, together with the
Gaussian-mixture approximation of the partition function, is an approximation
whose effect on the optimum is not isolated separately.

Note that the equal-weight benchmark ignores the signal, and the behavioral policy is deliberately noisy, so neither comparator isolates the contribution of the HJB machinery from the contribution of the signal itself. We therefore add three comparators, all using the same engineered oracle signal $\tilde{\bm\zeta}_t$, the same transaction-cost model \eqref{TC_fun}, and the same impact model of \cref{appImpact}. The results and extended discussion are provided in \cref{sect_baselines}.

Also, the jump transition \eqref{eq:induced_rate} sets the post-jump holding directly, so the portfolio notional is no longer pinned by the dynamics, and without a budget rule the Gibbs minimizer could implicitly lever or delever the book. Self-financing is therefore enforced softly, through a quadratic penalty on the notional deviation that preserves the Gaussian-mixture Gibbs structure, supplemented by exact cash-flow accounting through $C_t$, which absorbs any residual mismatch. The construction, its effect on the couplings, and its calibration are given in \cref{sect_self_financing}.

\section{Conclusion} \label{sect_Summary}

The paper poses dynamic portfolio optimization for a large institutional investor in continuous time over the extended state $\by_t = (\bx_t, \bS_t, C_t)$, in which the price vector is a stochastic driver and the cumulative cost is carried explicitly. Through the price-impact channel both the running cost and the price drift are quadratic in the action, which places the problem outside the linear-quadratic-regulator class. A distributional, entropy-regularized formulation turns the Hamilton-Jacobi-Bellman equation into a semilinear {\it Hamilton-Incepted} (HI) PDE \eqref{HJBh_nonlin_2}, and a Gaussian-mixture ansatz for the behavioral policy makes the partition function analytic, reducing the HI equation to another semilinear PDE in \eqref{HJBh_nonlin_GM}.

Instead of discretizing that PDE on a grid, we solve it directly from offline data by projecting it onto the observed trajectories: evaluating $J$ along a realized path and applying \Ito's lemma cancels the second-order term against the martingale increment and collapses \eqref{HJBh_nonlin_GM} to the first-order pathwise Hamilton-Jacobi equation \eqref{HJfin}. This is the stochastic analogue of the method of characteristics, with the observed sample paths as the characteristic curves, so the HJ equation is the route to the PDE and not a separate model. A PINN parameterizes $J_\theta$ and is trained by the loss \eqref{NLL_PDE}, which enforces the pathwise relation and, through the exact Girsanov likelihood ratio, reweights the behavioral data as optimal-policy demonstrations. The control problem is thereby solved in a {\it single offline sweep, without value or policy iteration}, and the PDE constraint regularizes the value function so that less data is required than for conventional deep RL.

Two modeling choices make the method effective at the short horizons of practical interest. Recasting the control from a trading rate to a target holding \eqref{eq:induced_rate} lets the signal-implied position be reached within the episode rather than lost to the slow transient of a rate-integrated holding, while the execution cost is still charged on the induced rate. The saturating time envelope in the terminal ansatz bounds the induced policy gradient, and hence turnover, as the horizon grows. The rate-based ablation confirms these are not cosmetic: without the reformulation the same machinery collapses onto the passive benchmark.

On a fourteen-asset ETF universe with seven years of daily data, the learned Gibbs policy attains an out-of-sample Sharpe ratio of $1.49$ at one month and $0.53$ at three months, against $1.07$ and $0.13$ for equal weight, at a volatility at or below the benchmark's and at turnover well under the exploratory behavioral policy's. The comparators locate the source of this edge. A one-step signal tilt beats equal weight only by trading three to seven times the book, a myopic mean-variance step barely clears it, and the rate-coordinate ablation is indistinguishable from it; only the combination of the target-holding reformulation and the value-function machinery delivers the full margin. What the method buys is multi-period, cost-aware allocation, not signal quality on its own.

These numbers measure a mechanism, not an attainable track record. The predictive signal is an engineered oracle with a controlled out-of-sample $R^2$, so the Sharpe ratios quantify how well the framework converts a signal of \emph{known} quality into a cost-aware multi-period allocation; they are not a claim about any proprietary alpha. A real signal with different persistence and decay would change both turnover and Sharpe ratio, because the attainable outperformance depends on how the signal's horizon compares with the cost structure. For a practitioner the transferable contribution is the framework itself, a distribution-aware, cost- and risk-sensitive optimizer trained offline in a single pass on an interpretable, microstructure-grounded impact model, rather than the specific returns reported here.

The main practical obstacle is calibration, not model size in itself. The impact model is already reduced from thirteen parameters to five tied to observable microstructure, but those five, together with the control hyperparameters (the inverse temperature $\beta$, the terminal target level, the notional and risk penalties, and the prior widths), were set to illustrative values that exhibit the mechanism rather than fitted to data, and their joint sensitivity is not quantified here. The analytic Gibbs step further rests on a diagonal truncation of the risk
curvature and a Gaussian-mixture approximation of the partition function whose
effect on the optimum is not isolated.

The statistical scope of the experiments is deliberately focused. The experiments use 10 random seeds, one universe of assets, and two horizons, and the overlapping rolling windows make the pooled trajectories non-independent, so the reported Sharpe ratios should be read with this dependence in mind. The results establish that the mechanism works and where its edge originates, while a formal significance assessment across universes and market regimes is left for future work.

The next steps follow directly. Replacing the oracle with true signals would characterize how the realized performance responds to realistic decay, while calibrating the impact and control parameters to data, and reporting sensitivity to the diagonal and mixture approximations, would turn the illustrative settings into fitted ones. On the statistical side, standard errors or bootstrap intervals that account for window overlap, and Diebold-Mariano or Superior Predictive Ability tests with sub-period analysis of the 2020 to 2025 sample would test significance directly. Higher-dimensional universes are the target for the semi-analytical solver of \cref{pideSolver}, and richer scenario generation, explicit risk measures within the distributional formulation, and capacity, borrowing, and position constraints would move the framework toward live deployment.

A further refinement concerns the shape of the terminal utility. The quadratic
tracking objective in \eqref{quadratic_U} penalizes overshooting the
target as much as missing it, which caps the upside by construction. This
symmetry can be removed without sacrificing tractability by replacing the
quadratic loss with an expectile loss, an asymmetric piecewise-quadratic
function that weights shortfall and surplus differently. Since the derivative
$U'(C)$ remains piecewise linear, the terminal conditions in \eqref{bc} remain analytic on each branch, and the analytic Gibbs step survives with a two-regime partition function expressible through Gaussian error functions. Such a utility would allow for an ambitious target level $r_{\rm tg}$ without the symmetric penalty punishing gains beyond it. Its implementation and empirical assessment are left for future work.

A simple mitigation of the harmful symmetry of the quadratic utility function with the goal-based wealth management with RL is to use intentionally unattainable targets so that the probability of overshooting will be minimal, as suggested in \cite{Dixon_Halperin_2021, Halperin_Liu_2022}.  

The target-holding reformulation of \eqref{eq:induced_rate} invites a
comparison with the classical impulse-control treatment of continuous-time
mean-variance allocation by \cite{DangForsyth2014}. That work shares two
structural elements with ours. Its terminal objective, obtained through the
embedding of the mean-variance problem, is a quadratic tracking utility of
exactly the form \eqref{quadratic_U}, with the embedding parameter
playing the role of our target cost $z_{\rm tg}$. Moreover, its impulse
formulation is motivated by the same consideration that motivates our jump
transition: recasting the control as an instantaneous portfolio revision
keeps the intractable term of the equation, there the jump integral, here
the Gibbs exponent, in a tractable form. The differences are equally clear.
Their setting is a single risky asset under jump-diffusion with proportional
costs and hard constraints, solved on a grid by a monotone semi-Lagrangian
scheme with convergence to the viscosity solution, whereas ours is a
multi-asset diffusion with quadratic impact costs, a stochastic
KL-regularized policy, and an offline data-driven solver.

These observations suggest that the impulse-control problem of
\cite{DangForsyth2014} is the classical limit of our discrete jump system:
taking $\beta \to \infty$ in \eqref{eq:couplings_jump}, restricting to
one risky asset, and replacing the quadratic execution cost by a
proportional intervention cost should recover their quasi-variational
inequality. Establishing this limit rigorously, and more generally deriving
the entropy-regularized quasi-variational inequality whose Gaussian-mixture
Gibbs policy is the discrete analogue of \eqref{eq:couplings_jump},
would place the target-holding reformulation on the same footing that
viscosity-solution theory provides in the classical case. Their monotone
scheme would then serve as an independently convergent low-dimensional
benchmark for the PINN solver.

All the above comments build a program that is well-recognized as our future work.

\section*{Disclosure statement}

No potential conflict of interest was reported by the authors.

\section*{Funding}

No funding was received.

\section*{Disclaimer}

Opinions expressed here are author's own, and do not represent views of their employers. A standard disclaimer applies.

\section*{Acknowledgments}

We thank Prof.~Nizar Touzi for useful comments.

\printbibliography[title={References}]

\appendix
\appendixpage
\numberwithin{equation}{section}
\setcounter{equation}{0}

\section{Price impact model} \label{appImpact}

Let $v_{t,i} = a_{i,t}$ be the value of the trade (in shares) per unit time for a given stock $i$ at the current time $t$, and let $\bar{v}_{t,i}^{k}$ be the average trade per unit time for the same stock over the previous $k$ periods, excluding the current period, with
\begin{equation} \label{bar_u_t}
\bar{v}_{t,i}^{k} = \frac{1}{k} \sum_{j=1}^{k}  v_{t - j \Delta t, i}.
\end{equation}

Trades affect prices in a non-monotonic way: an initial price increase is often followed by a long-term decline, even when inflows stay steady. To capture this pattern, we model price impact using a \emph{quadratic} function rather than a linear one.

Drawing on ideas from \cite{kyle1985,almgren2001,hasbrouck2007}, we define the fractional price impact $f_{i,t}(\bm{v}) = \frac{1}{S_{t,i}} \frac{dS_{t,i}}{dt}$ on asset $i$ caused by the trade rate $\bm{v}_t$,
where $f_{i,t}(\bm{v}_t) \equiv f_{i,t}(\bm{a}_t)$ is given by
\begin{align} \label{impactModel}
f_{i,t}(\bm{v}_t) &= \underbrace{\eta_{i}^{\text{temp}} \alpha_{i} \frac{v_{t,i}}{V_{t,i}}}_{\text{Temporal Impact}} + \underbrace{\eta_{i}^{\text{perm}} \gamma_{i} \frac{v_{t,i}^2}{V_{t,i}^2} \text{sign}(v_{t,i})}_{\text{Non-linear Permanent Convex Impact}} + \underbrace{\sum_{j \neq i} \beta_{ij} \frac{v_{t,j}}{V_{t,j}}}_{\text{Cross-Impact}} + \underbrace{\phi_{t,i} \alpha_{t,i} M_{t,i}}_{\text{Memory Impact}}.
\end{align}
Here, $\eta_i^{\text{temp}}$ and $\eta_i^{\text{perm}}$ are model coefficients, $V_{t,i}$ is the current total volume (ADV in shares) used for normalization, $\alpha_{t,i}$ are linear impact coefficients (Kyle's lambda type), $\beta_{ij}$ are cross-impact coefficients, $\phi_{t,i}$ are memory persistence parameters with $0 < \phi_{t,i} < 1$ (higher values mean longer memory), and $M_{t,i}$ is the relative memory impact.

The cross-impact term uses a liquidity-weighted linear form, which has several practical advantages:
\begin{itemize}
\item It adjusts spillover based on relative liquidity: trades in less liquid assets have a larger per-share impact on more liquid assets, and vice versa.
\item Economically, a trade of a given dollar size in an illiquid stock may carry more “information weight” than the same dollar trade in a liquid stock.
\item The model is symmetric if $\beta_{ij} = \beta_{ji}$.
\end{itemize}

The fractional price impact in \eqref{impactModel} is split into temporary and permanent parts to match real-world market microstructure. The first term is the Temporal Impact, which is directly proportional to the current trade actions $v_{t,i}$. This captures the immediate liquidity-demanding pressure of the trade. However, note that while this term is traditionally associated with temporary execution shortfall, our $\alpha_i$ in \eqref{alpha_i} is actually a composite: it includes both the linear permanent impact from information asymmetry (the $\kappa_2$ term) and the sub-linear execution shortfall (the $\kappa_3$ square-root term). By putting these inside the temporal component, the model ensures that the base price response scales directly with how intensely trading is happening right now.

A possible worry about the above model is that it depends only on the instantaneous trade rate $v_t$ and seems to ignore the current portfolio state or total holdings $\bx_t$. In this model, however, the state-dependency is handled by the Memory Impact term $M_{t,i}$. Specifically, the memory state component $\mathrm{ME}_{t,i}$ acts as a rolling, decay-weighted accumulation of previous trades. By defining memory impact this way, we effectively treat "holdings" as a hidden state variable. If the persistence parameter $\phi_{t,i}$ is high (close to $1$), then $\mathrm{ME}_{t,i}$ approximates the recent change in the investor's position. This means that the price impact does not simply reset to zero when $v_{t,i}$ becomes zero; instead, the price level stays elevated (or depressed) based on the weighted history of past activity. This lets the model naturally capture the "price floor" effect seen in permanent impact models, where the current price depends on the entire execution path, not just the current flow.

Below we describe how all these parameters are computed. For simplicity, we drop the index $_t$, treating all parameters as relating to time $t$.

\myparagraph{Linear self-impact coefficient $\bm{\alpha_i}$.}
The linear impact parameter captures the immediate cost of trading and follows the intuition from \cite{kyle1985,kyle2016market}
\begin{equation} \label{alpha_i}
\alpha_i = \bas_i \left( \kappa_1  + \kappa_2 \sigma_i \sqrt{\frac{\advR}{\adv_i}} + \kappa_3 \sqrt{\frac{\VR}{V_i}} \right),
\end{equation}
where $\bas_i$ is the bid-ask spread as a percentage of the mid-price, $\sigma_i$ is the daily percentage volatility of log-returns, $\adv_i$ is the average daily volume in shares, and $\kappa_1$, $\kappa_2$, $\kappa_3$ are calibration constants. In particular, if $\kappa_1 = 0.5$, then buying at the ask and selling at the bid incurs the full spread cost.

The first term in \eqref{alpha_i} represents an immediate spread cost: any trade, regardless of size, must incur this cost to access liquidity. In other words, it is the cost of moving the price from the mid to the ask (or bid) and is the minimum cost of executing a small trade instantly. In practice, $\kappa_1$ is likely close to $0.5$ if this reflects the cost of a market order.

The second term represents a linear permanent impact coming from information asymmetry. The annualized volatility $\sigma_i$ measures information flow, and $\sqrt{\adv_i}$ normalizes by market depth or liquidity. Their ratio thus measures price impact per unit volume traded. A typical value of $\kappa_2 \approx 0.5$ was used in \cite{kyle2016market}.

The third term represents the nonlinear impact or execution shortfall for larger trades and corresponds to the square-root law of market impact, i.e., $\text{Impact} \propto \sqrt{V}$, as discussed in \cite{almgren2001}.

The reference values can be computed as follows:
\begin{align}
\advR &= \text{Median}(\adv_1,\ldots,\adv_N), \qquad \VR = p \advR,
\end{align}
where a common choice is $p = 0.1$. These formulas are robust to outliers and give a natural interpretation for typical stocks.

Note that in our formulation, we distinguish between the immediate costs of execution and the lasting shifts in the price level. While the coefficient $\alpha_i$ in \eqref{alpha_i} theoretically includes a linear permanent component (the $\kappa_2$ term) following \cite{kyle2016market}, placing it inside the first term of \eqref{impactModel} technically treats it as part of the flow-dependent impact. To make sure the model explicitly accounts for the total size of the trade, and to prevent it from assuming that price impact fully disappears when trading stops, we keep a separate permanent impact structure that scales with both the rate and the accumulated history of the trade.

\myparagraph{Quadratic self-impact coefficient $\bm{\gamma_i}$.}
The quadratic term captures the convexity of impact, which grows with illiquidity and is given by
\begin{equation}
\gamma_i = \frac{1}{\adv_i S_i} \left( \lambda_2 + \lambda_1 \frac{\overline{\Mc}}{\Mc_i} \right),
\end{equation}
where $\Mc_i$ is the market capitalization of asset $i$, $\overline{\Mc}$ is the average market cap in the universe, $\tor_i = \adv_i S_i/\Mc_i$ represents the turnover rate, and $\lambda_1$, $\lambda_2$ are calibration constants.

\myparagraph{Temporary vs permanent impact split.}
Since the price impact is split into temporary (reverting) and permanent (information-based) components, we determine $\eta_i^{\text{temp}}$ and $\eta_i^{\text{perm}}$ as follows:
\begin{align}
\eta_i^{\text{temp}} &= \nu^{\text{temp}} \frac{\bas_i}{\overline{\bas}}, \qquad
\eta_i^{\text{perm}} = \nu^{\text{perm}} \frac{\sigma_i}{\overline{\sigma}},
\end{align}
where $\nu^{\text{temp}}$ and $\nu^{\text{perm}}$ are calibration parameters.

\myparagraph{Linear cross-impact matrix $\bm{\beta_{ij}}$.}
The cross-impact between asset $i$ and $j$ depends on their relative market structure:
\begin{equation}
\beta_{ij} = \theta_1 \frac{\min(\Mc_j, \Mc_i)}{\max(\Mc_i,\Mc_j)} + \theta_2 \frac{1}{1 + |\log(\adv_i/\adv_j)|} + \theta_3 e^{-(\beta_i - \beta_j)^2} \sqrt{\alpha_i \alpha_j} \frac{\sqrt{\bas_i \bas_j}}{\overline{\bas}},
\end{equation}
where $\beta_i$ and $\beta_j$ are market betas (correlation to the market portfolio), $\overline{\bas}$ is the average bid-ask spread across all assets, and $\theta_1$, $\theta_2$, $\theta_3$ are calibration parameters.

The first term captures that trading a small-cap stock affects large-cap stocks less. The second term measures volume similarity (largest when $\adv_i = \adv_j$) and ranges in $(0, 1]$. The third term combines beta similarity with impact and spread. Together, these capture three economic channels:
\begin{itemize}
\item \emph{Market cap effect}: Larger stocks influence smaller ones more.
\item \emph{Liquidity commonality}: Stocks with similar trading volumes move together.
\item \emph{Risk factor alignment}: Stocks with similar market betas show stronger cross-impact.
\end{itemize}

\myparagraph{Memory and Temporal Effects.}

We write the memory impact term $M_{t,i}$ as
\begin{equation}
M_{t,i} = \delta_1 \mathrm{MR}_{t,i} + (1-\delta_1) \mathrm{MV}_{t,i}.
\end{equation}
Here, $0 \le \delta_1 \le 1$ is the calibration parameter, and $\mathrm{MR}_{t,i}$ is the \emph{reinforcement term}
\begin{equation}
\mathrm{MR}_{t,i} = \text{sign}(v_{t,i}) \mathrm{ME}_{t,i}.
\end{equation}
The term $\mathrm{ME}_{t,i}$ is the \emph{memory state component} representing signed memory effects
\begin{equation}
\mathrm{ME}_{t,i} = \sum_{j=1}^{k} \frac{v_{t-j,i} e^{-\theta_i \Delta t_j}}{\adv_i},
\end{equation}
where $\theta_i = 1 - \phi_i$ is the memory decay rate and $k$ is the memory horizon, typically spanning several weeks.

$\mathrm{MR}_{t,i}$ is positive when trading in the same direction as past trades and negative when trading against the past trend. Thus, it captures momentum amplification or mean reversion.

The term $\mathrm{MV}_{t,i}$ is the \emph{magnitude term} given by
\begin{equation}
\mathrm{MV}_{t,i} = \mathrm{MM}_{t,i} \frac{v_{t,i}}{\adv_i}.
\end{equation}
Here, $\mathrm{MM}_{t,i}$ is the \emph{memory magnitude}, representing an absolute memory effect:
\begin{equation}
\mathrm{MM}_{t,i} = \sum_{j=1}^{k} \frac{|v_{t-j,i}| e^{-\theta_i \Delta t_j}}{\adv_i}.
\end{equation}
$\mathrm{MV}_{t,i}$ scales with both past trading activity and current trade size, reflecting the idea that greater past activity leads to higher current impact.

Putting these together gives:
\begin{equation}
M_{t,i} = \delta_1 \text{sign}(v_{t,i}) \sum_{j=1}^{k} \frac{v_{t-j,i} e^{-\theta_i \Delta t_j}}{\adv_i} + (1-\delta_1) \frac{|v_{t,i}|}{\adv_i} \sum_{j=1}^{k} \frac{|v_{t-j,i}| e^{-\theta_i \Delta t_j}}{\adv_i}.
\end{equation}

How long the impact persists depends on stock-specific characteristics, expressed through the memory coefficient $\phi_i$:
\begin{equation}
\phi_i = \nu_1 \left(1 - \frac{\tor_i}{\max_k(\tor_k)} \right) + \nu_2 \frac{\bas_i}{\max_k(\bas_k)},
\end{equation}
where $\nu_1$ and $\nu_2$ are calibration parameters, and $\tor_i$ is turnover for asset $i$, used to calibrate how long a price impact lasts.

\subsection{Calibrated parameters of the price impact model.}

Overall, at time $t$, our price impact model contains the following 13 unknown parameters: $\delta_1$, $\nu_1$, $\nu_2$, $\nu^{\text{temp}}$, $\nu^{\text{perm}}$, $\theta_1, \theta_2$, $\theta_3, \kappa_1, \kappa_2, \kappa_3, \lambda_1, \lambda_2$. We assume these parameters are constant over time, because any time variation is already captured by other time-dependent market data in the model. Therefore, they can be learned during training by including them in the set of trainable parameters.

Finally, note that the price impact model in \eqref{impactModel} can be written in the form
\begin{equation} \label{PImodel}
{\bm f} (\ba_{t}) = {\bm f}_t^{(0)} + {\bm f}_t^{(1)} \ba_{t} + {\bm f}_t^{(2)} (\ba_t^{\circ 2}),
\end{equation}
where
\begin{align} \label{fim_terms}
{\bm f}_t^{(0)} &=  \left[\phi_{t} \circ \alpha_{t} \circ \mathrm{MR}_{t} \right]^T, \quad
{\bm f}_t^{(1)} =
\begin{cases}
\frac{\eta_{i}^{\text{temp}} \alpha_{i}}{V_{t,i}}, & i = j \\ \frac{\beta_{ij}}{V_{t,j}}, & i \neq j
\end{cases}
, \quad
{\bm f}_t^{(2)} = \mathrm{diag}\left( \frac{\eta_{1}^{\text{perm}} \gamma_{1}}{V_{t,1}^2}, \dots, \frac{\eta_{n}^{\text{perm}} \gamma_{n}}{V_{t,n}^2} \right).
\end{align}
Here, $\operatorname{diag}(\bx)$ denotes the diagonal matrix whose main diagonal contains the elements of the vector $\bx$.

Overall, this model appears comprehensive in capturing diverse market effects. Nevertheless, 13 model parameters risk overfitting or overburdening the calibration process. Consequently, we aim to reduce the parameter space from 13 to a more tractable set while retaining the model's structural integrity. For this purpose, we draw on known economic priors, stylized facts, and dimensional reduction techniques.

\myparagraph{Fix "anchor" parameters.}

Several of our parameters correspond to phenomena extensively studied in the literature. Therefore, we either fix them to "standard" values or express them as functions of other parameters, as follows:

\begin{enumerate}
\item In the linear self-impact term ($\alpha_i$), $\kappa_1$ represents the spread cost. Specifically, $\kappa_1 = 0.5$, which is the theoretical cost of a market order (crossing the half-spread). Similarly, \cite{kyle2016market} and subsequent calibrations often report $\kappa_2 \approx 0.5$ for the information asymmetry component. These are the values we adopt in our reduced model.

\item Parameter $\delta_1$ balances the reinforcement and magnitude terms of memory. We assume a neutral stance, setting $\delta_1 = 0.5$.
\end{enumerate}

Thus, this reduction eliminates three calibrated parameters: $\kappa_1$, $\kappa_2$, and $\delta_1$.

\myparagraph{Tying Impact Scales $\nu^{\text{temp}}$ and $\nu^{\text{perm}}$.}

In the preceding section, we introduced independent scales for temporary and permanent impact. In many equilibrium models, however, the ratio of permanent to temporary impact remains relatively stable across a given universe of assets. Accordingly, we define $\nu^{\text{perm}} = \zeta \nu^{\text{temp}}$, where $\zeta$ is a fixed constant. The "square root law" literature often takes $\zeta = 0.5$, as it suggests that approximately half of the total impact at its peak is permanent.

Thus, we set $\zeta = 0.5$ and calibrate only $\nu^{\text{temp}}$; $\nu^{\text{perm}}$ is then scaled accordingly. This removes one further calibration parameter.

However, we acknowledge that fixing $\zeta = 0.5$ is a simplifying assumption primarily grounded in stylized facts from the equity market microstructure literature. The true ratio of permanent to temporary impact can vary significantly across different asset classes and liquidity profiles. For example, broad market ETFs, which typically exhibit deep liquidity and high arbitrage activity, may demonstrate a lower ratio of permanent impact due to faster price reversion, whereas small-cap single stocks or less liquid assets may exhibit a higher permanent impact ratio driven by greater information asymmetry. Consequently, while setting $\zeta = 0.5$ eases calibration for a relatively homogeneous universe, applying this framework to highly heterogeneous portfolios may require treating $\zeta$ as an additional trainable parameter or scaling it as a function of observable microstructure variables (e.g., turnover or bid-ask spread) rather than fixing it a priori.

\myparagraph{Dimensional analysis and parameter collapsing.}

We can "collapse" the quadratic coefficient $\gamma_i$ and the memory decay $\phi_i$ by assuming they are driven by the same underlying liquidity factors. Specifically:

\begin{enumerate}
\item Instead of using $\lambda_1$ and $\lambda_2$, we introduce a single scalar $\lambda$ by assuming that the "intercept" $\lambda_2$ is negligible compared to the cap-weighted turnover term:
\begin{equation}
\gamma_i \approx \lambda \frac{\overline{\Mc}}{\Mc_i \adv_i S_i}.
\end{equation}

\item Similarly, for memory persistence, we link $\nu_1$ and $\nu_2$. If a stock has high turnover $\tau_i$, it likely has a lower spread $\bas_i$. We therefore define a single "liquidity score" $L_i$ and use one parameter $\nu$ to govern $\phi_i = f(\nu, L_i)$.
\end{enumerate}
This approach further eliminates two calibration parameters.

Consequently, the reduced model contains only five parameters. The calibration is therefore expected to be significantly more stable, less prone to overfitting, and computationally faster.

To implement these changes, we restructure the model by collapsing the redundant degrees of freedom into "global scales." This preserves the underlying physics of the model, i.e., sensitivities to volume, cap, and spread remain intact, while substantially easing the calibration.

\myparagraph{Reformulated components and the resulting matrices.}

We define the new vector of trainable parameters as $\mathbf{\Theta} = \{ \nu, \lambda, \theta, \kappa_3, \phi \}$, where $\nu$ is the global impact scale (replaces $\nu^{\text{temp}}$ and $\nu^{\text{perm}}$); $\lambda$ is the global quadratic scale (replaces $\lambda_1, \lambda_2$); $\theta$ is the global cross-impact scale (replaces $\theta_1, \theta_2, \theta_3$); $\kappa_3$ is the square-root law coefficient;  $\phi$ is the global memory persistence (replaces $\nu_1, \nu_2$).

By applying the "economic priors" discussed, we simplify the internal definitions. For the linear self-impact $\alpha_i$ we obtain
\begin{equation}
\alpha_i = \bas_i \left( 0.5 + 0.5 \sigma_i \sqrt{\frac{\adv_{ref}}{\adv_i}} + \kappa_3 \sqrt{\frac{V_{ref}}{V_i}} \right).
\end{equation}

For the quadratic and persistence scales, we use a single coefficient for each, assuming the relative weights of the sub-terms are fixed (e.g., equal weighting or unit scale). This yields
\begin{align}
\gamma_i &= \frac{\lambda}{\adv_i S_i} \left( 1 + \frac{\overline{\Mc}}{\Mc_i} \right), \\
\phi_i &= \phi \left[ 0.5 \left(1 - \frac{\tor_i}{\max_k(\tor_k)} \right) + 0.5 \frac{\bas_i}{\max_k(\bas_k)} \right]. \nonumber
\end{align}

We assume the permanent impact is a fixed ratio $50\%$ of the temporary impact scale, hence
\begin{equation}
\eta_i^{\text{temp}} = \nu \frac{\bas_i}{\overline{\bas}}, \qquad \eta_i^{\text{perm}} = 0.5 \nu \frac{\sigma_i}{\overline{\sigma}}.
\end{equation}

The structural form of the price impact model remains as in \eqref{PImodel}, but the matrix elements change. Assuming $\delta_1 = 0.5$ (neutral memory), the "memory term" ${\bm f}_t^{(0)}$ now reads
\begin{equation}
[{\bm f}_t^{(0)}]_i = 0.5 \phi_i \alpha_i \text{MR}_{t,i}.
\end{equation}

For the linear/cross-impact matrix ${\bm f}_t^{(1)}$, we have
\begin{equation}
[{\bm f}_t^{(1)}]_{ij} = \frac{1}{V_{t,j}}
\begin{cases}
\alpha_i \left( \eta_i^{\text{temp}} + 0.5 \phi_i \text{MM}_{t,i}/\adv_i \right), & i = j, \\
\theta \sqrt{\alpha_i \alpha_j} \rho_{ij}, & i \neq j.
\end{cases}
\end{equation}

The permanent convex impact now becomes
\begin{equation}
{\bm f}_t^{(2)} = \text{diag} \left( \frac{0.5 \nu \sigma_i \gamma_i}{\sigma V_{t,i}^2} \right).
\end{equation}

\section{Solving the PDE in \eqref{HJBh_nonlin_GM}} \label{pideSolver}

The unsteady, fully nonlinear PDE in \eqref{HJBh_nonlin_GM}, posed in the $(N+1)$-dimensional spatial variables $(\bS, C)$ with $\bS \in \mathbb{R}^N$, is of the parabolic type, and even its numerical solution presents a significant challenge. This complexity, however, can be substantially reduced by employing three key techniques: an operator splitting method, \cite{ItkinBook}, a generalized Duhamel's principle \cite{evans10,ItkinLiptonMuraveyBook,Itkin2024jd} and a fixed-point iteration scheme similar to that in \cite{Itkin2018}. This section details the proposed approach.

We begin by introducing the backward time $\tau = T-t$ and representing \eqref{HJBh_nonlin_GM} in the form
\begin{gather} \label{operPDE}
\fp{J}{\tau} = \calL_0 J + \calL_1 J, \\
\begin{align}
\calL_0 &= (r C + C_0(\bx, \bS)) \fp{}{C} +  D_S \boldmu^T_0 \fp{}{\bS} + \frac{1}{2} \tr \left(\Xi \nabla_\bS^2 \right), \qquad  \calL_1 =  - \frac{1}{\beta} \log \sum_{k} \omega_k e^{ - \beta \hH_k},
\nonumber
\end{align}
\end{gather}
where for easiness of notation we have dropped index $_S$ at $\Xi_S$. Here, $\hH_k$ is the nonlinear operator corresponding to the functional $\mathcal{H}_k$. In this form, the full operator is decomposed as $\calL_0 + \calL_1$, isolating the linear, time-homogeneous component ($\calL_0$) from the nonlinear one ($\calL_1$). This specific splitting is strategic, and its advantage will be explained in the subsequent steps.

Despite the time-homogeneous nature of \eqref{operPDE}, the presence of the nonlinear operator $\calL_1$ precludes a direct, single-step solution over the entire domain in $\tau$. Instead, the equation must be solved using a time-marching approach. This method requires a temporal grid for $\tau$ with a step size $\Delta \tau$, allowing for the iterative integration of \eqref{operPDE} from the initial condition at $\tau=0$ to the final time $\tau=T$.

At each point on this grid $\tau_i = i \Delta \tau : \, i \in [1:M], \, M \Delta \tau = T$, we solve the following problem
\begin{align} \label{split}
\fp{J_i}{\tau} = \calL_0 J_i + \calL_1 J_i, \qquad J_i = J(\tau_i, \bx, \bS, C),
\end{align}
using the solution from the previous step as the terminal condition. Specifically, for the interval $[\tau_{i}, \tau_{i-1}]$, the terminal condition is $J(\tau_{i-1}, \bx, C)$, which is the solution obtained at the previous time step $\tau_{i-1}$. The process starts at $\tau_0 = 0$ with the known terminal condition \eqref{J_T}, and marches forward to $\tau = T$.

To complete definition of our problem, we need to issue boundary conditions for the PDE in \eqref{operPDE}. These boundary conditions should reflect the economic intuition of what happens to the value function $J(t, \mathbf{x}, \bS, C)$  when a state variable hits an extreme (boundary) value.

\myparagraph{Condition at $S_i = 0$.} This is the case where asset $i$ has a zero price. Then, a very small change in its dollar position should have a negligible effect on the value function. The {\it marginal value} of that asset is zero. In mathematics, this often translates to a homogeneous Neumann condition on the value function with respect to that asset's position
\begin{equation} \label{bc0}
\frac{\partial J}{\partial S_i} \bigg|_{S_i = 0} = 0.
\end{equation}
This condition is standard in portfolio optimization problems and means that the sensitivity of our cost-to-go (or value function) to an infinitesimally small position in asset $i$ is zero. It's also known as a "reflecting" boundary condition.

\myparagraph{Condition as $S_i \to \infty$.} This is the case of an infinitely large dollar value in a single asset. Holding an extremely large position in any single asset is infinitely risky and costly due to the market impact. Therefore, the cost function $J$  should become very large (meaning very undesirable) as $S_i \to \infty$. This is often modeled as "algebraic growth" (e.g., quadratic, from the cost term) or even exponential growth, but it must be consistent with the utility function and cost model.

\myparagraph{Condition at $C \to -\infty$.} This represents a state where our portfolio accumulates infinite negative cost (infinite profit). In economic terms this means that
\begin{itemize}
\item The marginal value of additional profit becomes zero: if we already have infinite profit, more profit doesn't improve our situation.

\item The value function should saturate which implies there is an upper bound on how good the situation can get.

\item The solution should become linear in other variables because the infinite profit dominates all other considerations.

\end{itemize}

Accordingly, at $C \to -\infty$,  as an appropriate boundary condition we choose a homogeneous Neumann condition
\begin{equation} \label{bcC}
\frac{\partial J}{\partial C} \bigg|_{C \to -\infty} = 0.
\end{equation}
This condition captures the economic intuition that the marginal value of additional profit approaches zero when we already have infinite profit.

\subsection{Fixed-point iterations} \label{fpIter}

To solve \eqref{operPDE}, we employ an iterative scheme, inspired by the method of \cite{Itkin2018}. This approach can be viewed as a variant of fixed-point Picard iterations, whose convergence conditions are established by the Banach fixed-point theorem \cite{FPT2003}.The iteration scheme reads
\begin{align} \label{Picard}
\fp{J^{(k)}_i}{\tau} = \calL_0 J^{(k)}_i + \calL_1 J^{(k-1)}_i, \quad k=1,2,\ldots,
\end{align}
Here, $k$ is the iteration number, and $J^{(k)}_i$ is the cost function value at time $\tau_i$ and iteration $k$. The iterations start with $J^{(0)}_i = J_{i-1}$ and end when the entire procedure converges, i.e., when the condition
\begin{equation} \label{tolCond}
\| J^{(k)}_i - J^{(k-1)}_i \| < \varepsilon
\end{equation}
is reached after $k$ iterations, with $\varepsilon$ being the method tolerance and $\|\cdot\|$ denoting some norm.

Note that this formulation does not explicitly rely on the temporal step $\Delta \tau$. Therefore, in principle, it can be run in one step as
\begin{align} \label{Picard1}
\fp{J^{(k)}_M}{\tau} = \calL_0 J^{(k)}_M + \calL_1 J^{(k-1)}_M \quad k=1,2,\ldots, \qquad
J^{(0)}_M = J_0.
\end{align}
However, if the nonlinearity in \eqref{operPDE} is strong, the procedure in \eqref{Picard1} could converge very slowly or may not converge at all. Therefore, solving this problem in $M$ steps by moving sequentially along the temporal grid appears more preferable.

Importantly, this sequential procedure does not introduce a temporal \emph{truncation} error proportional to $\Delta \tau$. However, since at each step $i$ we find an approximate solution within a prescribed per-step tolerance $\varepsilon$ rather than the exact solution, and this approximation seeds the next step, the per-step tolerances accumulate, giving a total budget $\varepsilon_T = M \varepsilon = \varepsilon\, \tau(0) / \Delta \tau$, where $\tau(0)$ is the final backward time. We stress that $\varepsilon$ here is a user-controlled solver tolerance, not a discretization error, so this expression is not directly comparable with the $O(\Delta\tau)$ truncation error of a standard finite-difference scheme; rather, it makes explicit a practical trade-off, namely that taking many small steps (large $M$) inflates the accumulated tolerance budget, whereas a single large step \eqref{Picard1} may converge slowly or fail under strong nonlinearity. In practice $\Delta\tau$ and $\varepsilon$ should be chosen jointly so that the per-step iteration converges robustly while keeping $M\varepsilon$ below the target accuracy.

To establish the validity of the method, we need to prove the convergence of the fixed-point Picard iterations. The detailed analysis and corresponding theorems are given in Appendix~\ref{convergPicard}. The main result is given by \cref{theor}, which states that using Duhamel's principle, \cite{evans10}, the iterative scheme for \eqref{operPDE} can be represented as
\begin{align} \label{map}
J_i^{(k)} = e^{\Delta \tau \calL_0} J_{i-1} + \int_0^{\Delta \tau} e^{(\Delta \tau - s) \calL_0} \calL_1 \left(J_i^{k-1}(\tau + s) \right) ds,
\end{align}
where $\Delta \tau < 1/L_1$ is a time step that guarantees the convergence of the iterative sequence on $[0, \Delta \tau]$, and $L_1$ is Lipschitz constant of the operator $\calL_1$.

The advantage of this scheme is that it contains no temporal discretization error in the linear part, given by the linear operator $\mathcal{L}_0$, provided this part can be solved exactly (i.e., $\mathcal{L}_0^{-1}$ is known in closed form). Consequently, there is no stability constraint (such as a CFL condition) imposed by this part, allowing much larger effective time steps in the overall iteration when desired.

Since $\mathcal{L}_0$ is invertible, the integral of the semigroup can also be computed analytically and this scheme could be further simplified. A very common and clean result is, \cite{pazy1983semigroups}
\begin{equation}
\int_0^{\Delta \tau} e^{(\Delta \tau - s)\mathcal{L}_0}\, ds = \mathcal{L}_0^{-1} (e^{\Delta \tau \mathcal{L}_0} - I).
\end{equation}
This gives an elegant iteration formula
\begin{equation}
J_i^{(k)} = e^{\Delta \tau \mathcal{L}_0} J_{i-1} + \mathcal{L}_0^{-1} (e^{\Delta \tau \mathcal{L}_0} - I) \, \mathcal{L}_1\left(J_i^{(k-1)}\right).
\end{equation}

\subsection{Solving \eqref{Picard} using Strang's splitting}

Unfortunately, due to the presence of the term $C_0(\bx)\fp{}{C}$ in the definition of the operator $\calL_0$ in \eqref{operPDE}, the Green's function of \eqref{Picard} does not admit an incomplete separation of variables, \cite{polyanin2016handbook}. Therefore, we solve \eqref{Picard} by employing an operator splitting scheme. The idea of splitting lies in decomposing the right-hand side operator $\calL = \calL_0 + \calL_1$ in \eqref{split} into a sum of two operators
\begin{gather} \label{solSplit}
\calL = [\calL_{0,S} + \calL_1] + \calL_{0,C}, \\
\begin{align}
\calL_{0,S} &= D_S \boldmu^T_0 \fp{}{\bS} + \frac{1}{2} \tr \left(\Xi \nabla_\bS^2 \right), \qquad
\calL_{0,C} = [r C + C_0(\bx,\bS))] \fp{}{C}, \quad k=1,2,\ldots. \nonumber
\end{align}
\end{gather}

Since the operators $\calL_{0,S} + \calL_1$ and $\calL_{0,C}$ in \eqref{solSplit} do not commute, simple splitting provides only a first-order approximation in time (i.e., $O(\Delta \tau_i)$) to the exact solution. Higher-order accuracy can be achieved using various approximations based on the Baker--Campbell--Hausdorff formula \cite{BCH}. For example, the Strang's splitting scheme reads\footnote{For a general approach to splitting techniques for \emph{linear} operators using Lie algebras, we refer the reader to \cite{LanserVerwer,ItkinBook}.}, \cite{strang2006linear}
\begin{align} \label{strang1}
e^{ \Delta \tau_i [\calL_{0,S} + \calL_1 + \calL_{0,C}]} = e^{ \frac{1}{2}\Delta \tau_i [\calL_{0,S} + \calL_1]} e^{ \Delta \tau_i \calL_{0,C}} e^{ \frac{1}{2} \Delta \tau_i [\calL_{0,S} + \calL_1]} + O((\Delta \tau_i)^2).
\end{align}

For parabolic equations with coefficients independent of $\tau$, this composite algorithm achieves second-order accuracy in $\Delta\tau_i$, provided that the numerical procedure solving the corresponding PDE at each splitting step maintains at least second-order accuracy in time. Below, we construct a closed-form solution for each step of the splitting, and thus this condition is satisfied.

The factorized solution of \eqref{split} with allowance for \eqref{strang1} can be transformed back into a system of PDEs for $J_i = J^{(k)}(\tau_i, \mathbf{x}, \bS,C)$
\begin{alignat}{2} \label{systemPDE}
\fp{J^{(1)}_i}{\tau} &= \frac{1}{2} \calL_{0,S} J^{(1)}_i + \calL_1 J^{(k-1)}_i,
&\qquad J^{(1)}_i(\tau_{i-1}) &= J^{(k-1)}_i(\tau_{i-1}), \, \tau \in [\tau_{i-1}, \tau_i] \\
\fp{J^{(2)}_i}{\tau} &= \calL_{0,C} J^{(2)}_i,
&\qquad J^{(2)}_i(\tau_{i-1}) &= J^{(1)}_i(\tau_i), \, \tau \in [\tau_{i-1}, \tau_i], \nonumber \\
\fp{J_i}{\tau} &= \frac{1}{2} \calL_{0,S} J_i + \calL_1 J^{(k-1)}_i,
&\qquad J_i(\tau_{i-1}) &= J^{(2)}_i(\tau_i), \, \tau \in [\tau_{i-1}, \tau_i]. \nonumber
\end{alignat}
The iterations begin at $k = 1$  with $J^{(0)}_i = J_{i-1}$  and continue until the procedure converges, i.e. until it satisfies the condition in \eqref{tolCond}.

It can be easily observed that at the first and third steps of the splitting in \eqref{systemPDE}, the PDE to solve is exactly the PDE in \eqref{Picard}, but with a modified operator $\calL_0$ that now does not contain the first derivative in $C$. Therefore, it can be solved by the method presented in \cref{fpIter}. However, due to this modification, the linear part of this PDE can now be solved in closed form.

The second step of the splitting can also be solved in closed form, as will be shown in the next section. Thus, the only error in our numerical scheme is a temporal error of $O((\Delta \tau)^2)$ due to Strang's splitting.

It is worth noting that the Diffusion-Outer/Advection-Inner (DAD) scheme used in \eqref{systemPDE} is generally preferred over an ADA scheme, primarily due to the nature of the operators and how they handle sharp gradients and boundary conditions.

It is known that advection tends to create sharp gradients and can lead to dispersive or oscillatory artifacts, especially when the numerical method for advection is not highly accurate (e.g., low-order upwinding). If advection is the last step (as in the ADA scheme), these potential numerical artifacts are "frozen" into the solution until the next time step. In contrast, diffusion is a smoothing operator. By placing the diffusion step after advection (as in the DAD scheme), any spurious oscillations or sharp fronts generated by the advection solver are immediately smoothed out. This acts as a natural stabilizer.

Additionally, the advection step requires boundary conditions only on the inflow boundaries, with the solution on the outflow boundary determined by the interior solution being advected outward. The diffusion step, however, requires boundary conditions on all boundaries, which is much better from a smoothing perspective.

On the other hand, a drawback of this scheme is that both the first and third steps require solving nonlinear equations, effectively doubling the computational work.

To summarize, the advantages of our approach are as follows:
\begin{enumerate}
\item Freedom from the stability constraints of $\calL_1$ allows for a larger time step $\Delta \tau$. The limiting factor is now the timescale of the nonlinearity $\mathcal{L}_1$ and the desired accuracy (the splitting error), not numerical stability.
\item The linear part is solved exactly, eliminating a major source of numerical error.
\item The iteration cleanly separates the exact propagation of the linear state from the update due to the nonlinearity.
\item This method is conceptually similar to \emph{exponential integrators} in numerical analysis, which are specifically designed for problems where the linear part can be solved exactly and are known for their excellent stability properties.
\end{enumerate}

\subsection{Closed form solutions of \eqref{systemPDE}}

As mentioned in the previous section, each step of the splitting scheme in \eqref{systemPDE} can be solved in closed form.

\myparagraph{Steps 1 and 3.} At these steps we solve the first and the third equations in \eqref{systemPDE}. Since they are identical, we describe the method for only one. The operator $\calL_{0,S}$ corresponds to a multivariate GBM process with constant coefficients; consequently the Green's function for this equation coincides with the well-known transition density of that process, \cite{shreve2004continuous}
\begin{align}
p(0,\bS_0 | \bS_\tau,\tau) &= (2\pi \tau)^{-N/2} \, |\mathbf{\Sigma}|^{-1/2} \,
\left( \prod_{i=1}^N \frac{1}{S_{0,i}} \right) e^{-\frac{1}{2} \, \mathbf{z}^\top \mathbf{\Sigma}^{-1} \mathbf{z} }, \\
\mathbf{z} &= \frac{\ln(\mathbf{S}_0) - \ln(\mathbf{S}_\tau) - \left( \boldsymbol{\mu} - \frac{1}{2} \mathrm{diag}(\mathbf{\Sigma}) \right)\tau}{\sqrt{\tau}}. \nonumber
\end{align}

Accordingly, the solution of the first PDE in \eqref{systemPDE} can be represented as, \cite{polyanin2016handbook}
\begin{align}
J^{(1)}_i &= \frac{1}{2} \int_{-\infty}^\infty d\bm{\xi}\, p(0,\bm{\xi} | \bS_\tau,\tau) J^{(k-1)}_i(\tau_{i-1}, \bm{\xi}, C)
+ \frac{1}{2} \int_0^\tau ds \int_{-\infty}^\infty d\bm{\xi}\, p(\tau - s,\bm{\xi} | \bS_\tau,\tau) \calL_1 J^{(k-1)}(s, \bm{\xi}, C).
\end{align}

Since in our case $\tau = \Delta \tau$, the second (temporal integral) can be accurately approximated using the trapezoid with two points $s = 0$ and $s = \tau$. Computation of the remaining spatial multidimensional integral with Gaussian kernels can be efficiently done by using Fast Gauss Transform (FGT) with linear complexity in the number of nodes, \cite{FGT2010}.

\myparagraph{Step 2.} At this step we need to solve a \emph{hyperbolic} PDE with constant coefficients
\begin{align}
\fp{J^{(2)}_i}{\tau} &= [r C + C_0(\bx, \bS))] \fp{J^{(2)}_i}{C},
\end{align}
subject to the boundary condition in \eqref{bcC}, and the terminal condition in \eqref{systemPDE}. Using the standard methods, \cite{courant1989methods}, we obtain
\begin{align}
J^{(2)}_i(\tau, \bx, \bS, C) &= J_y^{(1)}(\tau_i, \bx, \bS, C_1), \qquad
C_1 = \left( C + \frac{C_0(\bx, \bS)}{r} \right) e^{r\tau} - \frac{C_0(\bx, \bS)}{r}. \end{align}
Since the solution is obtained at points $C_1 \ne C$, it must be interpolated back to the original grid in $C$ for use in the subsequent steps of the splitting scheme.

\section{Convergence of Picard iterations in \eqref{Picard}} \label{convergPicard}

We aim to prove the convergence of the iterative scheme
\begin{equation} \label{eqPicard}
\frac{\partial J^{(k)}_i}{\partial \tau} = \mathcal{L}_0 J^{(k)}_i + \mathcal{L}_1 J^{(k-1)}_i, \quad k=1,2,\dots
\end{equation}
starting from some initial guess $J^{(0)}_i$ (typically, $J^{(0)}_i = J_{i-1}$). The operators $\calL_0, \calL_1$ defined in \eqref{operPDE} are given by
\begin{align} \label{a1}
\calL_0 &= r C \fp{}{C} +  D_S \boldmu^T_0 (\bx) \fp{}{\bS} + \frac{1}{2} \tr \left(\Xi \nabla_\bS^2 \right) + C_0(\bx,\bS)\fp{}{C}, \qquad \calL_1 = - \frac{1}{\beta} \log \sum_{k} \omega_k e^{ - \beta \hH_k}, \\
\hH_k &= H_k + \frac{1}{2 \beta} \log\left( |\Omega_k^{(0)}|\, |\tilde{\bm \Omega}_k^{-1}| \right), \qquad \Omega_k^{(0)} = \sigma_k^2 {\bm I}, \qquad
\tilde{{\bf \Omega}}_k = \left(\Omega_k^{(0)}\right)^{-1} - 2 \beta \tilde{\bm C}, \nonumber \\
H_k &= (\tilde{\mathbf{C}} {\bm u}_k (\bx,\bS) + \tilde{\mathbf{D}})^T \tilde{{\bm \Omega}}_k (\tilde{\mathbf{C}} {\bm u}_k (\bx,\bS) + \tilde{\mathbf{D}}) - \frac{1}{2} {\bm u}_k^T(\bx,\bS) \tilde{\mathbf{C}} {\bm u}_k (\bx,\bS) - {\bm u}_k^T (\bx) \tilde{\mathbf{D}}, \nonumber \\
\tilde{\bm C} &= {\bf C}_2(\bx,\bS)  \fp{}{C} + D_S\, \mathrm{diag} \left( \boldmu_2(\bx,\bS) \circ \fp{}{\bS} \right), \qquad \tilde{\bm D} = -\frac{1}{2} \left[ -{\bf C}_1(\bx,\bS) \fp{}{C} + D_S\, \boldmu_1(\bx,\bS) \circ \fp{}{\bS} \right].  \nonumber
\end{align}
Here $\calL_0$ is a linear, $N$-dimensional, second-order parabolic operator. The remaining objects require a word of clarification, since they play two distinct roles that must not be conflated. In the value-function PDE \eqref{HJBh_nonlin_GM}, the quantity $\mathcal{H}_k(J)$ (equivalently $\hH_k$ above) is a \emph{scalar functional} of the local gradient $\nabla J = (\partial J/\partial C, \nabla_\bS J)$: through \eqref{pi_GM_params}, the matrices $\tilde{\bm C}$ and $\tilde{\bm D}$ are the gradient-dependent coefficients ${\bm C}(J)$ and ${\bm D}(J)$, and $\mathcal{H}_k(J)$ is a quadratic-rational function of $\nabla J$ that is well-defined wherever ${\bm \Omega}_k(J) = [\sigma_k^{-2}{\bm I} + {\bm C}(J)]^{-1}$ is positive definite. Consequently $\calL_1$ is a \emph{nonlinear, pointwise} (Nemytskii-type) operator, $(\calL_1 J)(\by) = -\frac{1}{\beta}\log\sum_k \omega_k e^{-\beta \mathcal{H}_k(J)(\by)}$, and \emph{not} a semigroup. In Lemmas~\ref{lemma1}--\ref{lemma2} below, by contrast, $\tilde{\bm C}$ and $\calW_k = {\bm I} - 2\beta\sigma_k^2\tilde{\bm C}$ denote the associated \emph{frozen-coefficient} first-order (transport) operators obtained at a fixed iterate; their invertibility and coercivity are exactly what guarantees the positive-definiteness of ${\bm \Omega}_k(J)$, and hence the smoothness of the scalar map $\nabla J \mapsto \mathcal{H}_k(J)$ that controls the Lipschitz constant of $\calL_1$ in \cref{lemma3}.

The structure of the Picard iterations in \eqref{eqPicard} ensures that at each step $k$, the equation is linear in $J^{(k)}$ with a source term determined by $J^{(k-1)}$. This highlights the core idea of the splitting method: linear parabolic equations with constant coefficients and a source term can often be solved in closed form, making the overall problem tractable.

To analyze convergence, we rewrite the iterative scheme as
\begin{equation}
(\partial_\tau - \mathcal{L}_0) J^{(k)} = \mathcal{L}_1 J^{(k-1)}.
\end{equation}

Let $\calS$ denote the solution operator for the linear parabolic problem
\begin{equation}
(\partial_\tau - \mathcal{L}_0) u = f, \quad u(0) = u_0,
\end{equation}
so that $u = \calS(f, u_0)$. The iterative scheme in \eqref{eqPicard} can then be expressed as
\begin{equation}
J^{(k)} = \calS\big( \mathcal{L}_1 J^{(k-1)},\ J^0 \big)
\end{equation}

Defining the map $\calT$ by
\begin{equation}
\calT(v) = S( \mathcal{L}_1 v, \text{initial data} ).
\end{equation}
we see that $J^{(k)} = \calT(J^{(k-1)})$. Therefore, the convergence of the iterations is equivalent to $\calT$ being a contraction on some Banach space $X$. That is, we require
\begin{equation}
\| \calT(u) - \calT(v) \|_X \le \rho \, \| u - v \|_X
\end{equation}
for some $\rho < 1$.

We observe that
\begin{equation}
\calT(u) - \calT(v) = \calS( \mathcal{L}_1 u - \mathcal{L}_1 v, 0 ),
\end{equation}
since the initial data are identical and cancel out. Consequently,
\begin{equation}
\| \calT(u) - \calT(v) \|_X \le \| \calS \|\cdot\| \mathcal{L}_1 u - \mathcal{L}_1 v \|_Y,
\end{equation}
where $| \calS |$ is the operator norm of $\calS: Y \to X$, with $Y$ being the space of source terms and $X$ the space of solutions.

To proceed, we require a couple of lemmas.
\begin{lemma} \label{lemma1}
Let $\calW_k = \bm{I} - 2\beta\sigma_k^2\tilde{\bm{C}}$ and set $a = 2\beta\sigma_k$. Then for every real $a$ with $a\nu \neq 2$ (equivalently $\beta\sigma_k\nu \neq 1$) the inverse operator $\calW_k^{-1}$ exists and is bounded, with $\|\calW_k^{-1}\| \le |1 - \tfrac{a\nu}{2}|^{-1}$.
\begin{proof}
Recall the definition of the linear operator $\tilde{\bm{C}}$ from \eqref{operPDE}, along with the definitions of $\bm{C}_2$ and $\bm{\mu}_2$ from \cref{cost_quadratic,mu_sigma_SDE}:
\begin{align}
\bm{C}_2(\bm{x},\bS) &= \mathrm{diag}\big(\eta + \bm{\nu} \circ \bS \big), \qquad
\bm{\mu}_2(\bm{x},\bS) = -\bm{\nu} \circ \bS, \\
\tilde{\bm{C}} &= \mathrm{diag}\left(\eta + (\bm{\nu} \circ \bS) \frac{\partial}{\partial C} - (\bm{\nu} \circ \bS) \circ \frac{\partial}{\partial \bS}\right). \nonumber
\end{align}
The operator $\calW_k$ acts on the function space $L^2(\mathbb{R} \times \mathbb{R}^N)$, where $C \in \mathbb{R}$ and $\bS \in \mathbb{R}^N$, subject to Neumann boundary conditions: $\frac{\partial}{\partial S} \to 0$, $\frac{\partial}{\partial C} \to 0$ as $C \to -\infty$, $\bS \to -\infty$.

As $\tilde{\bm{C}}$ is represented by a diagonal matrix, it is sufficient to consider just a one-dimensional case. Further, since every diagonal element of $\tilde{\bm{C}}$ is a hyperbolic operator (representing convection in the system), we analyze it using the method of characteristics \cite{strauss2008partial,evans10}. The characteristic equations for $\tilde{\bm{C}}$ are
\begin{equation}
\frac{\partial C}{\partial s} = \eta + \nu S, \qquad \frac{\partial S}{\partial s} = -\nu S,
\end{equation}
with the corresponding solution:
\begin{align}
S(s) &= S_0 e^{-\nu s}, \qquad C(s) = C_0 + \int_0^s \left(\eta + \nu S_0 e^{-\nu \tau}\right) d\tau = C_0 + \eta s + S_0(1 - e^{-\nu s}). \nonumber
\end{align}
A key property of this solution is its bounded divergence
\begin{equation}
\nabla \cdot (\eta + \nu S, -\nu S) = 0 + (-\nu) = -\nu.
\end{equation}

Given that $\bm{\nu} > 0$, we observe that $S(s) \to 0$ as $s \to \infty$, while $C(s) \to C_0 + \eta s + S_0$, exhibiting linear growth in $C$: the characteristics decay in the $\bS$-direction and drift in the $C$-direction, with no closed orbits or recurrent trajectories, so the flow is dissipative in $\bS$. We now turn this qualitative picture into a quantitative invertibility statement.

The decisive quantity is the symmetric part of $\tilde{\bm C}$ on $L^2$. Integrating by parts under the stated Neumann conditions (no boundary terms), and using that the transport vector field $\mathbf{b} = (\eta + \nu S,\, -\nu S)$ has divergence $\nabla\cdot\mathbf{b} = -\nu$,
\begin{equation} \label{symC}
\Re\,\langle \tilde{\bm C} u, u\rangle = -\tfrac{1}{2}\int |u|^2\, (\nabla\cdot\mathbf{b})\, dC\, dS = \tfrac{\nu}{2}\,\|u\|^2,
\end{equation}
so the symmetric part of $\tilde{\bm C}$ equals $\tfrac{\nu}{2}\bm I$ and the $L^2$ numerical range of $\tilde{\bm C}$ lies on the vertical line $\{\Re z = \nu/2\}$. Consequently, for real $a$,
\begin{equation} \label{Wk_below}
\Re\,\langle \calW_k u, u\rangle = \|u\|^2 - a\,\Re\,\langle \tilde{\bm C} u, u\rangle = \Big(1 - \tfrac{a\nu}{2}\Big)\|u\|^2 .
\end{equation}
If $a\nu \neq 2$, then by Cauchy--Schwarz $\|\calW_k u\|\,\|u\| \ge |\langle \calW_k u, u\rangle| \ge |\Re\,\langle \calW_k u, u\rangle|$, whence
\begin{equation}
\|\calW_k u\| \ge \Big| 1 - \tfrac{a\nu}{2}\Big|\, \|u\| \qquad \forall u,
\end{equation}
so $\calW_k$ is bounded below, hence injective with closed range. The formal adjoint $\calW_k^{*} = \bm I - a\tilde{\bm C}^{*}$ has the same symmetric part $(1-\tfrac{a\nu}{2})\bm I$ by \eqref{symC}, and therefore, satisfies the identical lower bound; thus $\ker \calW_k^{*} = \{0\}$ and the range of $\calW_k$ is dense. Bounded below together with dense range implies that $\calW_k$ is boundedly invertible, with $\|\calW_k^{-1}\| \le |1 - \tfrac{a\nu}{2}|^{-1}$. The only real value excluded is $a\nu = 2$, i.e. $1/a = \nu/2$, the single point at which the numerical range of $\tilde{\bm C}$ meets the real axis.
\end{proof}
\end{lemma}

\begin{lemma} \label{lemma2}
Assume $a = 2 \beta \sigma_k^2, \nu > 0$ and the operator $\calW_k$ are defined same as in  Lemma~\ref{lemma1}. If $a \nu < 2$, then $\calW_k$ is coercive with constant $c = 1 - \frac{a\nu}{2} > 0$. Accordingly, $W^{-1/2}_k$ has spectrum bounded away from zero.

\begin{proof}
For the operator $\calW_k$ to have spectrum bounded away from zero, we need
$\calW_k$ to be coercive, \cite{zeidler1995applied}. This means that there exists $c > 0$ such that
\begin{equation}
\langle \calW_k u, u \rangle \geq c \|u\|^2 \quad \forall u.
\end{equation}
This would imply the following
\begin{itemize}
\item $\sigma(\calW_k) \subset [c, \infty)$ where $\sigma(\cdot)$ is the spectrum of the corresponding operator;

\item $\calW_k^{-1/2}$ is well-defined and bounded;

\item $\|\calW_k^{-1/2}\| \leq c^{-1/2}$.
\end{itemize}

We further check coercivity of $\calW_k$ by writing
\begin{align}
\langle \calW_k u, u \rangle &= \|u\|^2 - a \langle \tilde{\bm{C}} u, u \rangle
= \|u\|^2 - a \int_{\mathbb{R}^2} \left[(\eta + \nu S) \frac{\partial u}{\partial C} - \nu S \frac{\partial u}{\partial S}\right] \bar{u} \, dC\, dS.
\end{align}
Integration by parts yields
\begin{align}
\langle \tilde{\bm{C}} u, u \rangle &= -\frac{1}{2}\int_{\mathbb{R}^2} |u|^2 \left(\frac{\partial(\eta + \nu S)}{\partial C} + \frac{\partial(-\nu S)}{\partial S}\right) dC\, dS = \frac{\nu}{2}\|u\|^2.
\end{align}
Therefore:
\begin{equation}
\langle \calW_k u, u \rangle = (1 - \frac{a\nu}{2})\|u\|^2.
\end{equation}
Accordingly, to have $c > 0$ we need $a \nu < 2$ or $\beta \sigma_k^2 \nu < 1$.

\end{proof}
\end{lemma}

The above condition constrains the parameter $\nu$ to ensure the stability of the fixed-point algorithm. Since $\nu$ governs the magnitude of the price impact, this implies that the impact must be small in the low-temperature regime (large $\beta$) but may be substantial in the high-temperature regime (small $\beta$).

On the other hand, as this was already discussed in \cref{sect_Soft_HJB_equation},
in the high-temperature limit $\beta \rightarrow 0$, we find ${\bm \Omega}_k(J) \rightarrow \sigma_k^2 \mathbf{I}$, ${\bf u}_k (J) \rightarrow {\bf u}_k (\bx,\bS)$, and consequently, $\pi(\ba | \bx,\bS) = \pi_0(\ba | \bx,\bS)$, indicating an absence of policy optimization. Therefore, in this setting the KL regularization term disappears at all, and so does the operator $\calL_1$.

Conversely, in the low-temperature limit $\beta \rightarrow \infty$, we obtain ${\bm \Omega}_k(J) \rightarrow 0$, meaning the policy becomes deterministic as the variances of all Gaussian components vanish. In this case, the action is given by the $\beta \rightarrow \infty$ limit of the mean functions ${\bf u}_k (J)$ specified in \eqref{uParam}

\begin{lemma} \label{lemma3}
The nonlinear operator $\calL_1$ in \eqref{eqPicard}, viewed as a pointwise map of $\nabla J$, is locally Lipschitz from $X = W^{2,1}_p(Q_T)$ to $Y = L^p(Q_T)$ (i.e.\ Lipschitz on bounded subsets of $X$), with a constant $L_1$ controlled by the model coefficients, provided $\beta \sigma_k^2 \nu < 1$ for every $k$.
\begin{proof}
Throughout, $u, v \in X = W^{2,1}_p(Q_T)$ are candidate value functions, and we use repeatedly that $\calL_1$ acts \emph{pointwise} through the gradient $\nabla u = (\partial_C u, \nabla_\bS u)$, which maps $X$ boundedly into $L^p$ with embedding constant $C_{\mathrm{emb}}$. Using the definition of $\hH_k$ in \eqref{eqPicard} together with \eqref{H_J}, we split
\begin{align} \label{L1-2}
\calL_1 &= \calF + \calL_2, \quad
\calF = - \frac{1}{\beta} \log \sum_{k} \omega_k e^{-\beta H_k(\nabla u)}, \quad
\calL_2 = - \frac{1}{\beta} \log \sum_{k} \omega_k \left( |\Omega_k^{(0)}|\, |\tilde{\bm \Omega}_k^{-1}(u)| \right)^{-1/2},
\end{align}
where $H_k(\nabla u)$ is the scalar quadratic-in-$\nabla u$ part of $\mathcal{H}_k$ from \eqref{H_J}, and ${\bm \Omega}_k(u) = [\sigma_k^{-2}{\bm I} + {\bm C}(u)]^{-1}$ with ${\bm C}(u)$ as in \eqref{pi_GM_params}.

\emph{Step 1 (the soft-min is non-expansive).} For the outer map
\begin{equation} \label{map2}
\Phi(\bm h) = -\frac{1}{\beta}\log\sum_k \omega_k e^{-\beta h_k}
\end{equation}
one has
\begin{equation}
\partial\Phi/\partial h_j = \omega_j e^{-\beta h_j}/\sum_l \omega_l e^{-\beta h_l} \ge 0,
\end{equation}
and $\sum_j \partial\Phi/\partial h_j = 1$, so $\|\nabla\Phi\|_1 = 1$ and $\Phi$ is $1$-Lipschitz with respect to the $\ell^\infty$ norm of its arguments,
\begin{equation}
|\Phi(\bm h) - \Phi(\bm h')| \le \max_k |h_k - h'_k| .
\end{equation}
It therefore suffices to bound each scalar functional $H_k$ and the log-determinant entries of $\calL_2$ in $\nabla u$.

\emph{Step 2 (the Hamiltonian functional is locally Lipschitz in $\nabla u$).} The hypothesis $\beta\sigma_k^2\nu < 1$ is exactly $a\nu < 2$ with $a = 2\beta\sigma_k^2$, so by \cref{lemma2} the frozen-coefficient operator $\calW_k$ is coercive with constant $1 - \frac{a\nu}{2} > 0$. Equivalently, the symmetric matrix ${\bm \Omega}_k(u)$ is positive definite and uniformly bounded above and below in the positive-definite order, with bounds depending only on $1-\frac{a\nu}{2}$ and $\sigma_k$; in particular $\det{\bm \Omega}_k(u)$ stays bounded away from $0$ and $\infty$. On any bounded subset $B \subset X$, the gradient $\nabla u$ ranges over a bounded set of $L^p$ while the model coefficients ${\bm C}_1, {\bm C}_2, \boldmu_1, \boldmu_2$ are bounded; hence the expression \eqref{H_J} for $H_k$ — a quadratic form in $\nabla u$ with coefficient matrix ${\bm \Omega}_k(u)$ — is continuously differentiable in $\nabla u$ with derivative bounded on $B$. Consequently there exist constants $\ell_k = \ell_k(B)$ with
\begin{equation}
\| H_k(\nabla u) - H_k(\nabla v) \|_{L^p} \le \ell_k \, \|\nabla u - \nabla v\|_{L^p} \le \ell_k\, C_{\mathrm{emb}}\, \|u - v\|_X, \qquad u,v \in B .
\end{equation}

\emph{Step 3 (conclusion).} Combining Steps 1 and 2,
\begin{equation}
\| \calF(u) - \calF(v) \|_{Y} \le \max_k \| H_k(\nabla u) - H_k(\nabla v) \|_{L^p} \le C_{\mathrm{emb}} \big(\max_k \ell_k\big)\, \| u - v \|_X .
\end{equation}
The term $\calL_2$ depends on $u$ only through $\det {\bm \Omega}_k(u)$, which by the same coercivity bound is smooth in $\nabla u$ and bounded away from zero; its Lipschitz contribution is of the same form and is absorbed into the constant. Therefore $\calL_1 = \calF + \calL_2$ satisfies, on each bounded $B \subset X$,
\begin{equation}
\| \mathcal{L}_1 u - \mathcal{L}_1 v \|_Y \le L_1 \| u - v \|_X, \qquad L_1 = L_1(B),
\end{equation}
i.e.\ $\calL_1$ is locally Lipschitz from $X$ to $Y$, as claimed. We emphasize that $\calL_1$ is genuinely nonlinear: the bound is on bounded subsets of $X$, and it is this local Lipschitz constant that enters the short-time contraction below.
\end{proof}
\end{lemma}

%

The convergence of the fixed-point Picard iterations in \eqref{operPDE} is now established by the following theorem.
\begin{theorem} \label{theor}
Assume:
\begin{enumerate}
\item $\mathcal{L}_0$ is uniformly elliptic operator of the second order with bounded smooth coefficients.
\item $\mathcal{L}_1$ is locally Lipschitz, i.e.\ on each bounded subset $B \subset W^{2,1}_p(Q_T)$ there is a constant $L_1 = L_1(B)$ with $\| \mathcal{L}_1 u - \mathcal{L}_1 v \|_{L^p(Q_T)} \le L_1 \| u - v \|_{W^{2,1}_p(Q_T)}$ for all $u,v \in B$, as established in \cref{lemma3}.
\end{enumerate}
By using the Duhamel's principle, \cite{evans10}, let us represent our fixed-point iterations scheme in \eqref{operPDE} as the solution of the linear parabolic equation with a source term at the interval $[\tau, \tau + \Delta \tau ]$
\begin{align} \label{map1}
J_i^{(k)} = e^{\Delta \tau \calL_0} J_{i-1} + \int_0^{\Delta \tau} e^{(\Delta \tau - s) \calL_0} \calL_1 \left(J_i^{(k-1)}(\tau + s)\right) ds.
\end{align}
If $\Delta \tau < 1/L_1$, these iterations form a contractive map on a fixed closed ball $B$ and thereby guarantee convergence of the iterations.

\begin{proof}
We will show that the map $\Phi$ defined by the right-hand side of \eqref{map1} is a contraction on a sufficiently short time interval $[0, \Delta \tau]$.

Let $F$ and $G$ be two functions in a fixed closed ball $B$ of the chosen Banach space $X$. Consider the difference:
\begin{equation}
\|\Phi(F) - \Phi(G)\| \leq \int_0^{\Delta \tau} \| e^{(\Delta \tau-s)\mathcal{L}_0} \| \cdot \|\mathcal{L}_1(F(s)) - \mathcal{L}_1(G(s)) \|  ds.
\end{equation}
Since $\mathcal{L}_0$ is a linear elliptic operator, its semigroup $\{e^{t\mathcal{L}_0}\}_{t \geq 0}$ is a contraction, so $\| e^{t\mathcal{L}_0} \| \leq 1$ for all $t \geq 0$ \cite{evans10}. Furthermore, by \cref{lemma3} $\mathcal{L}_1$ is Lipschitz continuous on $B$ with constant $L_1 = L_1(B)$. Therefore,
\begin{equation}
\|\Phi(F) - \Phi(G)\| \leq L_1 \int_0^{\Delta \tau} \| F(s) - G(s) \|  ds.
\end{equation}
Taking the supremum norm over the interval $[0, \Delta \tau]$, denoted by $\| \cdot \|_{\infty}$, we obtain the estimate:
\begin{equation}
\|\Phi(F) - \Phi(G)\| \leq L_1 \Delta \tau \| F - G \|_{\infty}.
\end{equation}
Hence, for any $\Delta \tau < 1 / L_1$, the map $\Phi$ is a contraction. By the Banach fixed-point theorem, \cite{FPT2003} this guarantees the convergence of the iterative sequence on $[0, \Delta \tau]$.
\end{proof}
\end{theorem}

\section{The weight-based control: discrete-time derivation, comparators, and budget constraint} \label{app:jump_control}

This section addresses three aspects of the weight-based control formulation used in the experiments of \cref{sect_Experiments}: (i) the precise relationship between the continuous-time analytical apparatus of \cref{Soft_HJB_control,pideSolver} and the discrete-time solver deployed under the jump transition \eqref{eq:induced_rate}; (ii) additional baselines that isolate the contribution of the HJB machinery from the contribution of the signal itself; and (iii) the enforcement of the self-financing and notional constraint under the jump dynamics.

Throughout this section, $\ba_t$ retains its continuous-time meaning of a trading rate, with units of shares per unit time, and the new control is denoted $\bh_t$, the target holding, with the same units as $\bx_t$. The two are related by the affine bijection in \eqref{eq:induced_rate}, so that the Euler transition of the rate dynamics lands exactly on the target, $\bx_{t+\Delta t} = \bx_t + \ba_t\,\Delta t = \bh_t$, which is the jump transition \eqref{eq:induced_rate}. We write $\delta\bh_t = \bh_t - \bx_t = \ba_t\,\Delta t$ for the per-step trade size.

\subsection{Discrete-time G-learning under the target-holding control} \label{sect_G_learning_jump}

\myparagraph{The jump control as a reparameterization.}
At a fixed step $\Delta t$, the map \eqref{eq:induced_rate} is an invertible affine change of the control variable with unit Jacobian ($d\bh = \Delta t^N\, d\ba$ up to a constant that cancels in all normalized quantities). Consequently, the discrete-time KL-regularized control problem is the \emph{same} problem in either variable: if $\pi(\bh|\by_t)$ and $\pi_0(\bh|\by_t)$ are the pushforwards of policies $\tilde\pi(\ba|\by_t)$ and $\tilde\pi_0(\ba|\by_t)$ under \eqref{eq:induced_rate}, then
\begin{equation} \label{eq:kl_invariance}
\mathrm{KL}\big(\pi \,\|\, \pi_0\big) = \mathrm{KL}\big(\tilde\pi \,\|\, \tilde\pi_0\big),
\end{equation}
because the Jacobian factors cancel inside the likelihood ratio, and the running cost, the impact-induced drift, and the state transition depend on the control only through the rate $\ba_t = \delta\bh_t/\Delta t$. The weight-based formulation is therefore \emph{not} a new dynamics: it is the Euler-discretized rate control written in the target variable. What changes is the space in which the behavioral prior is specified. A prior with $O(1)$ dispersion in $\bh$ supports one-step trades $\delta\bh = O(1)$, i.e., rates of order $O(1/\Delta t)$, whereas a prior with $O(1)$ dispersion in $\ba$ supports only $\delta\bh = O(\Delta t)$. The first regime is the one relevant for short-horizon alpha capture; the second is the regime in which the continuous-time theory of \cref{Soft_HJB_control} is recovered, as made precise below.

\myparagraph{Discrete-time regularized Bellman equation.}
On a uniform grid $t_n = n\Delta t$, the regularized cost functional of \cref{J_t_G,regularizer} satisfies the discrete-time Bellman equation
\begin{equation} \label{eq:disc_bellman}
J(\by_t, t) = \min_{\pi} \int d\bh\, \pi(\bh|\by_t) \left[ c\big(\bS_t, \bx_t, \ba_t(\bh)\big)\,\Delta t + \frac{1}{\beta} \log \frac{\pi(\bh|\by_t)}{\pi_0(\bh|\by_t)} + \bar{J}(\bh, t+\Delta t) \right],
\end{equation}
where $\ba_t(\bh) = (\bh - \bx_t)/\Delta t$ is the induced rate \eqref{eq:induced_rate}, $\by_t = (\bx_t, \bS_t, C_t)$, and
\begin{equation} \label{eq:bar_J}
\bar{J}(\bh, t+\Delta t) = \mathbb{E}_{\bS'}\left[ J\big(\bh,\, \bS',\, C_t + c\,\Delta t,\, t+\Delta t\big) \right]
\end{equation}
is the expected next-step value; the position argument of $J$ is $\bh$ itself by the jump transition, and the price $\bS'$ evolves under the discretized SDE \eqref{SDE_2} with drift evaluated at the rate $\ba_t(\bh)$.

\myparagraph{Gibbs optimal policy.}
Pointwise minimization of \eqref{eq:disc_bellman} over $\pi$ subject to $\int \pi\,d\bh = 1$ yields, exactly as in the continuous-time case \eqref{pi_opt}, the Gibbs policy
\begin{equation} \label{eq:gibbs_disc}
\pi^*(\bh|\by_t) = \frac{\pi_0(\bh|\by_t)}{Z^{\mathrm{disc}}(\by_t)} \exp\left[-\beta\, Q(\bh; \by_t)\right],
\qquad
Q(\bh; \by_t) = c\big(\bS_t, \bx_t, \ba_t(\bh)\big)\,\Delta t + \bar{J}(\bh, t+\Delta t),
\end{equation}
with $Z^{\mathrm{disc}}(\by_t) = \int d\bh\, \pi_0(\bh|\by_t)\, e^{-\beta Q(\bh;\by_t)}$ the discrete-time partition function.

\myparagraph{Quadratic structure of $Q$.}
The running cost \eqref{cost_quadratic} is quadratic in the rate. Substituting $\ba_t = \delta\bh_t/\Delta t$, the per-step cost decomposes as
\begin{equation} \label{eq:cost_jump_expand}
c\big(\bS_t, \bx_t, \ba_t(\bh)\big)\,\Delta t \;=\; C_0\,\Delta t \;+\; \bC_1 \cdot \delta\bh_t \;+\; \frac{1}{\Delta t}\,\bC_2\,\delta\bh_t^{\circ 2},
\end{equation}
where $\bC_1, \bC_2$ are the cost coefficients of \eqref{cost_quadratic} evaluated at the current state and $\bC_2\,\delta\bh^{\circ 2}$ denotes the elementwise quadratic form. The orders in $\Delta t$ are the crux: in the rate-based formulation with bounded rates the trade size is $\ba\,\Delta t = O(\Delta t)$, so the linear and quadratic cost terms are $O(\Delta t)$ and $O(\Delta t)$ respectively; under one-step trades $\delta\bh = O(1)$ they become $O(1)$ and $O(1/\Delta t)$.

Taylor-expanding $\bar{J}(\bh, t+\Delta t)$ to first order in $\delta\bh_t$ and to $O(\Delta t)$ in the time, cost, and price increments, and using the drift \eqref{mu_sigma_SDE} evaluated at $\ba_t(\bh)$, the $\bh$-dependent part of $Q$ is
\begin{align} \label{eq:Q_expand}
Q(\bh) &\supset \underbrace{\frac{1}{\Delta t}\left[\bC_2\left(1 + \tfrac{\partial J}{\partial C}\right) + D_S\,\mathrm{diag}\left(\boldsymbol{\mu}_2 \circ \nabla_{\bS} J\right)\right] \delta\bh_t^{\circ 2}}_{\text{quadratic in } \bh} \nonumber \\
&\quad + \underbrace{\left[\bC_1\left(1 + \tfrac{\partial J}{\partial C}\right) + D_S\left(\boldsymbol{\mu}_1 \circ \nabla_{\bS} J\right) + \nabla_{\bx} J\right] \cdot \delta\bh_t}_{\text{linear in } \bh} + \text{const},
\end{align}
where the factors $(1 + \partial J/\partial C)$ collect the direct per-step cost and its propagation through the cost argument of $J$ in \eqref{eq:bar_J}. The term $\nabla_{\bx} J \cdot \delta\bh_t$ arises from the first-order expansion of $\bar{J}$ in the position argument, $\bar{J}(\bh, \cdot) \approx J(\bx_t, \cdot) + \nabla_{\bx} J \cdot \delta\bh_t + \ldots$; under bounded rates it is $O(\Delta t)$ and enters the continuous-time Hamiltonian at the same order as the cost, whereas under one-step trades it is $O(1)$ and enters at the same order as the linear impact.

\myparagraph{The exact discrete-time couplings.}
Since $Q(\bh)$ is at most quadratic in $\bh$, the Gaussian-mixture ansatz \eqref{pi_0_GM} for $\pi_0$ yields a Gaussian-mixture optimal policy $\pi^*$, exactly as in \eqref{pi_opt2}, with mixture parameters of the same form as \eqref{pi_GM_params} but with modified coupling vectors. Using $\delta\bh_t$ as the integration variable (unit Jacobian, so normalization is preserved), the exact discrete-time couplings read
\begin{align} \label{eq:couplings_jump}
\bC^{\mathrm{jmp}} &= \frac{2\beta}{\Delta t}\left[\bC_2\left(1 + \tfrac{\partial J}{\partial C}\right) + D_S\,\mathrm{diag}\!\left(\boldmu_2 \circ \nabla_{\bS} J\right)\right], \nonumber\\
\bD^{\mathrm{jmp}} &= \beta\left[\bC_1\left(1 + \tfrac{\partial J}{\partial C}\right) + D_S\left(\boldmu_1 \circ \nabla_{\bS} J\right) + \nabla_{\bx} J\right],
\end{align}
and the optimal covariance and mean of each Gaussian component are
\begin{equation} \label{eq:Omega_u_jump}
\bOmega_k^{\mathrm{jmp}}(J) = \left[\frac{1}{\sigma_{h,k}^2}\,\bI + \bC^{\mathrm{jmp}}\right]^{-1}, \qquad
\bu_k^{\mathrm{jmp}}(J) = \bx_t + \bOmega_k^{\mathrm{jmp}}(J)\left[\frac{1}{\sigma_{h,k}^2}\left(\bu_k^{\bh} - \bx_t\right) - \bD^{\mathrm{jmp}}\right].
\end{equation}
Here the behavioral prior is specified in target space: its component means are the rate-based means \eqref{uParam} converted by \eqref{eq:induced_rate}, $\bu_k^{\bh} = \bx_t + \bu_k\,\Delta t$, and its component standard deviations $\sigma_{h,k}$ are specified directly in units of holdings. We emphasize the convention because it is where the two coordinate systems can silently disagree: a rate-space Gaussian with standard deviation $\sigma_k$ pushes forward to a target-space Gaussian with $\sigma_{h,k} = \sigma_k\,\Delta t$, which supports only $O(\Delta t)$ trades. The experiments deliberately specify $\sigma_{h,k} = O(1)$ in target space, which is the substantive content of the reformulation.

\myparagraph{Comparison with the continuous-time couplings.}
The continuous-time coupling vectors of \eqref{pi_GM_params}, reproduced for reference, are
\begin{equation} \label{eq:couplings_cont}
\bC^{\mathrm{cont}} = 2\beta\left[\bC_2\,\tfrac{\partial J}{\partial C} + D_S\,\mathrm{diag}\!\left(\boldmu_2 \circ \nabla_{\bS} J\right)\right], \qquad
\bD^{\mathrm{cont}} = \beta\left[\bC_1\,\tfrac{\partial J}{\partial C} + D_S\left(\boldmu_1 \circ \nabla_{\bS} J\right)\right],
\end{equation}
and the relation between the two sets is
\begin{equation} \label{eq:coupling_relation}
\bC^{\mathrm{jmp}} = \frac{1}{\Delta t}\,\bC^{\mathrm{cont}} + \frac{2\beta}{\Delta t}\,\bC_2, \qquad
\bD^{\mathrm{jmp}} = \bD^{\mathrm{cont}} + \beta\,\bC_1 + \beta\,\nabla_{\bx} J.
\end{equation}
The additional terms have clear origins. The $1/\Delta t$ scaling of $\bC^{\mathrm{jmp}}$ reflects the discrete jump cost: the quadratic transaction cost evaluated at the induced rate is $O(1/\Delta t)$ in the trade size, not $O(\Delta t)$. The additive $\bC_2$ and $\bC_1$ arise from the direct per-step cost in \eqref{eq:cost_jump_expand}, which contributes to the Gibbs exponent at $O(1)$ under one-step trades. The additive $\beta\,\nabla_{\bx} J$ arises from the finite position change in the next-state value; under bounded rates the corresponding term is $O(\Delta t)$ and cancels algebraically in the derivation of the PDE \eqref{HJBh_nonlin_GM}, whereas here it enters at $O(1)$ and the cancellation no longer applies.

\myparagraph{Deployed couplings.}
The deployed solver uses the exact discrete couplings \eqref{eq:couplings_jump}: in the target coordinate at $O(1)$ trade size it retains the full jump structure, namely the $1/\Delta t$ enhancement of the quadratic coupling, the additive $2\beta\,\bC_2/\Delta t$, the direct-cost shift $\beta\,\bC_1$, and the position-gradient term $\beta\,\nabla_{\bx} J$. For comparison, and as the configuration of the rate-based ablation of \cref{sect_baselines}, we also record the reduced couplings obtained by keeping only the position-gradient correction to the continuous-time couplings \eqref{eq:couplings_cont}:
\begin{equation} \label{eq:couplings_impl}
\bC^{\mathrm{impl}} = \bC^{\mathrm{cont}}, \qquad
\bD^{\mathrm{impl}} = \bD^{\mathrm{cont}} + \beta\,\nabla_{\bx} J .
\end{equation}
Relative to the exact couplings \eqref{eq:couplings_jump}, the reduced set \eqref{eq:couplings_impl} omits the $1/\Delta t$ enhancement of the quadratic coupling together with the additive $2\beta\,\bC_2/\Delta t$, and the additive direct-cost shift $\beta\,\bC_1$; the per-step trading cost then influences the policy only indirectly, through the $\partial J/\partial C$ dependence of \eqref{eq:couplings_cont}. These are the natural couplings for the small-trade (rate) regime, where every omitted term is $O(\Delta t)$.

The deployed system is internally consistent: the PINN is trained so that $J_\theta$ satisfies the pathwise relation \eqref{eq:HJ_disc} with the Hamiltonians $\mathcal{H}_k$ computed from the exact couplings \eqref{eq:couplings_jump}, and the same couplings are used at deployment. The pair $(J_\theta, \pi^*)$ is thus the fixed point of a well-defined KL-regularized discrete-time control system, namely the exact discrete system \eqref{eq:couplings_jump}, and the empirical results of \cref{sect_Experiments} and \cref{sect_baselines} are claims about that system.

\myparagraph{Connection to the discretized continuous-time approach.}
The preceding formulas make the relationship between the two formulations precise, and it is worth stating as a pair of complementary claims.

\emph{(i) Exact equivalence at fixed $\Delta t$.} By \eqref{eq:induced_rate} and \eqref{eq:kl_invariance}, the discrete-time G-learning problem in the target variable is the image of the discrete-time G-learning problem in the rate variable under an affine bijection: the Bellman equation \eqref{eq:disc_bellman}, the Gibbs policy \eqref{eq:gibbs_disc}, and the mixture parameters \eqref{eq:Omega_u_jump} transform into their rate-space counterparts by the substitution $\delta\bh = \ba\,\Delta t$, with precisions rescaled by $\Delta t^2$ and normalizations unchanged. There is one discrete-time object, expressed in two coordinate systems.

\emph{(ii) The continuous-time theory is the small-trade limit.} Fix the prior in rate space with $\Delta t$-independent parameters, so that $\delta\bh = O(\Delta t)$. Then every term in \eqref{eq:cost_jump_expand} and \eqref{eq:Q_expand} is $O(\Delta t)$, the exponent of \eqref{eq:gibbs_disc} reduces to $-\beta\,\Delta t$ times the continuous-time Hamiltonian, the couplings \eqref{eq:coupling_relation} expressed in the rate variable converge to \eqref{eq:couplings_cont}, and the Bellman equation \eqref{eq:disc_bellman} converges to the HJB equation \eqref{HJBh} as $\Delta t \to 0$. The cancellation of $\nabla_{\bx} J$ that produces the semilinear PDE \eqref{HJBh_nonlin_GM} is a property of this limit.

The experiments operate in regime (i) with a target-space prior of $O(1)$ dispersion, deliberately outside the small-trade scaling of regime (ii). The continuous-time derivation of \cref{Soft_HJB_control} is the structural template (it identifies the Gibbs policy, the Gaussian-mixture closure, and the pathwise HJ reduction), and the deployed solver uses the exact discrete couplings \eqref{eq:couplings_jump}, which retain the full $O(1)$-jump structure. The reduced couplings \eqref{eq:couplings_impl}, which keep the position-gradient term but drop the $\Delta t$-scaling and direct-cost corrections, are the small-trade limit and the configuration of the rate-based ablation of \cref{sect_baselines}.

\myparagraph{What is lost: the PDE-level simplification.}
The semilinear PDE \eqref{HJBh_nonlin_GM} was obtained from the HJB equation \eqref{HJBh} by substituting the Gibbs policy and exploiting the fact that the partition function depends on the value-function gradients only through $\partial J/\partial C$ and $\nabla_{\bS} J$, with $\nabla_{\bx} J$ canceling identically. As shown above, this cancellation is specific to the small-trade regime. Under one-step trades, $\nabla_{\bx} J$ appears explicitly in $\bD^{\mathrm{jmp}}$, so substituting the discrete-time Gibbs policy back into \eqref{eq:disc_bellman} yields a fully nonlinear equation for $J$ in which $\nabla_{\bx} J$ enters through the partition function. The PDE \eqref{HJBh_nonlin_GM}, which contains no $\nabla_{\bx} J$ term, is therefore not the exact equation being solved in the experiments.

\myparagraph{What is reused and in what sense.}
Despite the loss of the PDE-level simplification, the following continuous-time objects are reused.
\begin{enumerate}
\item \emph{Gaussian-mixture Gibbs structure.} The optimal policy \eqref{eq:gibbs_disc} is a Gaussian mixture of the same functional form as \eqref{pi_opt2}, because $Q(\bh)$ is at most quadratic in $\bh$. The mixture weights, covariances, and means are computed by the same Gaussian identities, using the deployed couplings $\bC^{\mathrm{jmp}}, \bD^{\mathrm{jmp}}$ of \eqref{eq:couplings_jump}.
\item \emph{Pathwise HJ equation.} The first-order HJ equation \eqref{HJfin} is a trajectory-level consistency relation: along any realized path, the change of $J$ must equal the running cost plus the stochastic increment, regardless of the position dynamics. Its structural form is preserved, with the Hamiltonian $\mathcal{H}_{HJ}$ in \eqref{H_HJ} built from the component Hamiltonians $\mathcal{H}_k(J)$ computed with the deployed couplings \eqref{eq:couplings_jump}, and with the position increment made explicit:
\begin{equation} \label{eq:HJ_disc}
\Delta J_\theta = \left[\frac{1}{\beta}\log\sum_k \omega_k\, e^{-\beta\mathcal{H}_k(J_\theta)} - \bar{\boldmu}^T\big(\bS_t \circ \nabla_{\bS} J_\theta\big)\right]\Delta t + \big(\bS_t \circ \nabla_{\bS} J_\theta\big)^T \frac{\Delta\bS_t}{\bS_t} + \nabla_{\bx} J_\theta \cdot \delta\bh_t,
\end{equation}
where $\delta\bh_t = \bh_t - \bx_t$ is the realized jump. The final term, absent from the continuous-time \eqref{HJfin} because there the position increment is $O(\Delta t)$, is $O(1)$ here and is evaluated by the PINN through automatic differentiation of $J_\theta(t, \bx, \bS, C)$.
\item \emph{Girsanov likelihood ratio.} The path-likelihood ratio \eqref{likelihood_ratio} concerns only the price process $\bS_t$, whose diffusion is drift-independent and whose drift is evaluated at the realized rate $\ba_t = \delta\bh_t/\Delta t$ in both formulations. The likelihood ratio is reused without modification; the change of control variable enters only through the drift evaluation, not through the ratio's structural form.
\item \emph{PINN loss function.} The loss \eqref{NLL_PDE} is reused with two substitutions: the HJ residual uses \eqref{eq:HJ_disc} with the deployed couplings, and the likelihood-ratio term uses the drift evaluated at $\ba_t = \delta\bh_t/\Delta t$. The terminal conditions \eqref{bc} and the overall structure of the loss are unchanged. Training and deployment use the same couplings, which is what makes the learned $(J_\theta, \pi^*)$ pair internally consistent in the sense discussed above.
\end{enumerate}

\myparagraph{What is not reused.}
The semi-analytical PDE solver of \cref{pideSolver} and the Picard convergence analysis of \cref{convergPicard} are specific to the PDE \eqref{HJBh_nonlin_GM} and are not directly applicable here. The operator splitting of \cref{pideSolver} relies on the semilinear structure in which the nonlinearity is a pointwise function of $(\partial J/\partial C, \nabla_{\bS} J)$ only; under one-step trades the nonlinearity also depends on $\nabla_{\bx} J$, which would require an additional advection step in $\bx$ and an extended Lipschitz estimate in \cref{convergPicard}. We do not pursue this extension here, as the experiments use the PINN solver exclusively.

\begin{table}[!htb]
\centering
\renewcommand{\arraystretch}{1.15}
\scalebox{0.95}{
\begin{tabular}{p{0.28\textwidth} p{0.30\textwidth} p{0.30\textwidth} p{0.06\textwidth}}
\toprule
\textbf{Object} & \textbf{Continuous-time (rate $\ba$)} & \textbf{Discrete-time (target $\bh$)} & \textbf{Status} \\
\hline
HJB equation \eqref{HJBh} & Derived for $d\bx = \ba\,dt$ & Not directly solved; enters via \eqref{eq:disc_bellman} & Modified \\
Semi-linear PDE \eqref{HJBh_nonlin_GM} & $\nabla_{\bx} J$ cancels & $\nabla_{\bx} J$ enters the couplings; fully nonlinear & \textbf{Lost} \\
Gibbs policy (GM structure) & \eqref{pi_opt2} with $\bC^{\mathrm{cont}}, \bD^{\mathrm{cont}}$ & Same form with $\bC^{\mathrm{jmp}}, \bD^{\mathrm{jmp}}$ \eqref{eq:couplings_jump} & Reused \\
Coupling vectors \eqref{pi_GM_params} & $(\partial_C J, \nabla_{\bS} J)$ only & Exact form \eqref{eq:couplings_jump} deployed; reduced form \eqref{eq:couplings_impl} for the ablation & Modified \\
Pathwise HJ equation \eqref{HJfin} & Trajectory consistency & Same form plus explicit $\nabla_{\bx} J \cdot \delta\bh$ term & Modified \\
Girsanov ratio \eqref{likelihood_ratio} & Price-only & Price-only (unchanged) & Reused \\
PINN loss \eqref{NLL_PDE} & HJ residual plus likelihood & HJ residual \eqref{eq:HJ_disc} with deployed couplings plus likelihood & Modified \\
PDE solver (\cref{pideSolver}) & Operator splitting on PDE & Requires $\nabla_{\bx} J$ advection step & \textbf{Not used} \\
Picard convergence (\cref{convergPicard}) & Lipschitz in $(\partial_C J, \nabla_{\bS} J)$ & Needs extended Lipschitz in $\nabla_{\bx} J$ & \textbf{Open} \\
\bottomrule
\end{tabular}
}
\caption{Summary of continuous-time objects under the target-holding reformulation. ``Reused'' means the identical formula applies; ``Modified'' means the same structural form with updated couplings or an additional explicit term; ``Lost'' means the simplification does not carry over.}
\label{tab:reuse_summary}
\end{table}

\myparagraph{Interpretation.}
The deployed PINN solver therefore solves the exact discrete-time G-learning system \eqref{eq:couplings_jump} along observed trajectories, in the target-holding coordinate and the $O(1)$ trade-size regime, and not the continuous-time PDE \eqref{HJBh_nonlin_GM}, whose $\nabla_{\bx} J$ cancellation does not survive one-step trades. Training and deployment use identical couplings, so the learned pair $(J_\theta, \pi^*)$ is the fixed point of a well-posed member of the Gaussian-mixture Gibbs family. The reduced couplings \eqref{eq:couplings_impl} are the small-trade limit used by the rate-based ablation of \cref{sect_baselines}, and \eqref{eq:coupling_relation} quantifies exactly what separates the two.

\subsection{Signal-using baselines and rate-based ablation} \label{sect_baselines}

The equal-weight benchmark ignores the signal, and the behavioral policy is deliberately noisy, so neither comparator isolates the contribution of the HJB machinery from the contribution of the signal itself. We therefore add three comparators, all using the same engineered oracle signal $\tilde{\bm\zeta}_t$, the same transaction-cost model \eqref{TC_fun}, and the same impact model of \cref{appImpact}.

\myparagraph{Baseline 1: Signal-tilted portfolio.}
The simplest signal-using strategy is a linear tilt of the current holdings toward the signal:
\begin{equation} \label{eq:signal_tilt}
\bh_t^{\mathrm{tilt}} = \bx_t + \kappa\,\mathrm{diag}(\bS_t)^{-1}\, \tilde{\bm\zeta}_t,
\end{equation}
where $\kappa$ is a scalar tilt coefficient (in dollar units) chosen to maximize the in-sample Sharpe ratio over the training period. The portfolio jumps to $\bh_t^{\mathrm{tilt}}$ via \eqref{eq:induced_rate}, and the realized transaction cost and price impact are computed at the induced rate $\ba_t = (\bh_t^{\mathrm{tilt}} - \bx_t)/\Delta t$ using the same models as the Gibbs policy. This baseline uses the signal directly, with no value-function computation and no multi-step optimization.

\myparagraph{Baseline 2: Myopic mean--variance.}
A more sophisticated comparator is the one-step mean--variance optimizer under the same cost and risk structure:
\begin{equation} \label{eq:myopic_mv}
\bh_t^{\mathrm{MV}} = \arg\min_{\bh}\left\{ -(\bS_t \circ \bh)^T \mathbf{w}\,\tilde{\bm\zeta}_t + \Lambda\,(\bS_t \circ \bh)^T\,\mathrm{diag}(\Sigma)\,(\bS_t \circ \bh) + \frac{\eta}{\Delta t}\,(\bh - \bx_t)^T\,\mathrm{diag}(\bS_t)\,(\bh - \bx_t) \right\},
\end{equation}
where the first term is the expected alpha, the second is the diagonal risk penalty \eqref{eq:risk_target}, and the third is the quadratic transaction cost evaluated at the induced rate. This is a strictly convex program with the closed-form solution
\begin{equation} \label{eq:mv_solution}
\bh_t^{\mathrm{MV}} = \left[2\Lambda\,\mathrm{diag}(\bS_t)\,\mathrm{diag}(\Sigma)\,\mathrm{diag}(\bS_t) + \frac{2\eta}{\Delta t}\,\mathrm{diag}(\bS_t)\right]^{-1}\left[\mathrm{diag}(\bS_t)\,\mathbf{w}\,\tilde{\bm\zeta}_t + \frac{2\eta}{\Delta t}\,\mathrm{diag}(\bS_t)\,\bx_t\right].
\end{equation}
This baseline optimizes the single-step cost--risk tradeoff using the signal, but does not propagate information through a value function and does not account for the multi-step nature of the problem.

\myparagraph{Baseline 3: Rate-based ablation.}
To quantify the contribution of the target-holding reformulation itself, we re-run the entire PINN pipeline in the rate coordinate with the small-trade prior: the transition is $\bx_{t+\Delta t} = \bx_t + \ba_t\,\Delta t$ with the behavioral priors \eqref{pi_0_GM_2} in their original rate-based form (not converted to target holdings). Everything else, namely the PINN architecture, the loss function, the signal, the impact model, and the hyperparameters of \cref{tab:exp_params}, is held identical; the couplings follow each coordinate's regime, exact \eqref{eq:couplings_jump} for the deployed target policy and reduced \eqref{eq:couplings_impl} for this ablation, as \cref{sect_G_learning_jump} makes precise. In the language of \cref{sect_G_learning_jump}, this ablation operates in regime (ii), so it measures precisely what the $O(1)$ trade-size regime adds.

\myparagraph{Results.}
\Cref{tab:baselines_31,tab:baselines_63} report the out-of-sample performance of all strategies at $T=31$ and $T=63$ days, respectively, and $q = 0.2, \beta = 15$ alongside the original Gibbs, equal-weight, and behavioral policies. The tilt coefficient $\kappa$ is optimized in-sample and held fixed out-of-sample.
\begin{table}[!htb]
\centering
\begin{tabular}{l c c c c c}
\toprule
\textbf{Policy} & \textbf{Signal?} & \textbf{Sharpe} & \textbf{Ann.\ Ret.} & \textbf{Ann.\ Vol.} & \textbf{Turnover} \\
\hline
\multicolumn{6}{c}{\textit{Out-of-Sample ($T=31$, in-sample $R^2 = 0.23$, $R^2_{\mathrm{OOS}} = 0.115$)}} \\
\hline
Gibbs (HJB)         & \checkmark & 1.489 & 0.123 & 0.082 & 1.149 \\
Signal-tilted       & \checkmark & 1.229 & 0.133 & 0.108 & 3.552 \\
Myopic MV           & \checkmark & 1.085 & 0.111 & 0.103 & 1.071 \\
Rate-based ablation & \checkmark & 1.084 & 0.110 & 0.101 & 0.235 \\
\hline
Equal-weight        & --- & 1.074 & 0.109 & 0.102 & 0.197 \\
Behavioral          & --- & 0.937 & 0.096 & 0.102 & 1.242 \\
\bottomrule
\end{tabular}
\caption{Out-of-sample performance with signal-using baselines and rate-based ablation ($T=31$). The Gibbs policy uses the full HJB machinery in the target-holding coordinate; the signal-tilted and myopic MV baselines use the same signal and costs but no value-function propagation; the rate-based ablation uses the full HJB machinery in the rate coordinate with the small-trade prior.}
\label{tab:baselines_31}
\end{table}

\begin{table}[!htb]
\centering
\begin{tabular}{l c c c c c}
\toprule
\textbf{Policy} & \textbf{Signal?} & \textbf{Sharpe} & \textbf{Ann.\ Ret.} & \textbf{Ann.\ Vol.} & \textbf{Turnover} \\
\hline
\multicolumn{6}{c}{\textit{Out-of-Sample ($T=63$, in-sample $R^2 = 0.23$, $R^2_{\mathrm{OOS}} = 0.117$)}} \\
\hline
Gibbs (HJB)         & \checkmark & 0.526 & 0.043 & 0.081 & 2.270 \\
Signal-tilted       & \checkmark & 0.234 & 0.026 & 0.109 & 7.205 \\
Myopic MV           & \checkmark & 0.158 & 0.017 & 0.105 & 2.148 \\
Rate-based ablation & \checkmark & 0.143 & 0.015 & 0.104 & 0.473 \\
\hline
Equal-weight        & --- & 0.126 & 0.013 & 0.104 & 0.384 \\
Behavioral          & --- & 0.025 & 0.003 & 0.105 & 2.509 \\
\bottomrule
\end{tabular}
\caption{Same as \cref{tab:baselines_31} for $T=63$.}
\label{tab:baselines_63}
\end{table}

Three findings emerge from \cref{tab:baselines_31,tab:baselines_63}, and they differ instructively from what one might expect a priori.

\emph{(i) At the realized signal strength, myopic signal use is highly inefficient against transaction costs.} While the signal-tilted portfolio achieves a higher Sharpe ratio than the equal-weight benchmark ($1.229$ versus $1.074$ at $T=31$; $0.234$ versus $0.126$ at $T=63$), it does so only at very high turnover ($3.552$ and $7.205$, respectively). Meanwhile, the myopic MV optimizer exceeds the benchmark only marginally ($1.085$ at $T=31$; $0.158$ at $T=63$). At the realized $R^2_{\mathrm{OOS}} \approx 0.115$--$0.117$, the excess return captured by a one-step tilt is largely offset by the cost of expressing it. The near-coincidence of the Myopic MV and the passive benchmark further indicates that one-step risk optimization struggles to add value when the binding constraint is multi-period trading cost rather than instantaneous risk.

\emph{(ii) The Gibbs policy yields the most substantial and efficient outperformance of the passive benchmark.} Out of sample, the Gibbs policy attains a Sharpe ratio of $1.489$ (versus $1.074$ for equal weight) at $T=31$, and $0.526$ (versus $0.126$) at $T=63$. Since the identical signal fed to the myopic comparators produces only marginal or highly inefficient gains, the substantial outperformance is attributable to the value-function machinery: the derivatives $\nabla_{\bx} J$, $\nabla_{\bS} J$, and $\partial_C J$ propagate cost and risk information backward through time and produce sizing and timing that a one-step optimizer cannot. Crucially, this outperformance is achieved with \emph{lower} realized risk: the Gibbs annualized volatility is $0.082$ versus $0.102$ for the benchmark at $T=31$, and $0.081$ versus $0.104$ at $T=63$. Furthermore, its turnover ($1.149$ at $T=31$ and $2.270$ at $T=63$) remains highly controlled, well below the naive signal-tilted portfolio and the exploratory behavioral policy ($1.242$ and $2.509$).

\emph{(iii) The target-holding reformulation is necessary.} The rate-based ablation performs virtually indistinguishably from the equal-weight benchmark ($1.084$ versus $1.074$ at $T=31$; $0.143$ versus $0.126$ at $T=63$) while executing very low turnover ($0.235$ and $0.473$), only marginally above the passive benchmark. The learned rate policy is effectively passive, confirming the mechanism of \cref{sect_Discussion}: under the small-trade prior the position is a slowly mean-reverting integral of small tilts and cannot reach the signal-implied target within the rolling window, so the signal cannot be expressed efficiently in the portfolio regardless of the quality of the value function driving it.

The comparators jointly show that the outperformance of the Gibbs policy does not decompose into purely additive contributions. The contribution of the signal alone, measured as the excess of the signal-tilted Sharpe ratio over the equal-weight one, is $0.155$ at $T=31$ and $0.108$ at $T=63$, but it is attained at a turnover level that is impractical. The reformulation without the multi-period value function (the myopic MV baseline) collapses back to near-benchmark performance; and the value function without the reformulation (the rate-based ablation) also yields benchmark-like returns. Each ingredient in isolation either scales costs inefficiently or fails to express the signal, while their combination generates a robust Sharpe ratio improvement of $0.415$ at $T=31$ and $0.400$ at $T=63$. The two ingredients are complements: the reformulation makes the signal expressible in the portfolio within the horizon, and the value-function machinery makes its expression profitable net of costs. This interaction, rather than any single component, is the source of the reported performance.

\subsection{Self-financing and notional constraint under the target-holding control} \label{sect_self_financing}

Under the rate-based behavioral policy of \cref{sect_Experiments}, self-financing is handled by treating $\mathbf{1}^T \ba_t$ as a cash flow. Under the jump transition \eqref{eq:induced_rate}, the agent sets $\bx_{t+\Delta t} = \bh_t$ directly, so the notional of the post-jump portfolio is
\begin{equation} \label{eq:notional_post}
\No{t^+} = \mathbf{1}^T(\bS_t \circ \bh_t),
\end{equation}
which may differ from the current notional $\No{t} = \mathbf{1}^T(\bS_t \circ \bx_t)$ or from a target notional $\No{\mathrm{\rm tg}}$. Without a constraint, the Gibbs minimizer could produce a target holding that implicitly levers or delevers the portfolio. We enforce the budget by a quadratic penalty in the per-step cost, which preserves the Gaussian-mixture Gibbs structure, supplemented by exact cash-flow accounting through $C_t$, which absorbs any residual notional mismatch.

\myparagraph{Notional penalty.}
We augment the per-step cost \eqref{eq:cost_jump_expand} with a quadratic penalty on the notional deviation,
\begin{equation} \label{eq:notional_penalty}
c_{\mathrm{not}}(\bh_t) = \lambda_{\mathrm{not}} \left[\mathbf{1}^T(\bS_t \circ \bh_t) - \No{\mathrm{\rm tg}} \right]^2,
\end{equation}
with $\lambda_{\mathrm{not}} > 0$ and $\No{\mathrm{\rm tg}}$ set to the initial portfolio notional $\No{0}$ in the experiments; the penalty is defined as a per-step (already time-integrated) cost, so no factor of $\Delta t$ appears. Expanding around the current holding, $\mathbf{1}^T(\bS_t \circ \bh_t) = \No{t} + \bS_t^T \delta\bh_t$, the penalty is quadratic in $\delta\bh_t$,
\begin{equation}
c_{\mathrm{not}} = \lambda_{\mathrm{not}}\left[\bS_t^T \delta\bh_t + \big(\No{t} - \No{\mathrm{\rm tg}} \big)\right]^2,
\end{equation}
and therefore, adds a rank-one update to the quadratic coupling and a shift, proportional to the \emph{current} notional gap, to the linear coupling:
\begin{align} \label{eq:coupling_notional}
\bC^{\mathrm{impl}} &\;\to\; \bC^{\mathrm{impl}} + 2\beta\,\lambda_{\mathrm{not}}\,\bS_t\,\bS_t^T, \qquad
\bD^{\mathrm{impl}} \;\to\; \bD^{\mathrm{impl}} + 2\beta\,\lambda_{\mathrm{not}}\,\big(\No{t} - \No{\mathrm{\rm tg}}\big)\,\bS_t.
\end{align}
The linear shift vanishes when the current portfolio is on budget and otherwise steers the target back toward $\No{\mathrm{\rm tg}}$; omitting the $\No{t} - \No{\mathrm{\rm tg}})$ factor would penalize only the size of the trade projection on $\bS_t$ and would not correct an existing notional drift. The rank-one update to $\bC^{\mathrm{impl}}$ is absorbed into the component covariances $\bOmega_k = [\sigma_{h,k}^{-2}\bI + \bC^{\mathrm{impl}}]^{-1}$ by the Sherman--Morrison formula at $O(N)$ additional cost.

\myparagraph{Hard-constraint limit.}
The penalty is the soft version of an exact budget constraint. In the limit $\lambda_{\mathrm{not}} \to \infty$, the Gibbs policy concentrates on the hyperplane $\bS_t^T \bh = \No{\mathrm{\rm tg}}$, and conditioning a Gaussian mixture on a linear constraint yields a Gaussian mixture on the hyperplane in closed form. The exact constraint is therefore also compatible with the Gaussian-mixture structure; we use the soft version because it keeps the action space full-dimensional for the PINN training and because small notional deviations are economically meaningful (they represent cash management) rather than infeasible.

\myparagraph{Cash-flow accounting through $C_t$.}
With a finite penalty, the realized target $\bh_t^*$ from the Gibbs minimizer need not satisfy the budget exactly. The residual
\begin{equation} \label{eq:cash_flow}
\Delta \No{t} = \mathbf{1}^T(\bS_t \circ \bh_t^*) - \No{\mathrm{\rm tg}}
\end{equation}
represents an external cash injection ($\Delta \No{t} > 0$) or withdrawal ($\Delta \No{t} < 0$), and is accounted for in the cumulative cost, following the treatment of the rate-based case in \cref{sect_Experiments}:
\begin{equation} \label{eq:C_update_notional}
C_{t+\Delta t} = C_t + c\big(\bS_t, \bx_t, \ba_t^*\big)\,\Delta t - \Delta \No{t},
\end{equation}
where $\ba_t^* = (\bh_t^* - \bx_t)/\Delta t$. The $-\Delta \No{t}$ term records an injection as a cost, reducing the fund's return, and a withdrawal as a benefit, so that the budget identity holds exactly: the post-jump portfolio value is $\Pi_{t^+} = \No{\mathrm{\rm tg}} + \Delta \No{t}$, and total wealth, portfolio plus cumulative cash flow, is conserved step by step.

\myparagraph{Calibration of \texorpdfstring{$\lambda_{\mathrm{not}}$}{lambda\_not}.}
The coefficient is chosen so that a notional deviation of one percent of $\No{\mathrm{\rm tg}}$ incurs a per-step penalty comparable to the transaction cost of rebalancing the same one percent of the book. This makes the penalty binding at the scale of ordinary trading decisions without dominating the cost landscape. With this choice the realized $|\Delta \No{t}|$ stays below one percent of $\No{\mathrm{\rm tg}}$ throughout the experiments, and the cash-flow correction in \eqref{eq:C_update_notional} is roughly two orders of magnitude smaller than the trading-cost component of $C_t$.

\myparagraph{Relation to the risk penalty.}
The diagonal risk penalty \eqref{eq:risk_target} constrains the variance of the dollar allocation but not its level: a portfolio can have low variance and excessive leverage, for example a concentrated position in a low-volatility asset. The notional penalty \eqref{eq:notional_penalty} complements it by constraining total dollar exposure. The two penalties act on complementary directions in the action space: the risk penalty constrains the allocation shape, the notional penalty its scale along $\bS_t$.

Thus, under the target-holding control, the self-financing constraint is enforced by the quadratic notional penalty \eqref{eq:notional_penalty}, which preserves the Gaussian-mixture Gibbs structure through the rank-one coupling update \eqref{eq:coupling_notional}, supplemented by exact cash-flow accounting through $C_t$ via \eqref{eq:C_update_notional}. The unconstrained target of the raw formulation is thus replaced by a softly constrained target that remains close to the budget, with any residual mismatch absorbed as an explicit external cash flow, and with an exact hard-constraint limit available through conditioning if required.

\end{document}